\documentclass[%
reprint,
superscriptaddress,
 amsmath,amssymb,
 aps,
pra,
]{revtex4-2}

\usepackage{xcolor}

\definecolor{FG}{HTML}{ff9b54}

\usepackage{graphicx}
\usepackage{dcolumn}
\usepackage{comment}\usepackage{bm}

\usepackage{url}

\usepackage{hyperref}
\hypersetup{hidelinks}

\usepackage{etoolbox}
\usepackage{soul}

\usepackage[mathlines]{lineno}

\usepackage[normalem]{ulem}
\begin{document}


\title{Geometry-Derived Capacitance Dyadics with Floquet Sideband Correction \\ for Anisotropic Periodic Patch Metasurfaces}
\author{Arghyadeep Pal}
\email{a.pal@student.unisi.it}

\author{Federico Giusti}
\email{federico.giusti2@unisi.it}

\author{Stefano Maci}
\email{macis@dii.unisi.it}

\affiliation{
Department of Information Engineering and Mathematics,
University of Siena,
53100 Siena, Italy
}

\date{November 15, 2026}

\begin{abstract}
 Homogenised sheet models describe periodic metallic patch layers using a scalar sheet capacitance, which is only valid when the effective response is isotropic. An isotropic response exists for unit cells with patch geometries that have rotational symmetry of order three or higher. Lower-symmetry unit cells with irregular or random geometries generically require a full $2\times2$ surface-capacitance dyadic. We present a framework that computes this dyadic directly from the arbitrary gap-connected patch geometry on a unit cell of a hexagonal lattice. The two-dimensional Bravais periodicity fixes the two independent macroscopic loading directions, the connectivity of the gap ensures a full-rank response for the geometries considered here, and the point-group symmetry of the patch constrains the form of the dyadic. A quasi-static boundary value problem on a single periodic cell then determines its independent entries. We use the surface-capacitance dyadic in the anisotropic surface-wave dispersion relation, which includes the leading spatially-dispersive first shell Floquet-sideband correction. For each new patch geometry, the dyadic and the first-shell correction are obtained from a single quasi-static cell problem and from the reconstructed boundary. So anisotropic surface-wave dispersion follows without a per-geometry full-wave eigenmode band sweep or sheet-tensor retrieval. We validate this geometry-to-dispersion chain against full-wave eigenmode simulations across six representative symmetry classes, showing that the correction captures band flattening near the Brillouin-zone boundary and improves agreement relative to the local dyadic model.

\end{abstract}

\maketitle

\section{Introduction}

Metasurfaces (MTSs) are planar arrays of subwavelength scatterers that control the amplitude, phase, and polarization of electromagnetic waves through the engineered response of their unit cells \cite{Yu2011, Kildishev2013, YuCapasso2014, Kuznetsov2016, Chen2020}. Beyond shaping free-space wavefronts, suitably designed metasurfaces can also support and guide bound surface waves \cite{Sun2012}. In a homogenised description, the MTS is represented by an effective sheet boundary condition rather than treated element by element \cite{Kuester2003,Holloway2012}, in which the entire unit-cell geometry is reduced to a small set of effective parameters. For capacitive metallic patch arrays, the isotropic response is described by a scalar sheet capacitance, for which simple and accurate closed forms are available for canonical square grids \cite{Tretyakov2003, Luukkonen2008}. Such a scalar description underlies many analysis and synthesis approaches for surface-wave MTSs, from high-impedance surfaces \cite{Sievenpiper1999} to modulated-MTS antennas \cite{Maci2011}. The more general homogenisation challenge, however, is not merely to reproduce a particular electromagnetic observable, such as a dispersion curve, but to establish a direct mapping from the unit-cell geometry to its effective sheet response. Such a constitutive description can then be employed independently of the particular electromagnetic problem in which the sheet is embedded.

A scalar sheet capacitance, however , presumes that the induced current is parallel to the applied tangential field for every in-plane direction. This holds when the unit cell is sufficiently symmetric \cite{Nye1985,Padilla07}. A patch with three-fold or higher-rotational symmetry responds isotropically, whereas any lower-symmetry cell couples the two tangential field components and requires a full $2\times2$ surface-capacitance dyadic. Anisotropic sheet responses are an established design resource across the spectrum. At microwave frequencies, tensor holographic impedance surfaces shape aperture fields through modulated anisotropy \cite{Fong2010,Quarfoth2013}, and printed-circuit tensor impedance surfaces have been characterised by retrieving the effective sheet tensor from full-wave simulations \cite{Patel2013}, and spatially varying sheet response has been proposed as  a platform for surface-wave transformation optics \cite{Vakil2011}. At optical frequencies, extreme sheet anisotropy produces hyperbolic surface-wave regimes with open isofrequency contours, topological transitions \cite{Hu2020,Chenn2020,Hu2022}, and strongly enhanced light--matter interaction \cite{Poddubny2013,GomezDiaz2015,High2015,Yermakov2015}, a regime realized not only in structured sheets but natively in anisotropic van der Waals materials \cite{Basov2016,Ma2018,Nemilentsau2016}. Self-complementary metasurfaces can also exhibit dual-mode hyperbolicity and supercanalization \cite{MartiniHyperbolicMTSs}. Closed-form dispersion models exist for canonical anisotropic elements such as elliptical patches \cite{Billow2003, Mencagli2015}.

Despite their different formulations, these approaches generally rely on canonical geometries, per-geometry full-wave characterization, or reduced models whose parameters must be determined for the specific structure under consideration. For general patch geometries, including irregular, multiply connected, or random shapes, a direct geometry-to-sheet relation is generally unavailable. A second limitation, common to both scalar and dyadic local sheet models, arises as the Bloch wavevector approaches the Brillouin-zone boundary. In this regime, the higher Floquet orders of the lattice are no longer deeply evanescent and the sheet response becomes nonlocal, i.e., spatially dispersive \cite{Belov2003,CorreasSerrano15,Shastri2023}, with related nonlocal homogenization effects also arising in structured metallic media such as space--time modulated wire media \cite{Yakovlev2026Fresnel}. Such non-locality is increasingly exploited as a design degree of freedom in its  own right \cite{Overvig2020}. The local dispersion branch then departs from the actual Bloch band \cite{Qiu2018Angular}. Multimodal equivalent-network models explicitly account for higher-order Floquet contributions to the response of periodic patch and aperture arrays \cite{RodriguezBerral2015,Mesa2016}. Accurate semianalytical Bloch approaches for square and hexagonal unit cells \cite{GiustiMMTMM,PetekMMTMM} rely on full-wave unit-cell characterization, which the present geometry-driven framework avoids. Such approaches provide an accurate description of the Bloch spectrum, but they do not directly establish a reduced geometry-to-sheet relation from which both the local anisotropic response and its leading spatial-dispersion correction can be obtained.

The practical consequence is that obtaining the surface-wave band diagram for a new irregular patch still typically requires a full-wave eigenmode computation, or an equivalent multi-modal Floquet analysis rebuilt for that geometry. The purpose of this work is to replace that step. Once the patch polygon and its point-group class are given, the local capacitance dyadic and the leading Floquet correction are computed from geometry alone, and the dispersion relation is solved without full-wave simulation. Full-wave eigenmode data are used only for validation.

In this paper, we present a geometry-driven framework that constructs the local anisotropic surface-capacitance response and its leading nonlocal correction directly from a general gap-connected patch geometry on a hexagonal lattice. The framework distinguishes three aspects of the homogenization problem. First, the two-dimensional Bravais periodicity defines two independent macroscopic tangential loading directions, associated with the two primitive translation directions of the lattice. In the periodically identified unit cell, these correspond to the two independent winding directions of the flat-torus representation \cite{Nakahara,Ozawa2019}. For the patch geometries considered here, the gap remains connected across both periodic directions, so that both macroscopic loadings produce non-degenerate electrostatic responses and the resulting surface-capacitance dyadic is full rank. Second, the admissible form of the dyadic is dictated by the point-group symmetry of the complete unit-cell geometry \cite{Nye1985,Dresselhaus2007}, spanning the symmetry classes from the fully symmetric hexagonal cell down to a patch with no point symmetry. Third, the numerical values of the allowed entries follow from a quasi-static boundary-value problem posed on a single periodic cell, based on homogenization theory for composite media \cite{Bergman1978,Milton2002}. These steps establish the local effective sheet response from geometry before any particular propagation problem is introduced.

Surface-wave propagation is then used to develop and validate the electromagnetic consequences of this effective description. The resulting dyadic drives the anisotropic surface-wave dispersion through its projection onto the transverse-electric (TE) polarization of the guided mode. Near the Brillouin-zone boundary, the six Floquet orders of the first reciprocal shell are retained explicitly \cite{RodriguezBerral2015,Mesa2016}. Their coupling to the fundamental guided mode is computed from the reconstructed patch boundary. A boundary Fourier overlap, motivated by coupled-mode theory \cite{Kogelnik1972,Yariv1973,Haus1984}, determines the coupling strength of each retained order, while the discrete turning of the boundary, where the induced edge charge concentrates \cite{Meixner1972}, provides its directional weight. The detuning of each order is determined by its own sheet-loaded decay constant. Mirror symmetries of the patch enter through the coherent pairing of the sidebands, as in the symmetry analysis of photonic crystal bands \cite{Sakoda2004,Ochiai2001}. Thus, both the local anisotropic response and its leading spatial-dispersion correction are derived from geometric quantities rather than fitted to the Bloch spectrum. The resulting chain from geometry to dispersion is validated against full-wave eigenmode simulations for representative structures spanning the symmetry classes considered in this work.

The paper is organized as follows. Section~\ref{sec:geometry_cp} develops the geometry-driven surface-capacitance dyadic. The two-dimensional periodicity defines the macroscopic loading space. The connectivity of the gap ensures a non-degenerate response. The tensor form is constrained by symmetry. A quasi-static cell problem determines the independent entries and their calibration to an equivalent sheet capacitance. Section~\ref{sec:dispersion} uses the effective sheet-capacitance dyadic to derive
the local anisotropic surface-wave dispersion relation and its
quasi-TE reduction. Section~\ref{sec:sideband} develops the geometry-derived
first-shell Floquet correction, incorporating the coupling to
higher-order harmonics through boundary Fourier overlaps and
directional turning measures. Section~\ref{sec:validation} validates the resulting model against full-wave eigenmode simulations across the symmetry classes considered, and Section~\ref{sec:conclusion} concludes.

\section{Geometry-driven capacitance dyadic for arbitrary gap-connected patch-type metasurfaces}
\label{sec:geometry_cp}

We represent the patch-type metasurface (MTS) in the quasi-static
regime by an infinitesimally thin, purely electric sheet at $z=0$.
The macroscopic fields satisfy the jump
conditions~\cite{Kuester2003,Holloway2012}
\begin{subequations}
\label{eq:jump}
\begin{align}
    \widehat{\mathbf z}\times[\mathbf H]
    &=\mathbf J_s
     =\overline{\overline Y}_s\cdot\mathbf E_t,
    \label{eq:jump_H_electric}
    \\
    \widehat{\mathbf z}\times[\mathbf E]
    &=\mathbf 0.
    \label{eq:jump_E_electric}
\end{align}
\end{subequations}
Here $[\mathbf F]=\mathbf F^{+}-\mathbf F^{-}$ denotes the field
jump, with $\mathbf F^{+}$ and $\mathbf F^{-}$ evaluated on the
upper and lower sides of the sheet, respectively.
The quantities $\mathbf J_s$ and $\overline{\overline Y}_s$ are
the equivalent electric surface-current density and the
sheet-admittance dyadic. Equation~\eqref{eq:jump_E_electric}
implies continuity of the tangential electric field,
$\mathbf E_t^{+}=\mathbf E_t^{-}\equiv\mathbf E_t$.

For a lossless reciprocal patch array in the quasi-static
capacitive regime, we write the induced tangential surface
polarization as
\begin{equation}
    \mathbf P_{s,t}
    =\overline{\overline C}_{\mathrm{eqv}}\cdot\mathbf E_t,
    \label{eq:surface_polarization}
\end{equation}
where $\overline{\overline C}_{\mathrm{eqv}}$ is the equivalent
surface-capacitance dyadic, in units of $\mathrm F$.
With the $e^{j\omega t}$ time convention, the associated
polarization current is $\mathbf J_s=j\omega\mathbf P_{s,t}$.
Substitution into Eq.~\eqref{eq:jump_H_electric} gives
\begin{equation}
    \overline{\overline Y}_s
    =j\omega\overline{\overline C}_{\mathrm{eqv}}.
    \label{eq:Ys_Cphys}
\end{equation}
Thus, $\overline{\overline C}_{\mathrm{eqv}}$ is the equivalent
sheet quantity entering the macroscopic electromagnetic boundary
conditions. In the following, we determine this dyadic directly
from the unit-cell geometry for gap-connected patch arrays on a hexagonal lattice.

\begin{figure}[t!]
\centering
\includegraphics[width=\columnwidth]{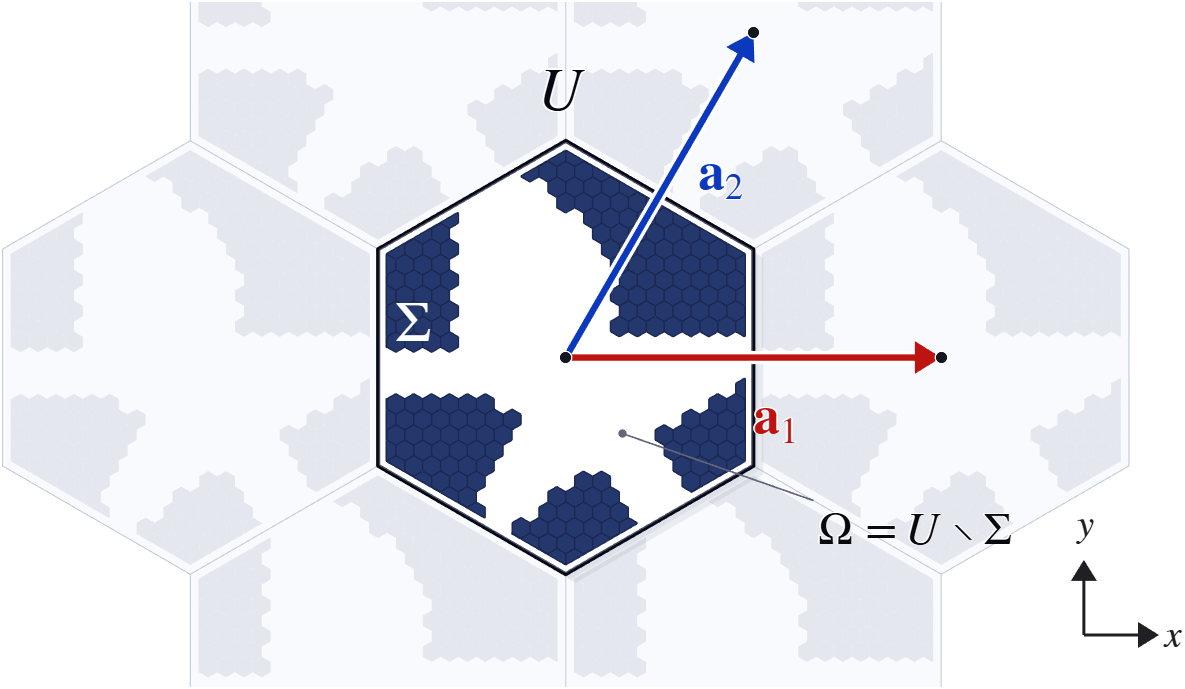}
\caption{An example of the geometry of the periodic patch
layer. The unit cell $U$ (black outline) repeats on a
hexagonal lattice generated by the primitive vectors
$\mathbf a_1$ and $\mathbf a_2$. Inside the unit cell, the
metallic patch $\Sigma$ (dark blue) is a random geometry, and
$\Omega=U\setminus\Sigma$ is the gap.}
\label{fig:geometry}
\end{figure}

An example of the unit-cell geometry considered in this work is
shown in Fig.~\ref{fig:geometry}. The metallic (i.e., patch)
region $\Sigma$ repeats on a hexagonal lattice generated by the
primitive vectors $\mathbf a_1$ and $\mathbf a_2$, and the gap is $\Omega=U\setminus\Sigma$, with $U$ being the
entire unit cell. For a patch having three-fold rotation symmetry
$\left(C_3\right)$ or higher, the sheet response is a scalar
surface capacitance~\cite{Tretyakov2003,Luukkonen2008}. For a
patch with lower symmetry, the induced charge depends on the
direction of the applied tangential field, and the response is
generally described by a symmetric $2\times2$
surface-capacitance dyadic~\cite{Fong2010,Quarfoth2013}.

To determine
$\overline{\overline C}_{\mathrm{eqv}}$ from this geometry, the
quasi-static cell problem first produces the auxiliary
two-dimensional dyadic
\begin{equation}
\overline{\overline C}
=
\begin{bmatrix}
C_{xx} & C_{xy}\\
C_{xy} & C_{yy}
\end{bmatrix},
\label{eq:C_dyadic}
\end{equation}
in units of $\left(\mathrm{F/m}\right)$. Its entries characterize
the electric energy stored in $\Omega$ under tangential
excitation along the two in-plane directions
(Sec.~\ref{subsec:diff_geom}). Because the quasi-static cell
problem is two-dimensional,
$\overline{\overline C}$ is a per-unit-length quantity and is
not yet the equivalent sheet capacitance appearing in
Eq.~\eqref{eq:Ys_Cphys}. The calibration introduced in
Sec.~\ref{subsec:calibration} converts it into equivalent
sheet-capacitance dyadic
$\overline{\overline C}_{\mathrm{eqv}}$, in units of
$\left(\mathrm{F}\right)$, which describes the electromagnetic
response of the patch layer. The auxiliary dyadic
$\overline{\overline C}$ and its calibrated counterpart
$\overline{\overline C}_{\mathrm{eqv}}$ therefore characterize
the effective quasi-static response of the unit-cell geometry,
with $\overline{\overline C}_{\mathrm{eqv}}$ providing the
geometry-dependent input to the local surface-wave dispersion
model of Sec.~\ref{sec:dispersion}.

The derivation relies on three complementary elements. The
two-dimensional Bravais periodicity defines two independent
macroscopic tangential loading directions, while connectivity of
the gap across both periodic directions ensures that
the corresponding quasi-static response is non-degenerate
(Sec.~\ref{subsec:topology}). The point-group symmetry of the
patch constrains the admissible form of the dyadic
(Sec.~\ref{subsec:group_theory}), and a quasi-static
boundary-value problem posed on a single periodic unit cell
determines its independent entries
(Sec.~\ref{subsec:diff_geom}). Once
$\overline{\overline C}$ has been obtained and calibrated,
$\overline{\overline C}_{\mathrm{eqv}}$ completely determines
the local homogenised response. The nonlocal Floquet correction
of Sec.~\ref{sec:sideband} additionally requires the
reconstructed boundary $\partial\Sigma$, which supplies the
first-shell overlaps and directional turning measures.

\subsection{Periodic loading space and non-degenerate capacitance response}
\label{subsec:topology}

\begin{figure}[t!]
  \centering
  \includegraphics[width=\linewidth]{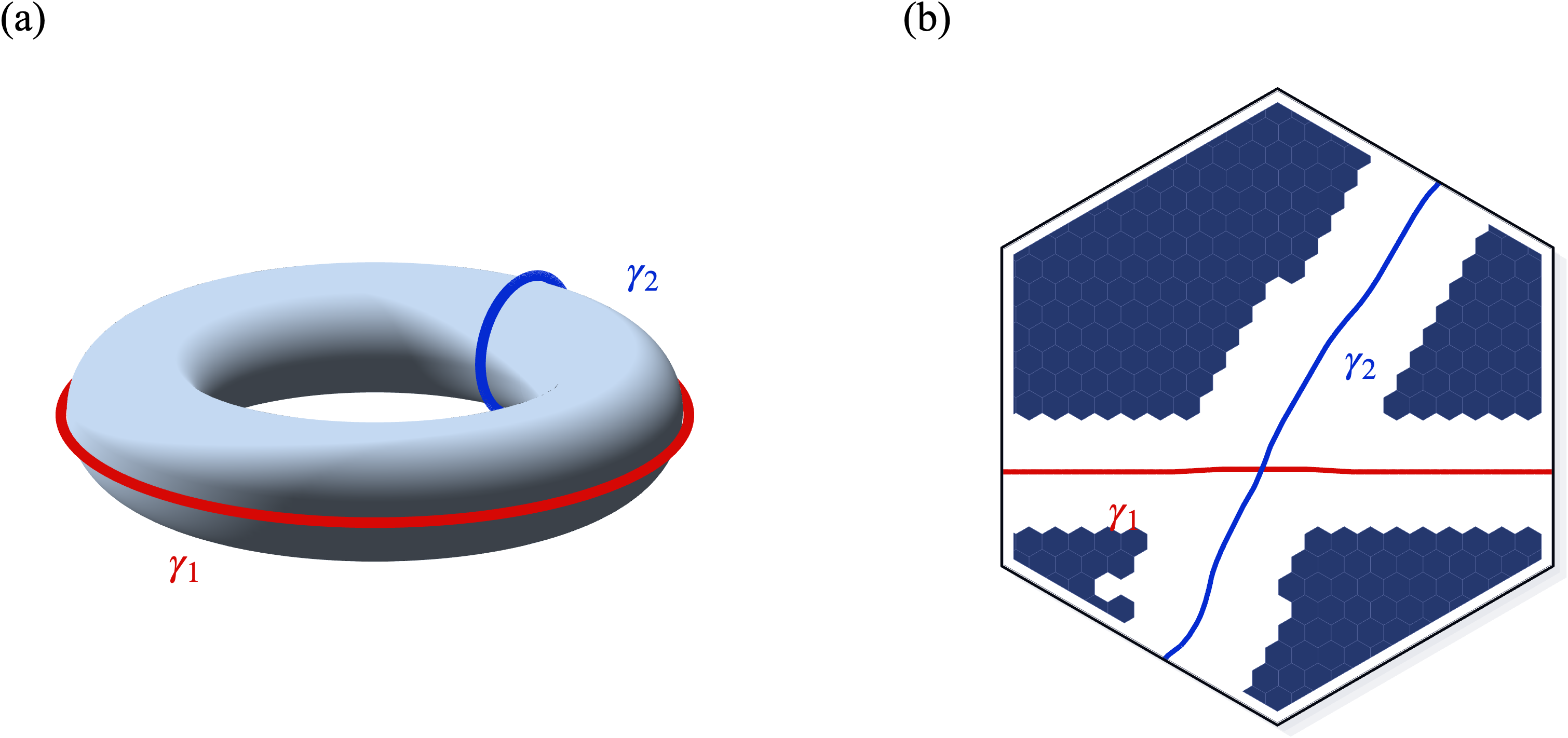}%
  \caption{Periodic loading directions and connectivity of the
  surface-capacitance problem.
  (a)~Identifying opposite edges of the unit cell maps the periodic plane
  onto a flat torus. The two highlighted winding cycles,
  $\gamma_1$ (red) and $\gamma_2$ (blue), correspond to the two primitive
  translation directions of the Bravais lattice and therefore to the two
  independent macroscopic tangential loadings.
  (b)~Representative hexagonal patch geometry for which the
  gap $\Omega=U\setminus\Sigma$ remains connected across both periodic
  directions. The existence of winding gap paths in both
  directions allows independent macroscopic potential variations to be
  sustained and gives a non-degenerate $2\times2$ capacitance response
  for the class of geometries considered here.}
  \label{fig:topology}
\end{figure}

We consider a periodic patch-type MTS invariant under the
two-dimensional Bravais lattice generated by the primitive translation
vectors $\mathbf{a}_1$ and $\mathbf{a}_2$,
\begin{equation}
\mathbf{R}_{n_1n_2}
=
n_1\mathbf{a}_1+n_2\mathbf{a}_2,
\qquad
n_1,n_2\in\mathbb{Z},
\end{equation}
as depicted in Fig.~\ref{fig:geometry}. The infinite periodic structure
can be represented exactly by a single primitive unit cell equipped with
periodic boundary conditions. Identifying opposite edges of the cell
maps the periodically repeated plane onto a flat torus~\cite{Nakahara},
providing a compact representation of the infinite periodic problem
[Fig.~\ref{fig:topology}(a)].

The significance of this construction for the present homogenisation
problem is that a two-dimensional periodic sheet admits two independent
macroscopic tangential loading directions. These may be represented by
the two primitive lattice directions, or equivalently by any two
independent Cartesian directions in the tangential plane. A prescribed
macroscopic electrostatic loading therefore has two independent
components, and the corresponding linear effective response is naturally
represented by a $2\times2$ surface-capacitance dyadic
$\overline{\overline{C}}$.

In the torus representation shown in Fig.~\ref{fig:topology}(a), the primitive periodic directions are
associated with two independent non-contractible winding cycles,
$\gamma_1$ and $\gamma_2$, corresponding respectively to translations
along $\mathbf{a}_1$ and $\mathbf{a}_2$~\cite{Nakahara}. These cycles identify the
two independent macroscopic directions in which a potential variation
may be imposed across the periodically identified cell.

The topology of the gap region enters separately through its
connectivity. Let
\begin{equation}
\Omega = U\setminus\Sigma
\end{equation}
denote the unmetallized area of the unit-cell gap, where $U$ is the complete unit cell and
$\Sigma$ is the metallic patch. A capacitive response to a macroscopic
loading in a given periodic direction requires the gap region to
remain connected across the cell in that direction, so that a
macroscopic potential variation can be sustained. For every geometry
considered in this work, $\Omega$ contains connected winding paths
across both primitive directions, as illustrated schematically in
Fig.~\ref{fig:topology}(b).

Consequently, neither of the two independent macroscopic loadings is
suppressed by the geometry, and the corresponding two column responses
of the capacitance dyadic remain linearly independent. Therefore for this class of
patch geometries considered here,
\begin{equation}
\operatorname{rank}
\left(
\overline{\overline{C}}
\right)
=
2.
\label{eq:rank}
\end{equation}
Thus, the two-dimensional Bravais periodicity fixes the dimension of the
macroscopic tangential loading space, whereas connectivity of the
gap across both periodic directions ensures that the
resulting $2\times2$ capacitive response is non-degenerate.

It is useful to distinguish these macroscopic winding directions from
the complete topology of the gap domain. Additional holes,
indentations, or disconnected metallic components may introduce further
closed paths within $\Omega$, but they do not increase the dimension of
the imposed macroscopic tangential loading space, which remains
two-dimensional. Such geometric features instead modify the solution of
the quasi-static cell problem and therefore the numerical entries of
$\overline{\overline{C}}$.

\begin{table*}[htbp]
\caption{\label{tab:tensor_forms}
Group-theoretically prescribed form of $\overline{\overline{C}}$
for the principal subgroups of $C_{6v}$.
Mirror symmetries pin the eigenvectors of $\overline{\overline{C}}$:
along the mirror axes for $C_{2v}$, and along and perpendicular to the
mirror (orientation $\varphi$) for $C_s$, with the two eigenvalues
free in both cases. For $C_1$ and $C_2$ both eigenvalues and
eigenvector orientation are free, since a two-fold rotation acts in
the plane as $-\overline{\overline{I}}$ and places no constraint on a
symmetric dyadic. Degrees of freedom are independent real scalars for a
lossless reciprocal sheet.
}
\small
\begin{ruledtabular}
\begin{tabular}{lccc}
\textrm{Point group $G$} &
\textrm{Form of $\overline{\overline{C}}$} &
\textrm{Degrees of freedom} &
\textrm{$C(\hat{\mathbf{n}})$} \\
\colrule
$C_{6v},C_{3v},C_6,C_3$
&
$C\,\overline{\overline{I}}$
&
1
&
isotropic \\
$C_{2v}$ (mirrors along the axes)
&
$\mathrm{diag}(C_{xx},C_{yy})$
&
2
&
mirror-axis aligned \\
$C_s$ (mirror at angle $\varphi$)
&
$
\begin{array}{c}
\begin{pmatrix}
C_{xx} & C_{xy}\\
C_{xy} & C_{yy}
\end{pmatrix} \\
\textrm{eigenvectors } \parallel/\perp \textrm{ to mirror}
\end{array}
$
&
2
&
axes parallel/perpendicular
to the mirror\\
$C_1,\,C_2$
&
$\begin{pmatrix}
C_{xx} & C_{xy}\\
C_{xy} & C_{yy}
\end{pmatrix}$
&
3
&
rotated, waisted \\
\end{tabular}
\end{ruledtabular}
\end{table*}

\begin{figure}[htbp]
  \centering
  \includegraphics[width=\columnwidth]
  {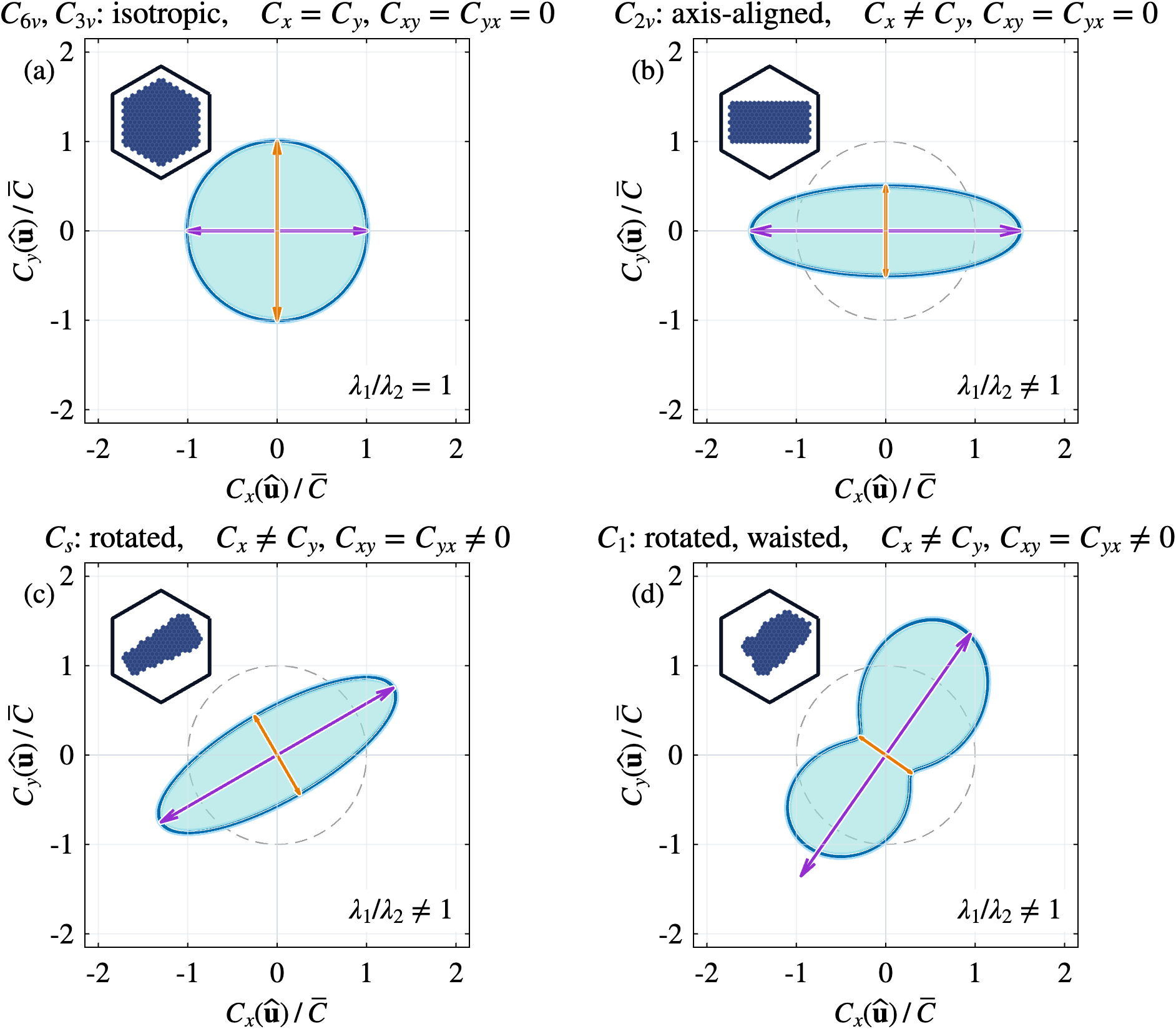}
  \caption{Directional-capacitance indicator
  $C(\hat{\mathbf{u}})
  =
  \hat{\mathbf{u}}\cdot\overline{\overline{C}}\cdot
  \hat{\mathbf{u}}/\bar{C}$,
  normalised by the mean eigenvalue
  $\bar{C}=\tfrac{1}{2}\mathrm{tr}(\overline{\overline{C}})$,
  for the four symmetry classes of Table~\ref{tab:tensor_forms}.
  The light-blue filled region bounded by the solid blue curve is the
  directional indicator; purple and orange arrows mark the two principal
  eigenvectors scaled by their normalised eigenvalues; the dashed grey
  circle is the isotropic reference ($\lambda_1/\lambda_2=1$).
  The inset in each panel shows a representative patch geometry $\Sigma$
  belonging to that symmetry class.
  (a)~$C_{6v}$/$C_{3v}$: full rotational symmetry forces
  $\overline{\overline{C}}=C\overline{\overline{I}}$,
  giving a perfect circle.
  (b)~$C_{2v}$: two distinct eigenvalues with eigenvectors
  aligned to the lattice axes, giving an axis-aligned directional-capacitance indicator.
  (c)~$C_s$: eigenvectors rotated to the mirror-axis orientation
  $\varphi$, giving a rotated directional-capacitance indicator.
  (d)~$C_1$: a non-zero off-diagonal entry $C_{xy}=C_{yx}$
  rotates the principal axes freely; at the strong anisotropy of this
  example the indicator develops a pronounced waist, reflecting the
  fully general three-parameter dyadic.}
  \label{fig:group_theory}
\end{figure}

Equation~\eqref{eq:rank} fixes only the number of independent
macroscopic responses; it does not prescribe the individual entries of
$\overline{\overline{C}}$. The admissible tensor form is instead
constrained by the point-group symmetry of the complete unit-cell
geometry, as discussed in Sec.~\ref{subsec:group_theory} and summarised
in Table~\ref{tab:tensor_forms} and Fig.~\ref{fig:group_theory}.
The numerical values of the independent entries are subsequently
determined by the quasi-static cell problem of
Sec.~\ref{subsec:diff_geom}.

\subsection{Capacitance dyadic and symmetry constraints: group-theoretic tensor form}
\label{subsec:group_theory}

Let $G$ denote the in-plane point group of the complete unit-cell geometry, including both the patch and the underlying lattice. The shape of the patch $\Sigma$ determines the symmetry of the structure. Since the hexagonal lattice can have at most six-fold symmetry, $G$ is always a subgroup of $C_{6v}$. The point groups relevant here are the subgroups of the regular hexagonal cell. The symmetry classes relevant to the present analysis are
$C_1$, $C_s$, $C_2$, $C_{2v}$, $C_3$, $C_{3v}$, $C_6$, and $C_{6v}$. In Sch\"onflies notation~\cite{Dresselhaus2007}, $C_n$ denotes a group generated by an $n$-fold rotation about the axis normal to the MTS, whereas $C_{nv}$ additionally contains vertical mirror symmetries. The groups $C_s$ and $C_1$ denote, respectively, a single-mirror group and the trivial group containing only the identity. Since $C_n$ denotes the cyclic group generated by a rotation of $2\pi/n$ about the axis normal to the sheet, $C_n = \{E, R_{2\pi/n}, \dots, R_{2\pi(n-1)/n}\}$ has order $n$, where $E$ is the identity. Adding a mirror plane containing that axis (a vertical mirror, $\sigma_v$) generates a group $C_{nv}$, of the order $2n$, which contains the $n$ rotations of $C_n$ together with $n$ vertical mirror reflections. The group $C_s = \{E,\sigma\}$ contains the identity and a single mirror, and $C_1 = \{E\}$ is the trivial group with no symmetry beyond the identity. The subgroup chain relevant to a hexagonal lattice is therefore
\begin{equation}
C_1 \subset C_2,\,C_s \subset C_{2v},\,C_3 \subset C_{3v},\,C_6 \subset C_{6v}
\end{equation}
where $C_3 \subset C_6 \subset C_{6v}$ and $C_3 \subset C_{3v} \subset C_{6v}$, so that every group containing a three-fold rotation inherits the isotropy constraint $\overline{\overline{C}} = C\,\overline{\overline{I}}$.

Because $\overline{\overline{C}}$ is an equivalent observable of the sheet, it must be invariant under every symmetry operation \mbox{$g \in G$}, which in the tangential plane acts as a $2\times 2$ orthogonal tensor
representation $\overline{\overline{D}}(g)$~\cite{Nye1985,Haussuhl,Padilla07}
\begin{equation}
  \overline{\overline{D}}(g)\cdot\overline{\overline{C}}\cdot\overline{\overline{D}}^T(g)\, = \overline{\overline{C}}\;,
  \qquad \forall\, g \in G\;.
  \label{eq:invariance}
\end{equation}
where the superscript $T$ denotes transpose. The explicit application of Eq.~\eqref{eq:invariance} to the symmetry
classes considered here, together with the resulting tensor constraints,
is given in Sec.~\ref{sec:S1_symmetry} of the Supplemental Material~\cite{SupplementalMaterial}.

When $G$ contains at least a three-fold rotation \mbox{($G\supseteq C_3$)},
rotational symmetry forces
\begin{equation}
  \overline{\overline{C}} = C\,\overline{\overline{I}}\;,
  \label{eq:schur}
\end{equation}
so that the sheet is isotropic and described by a single scalar $C$, with $\overline{\overline{I}}$ being the identity matrix. 
Reducing $G$ below $C_3$ allows the two eigenvalues of
$\overline{\overline{C}}$ to differ, and the principal axes to rotate, introducing at most two
additional degrees of freedom.
Table~\ref{tab:tensor_forms} lists the dyadic form prescribed by
group theory~\cite{Dresselhaus2007} for each subgroup of $C_{6v}$, together with the
resulting shape of the directional-capacitance indicator
$C(\hat{\mathbf{u}})=\hat{\mathbf{u}}\cdot\overline{\overline{C}}\cdot
\hat{\mathbf{u}}$, where $\hat{\mathbf{u}}=\left(\mathrm{cos}(\theta),\mathrm{sin}(\theta)\right)$ in the in-plane tangential vector along the applied tangential field [see Fig.~\ref{fig:group_theory}].

Note that these are prescribed forms, not values.
The explicit integral that evaluates each independent entry from the
unit-cell geometry is derived in Sec.~\ref{subsec:diff_geom}.

\subsection{Quasi-static computation of the dyadic entries}
\label{subsec:diff_geom}

For canonical geometries, such as circular and elliptical patches, and for arbitrary-shaped elements, the anisotropic MTS impedance can be obtained through reduced analytical or aggregate spectral-domain models~\cite{Mencagli2015, Giusti2026MoM}. Tensor impedance surfaces have also been modeled at the printed-circuit level by retrieving the effective tensor sheet impedance from full-wave simulations~\cite{Patel2013}. In contrast, the cell problem developed here computes the full quasi-static capacitive dyadic directly from the metal topology $\Sigma$, without assuming a canonical patch shape.

To compute the entries of $\overline{\overline{C}}$, we impress
over the whole unit cell a uniform static electric field of
amplitude $\mathrm{E_0}$, directed along $\hat{\mathbf{e}}_x$ or
$\hat{\mathbf{e}}_y$. The impressed field induces surface charge
on the metal patch, which redistributes until the tangential
field on the metal is canceled; each connected metal component
thereby becomes an equipotential surface. In the gap
$\Omega = U \setminus \Sigma$ the total field is the impressed
field distorted by the induced charge,
\begin{equation}
\mathbf{E}^{\mathrm{cell}}_{\xi}(\mathbf{r})
 = \mathrm{E_0} \left[\, \hat{\mathbf{e}}_{\xi}
 + \nabla w_{\xi}(\mathbf{r}) \,\right],
 \qquad \mathbf{r}\in\Omega, \quad \xi\in\{x,y\},
\label{eq:cellfield}
\end{equation}
%
%
in units of $\left(\mathrm{V/m}\right)$. Here $\hat{\mathbf{e}}_{\xi}$ is the
dimensionless unit vector of the imposed direction, and
$w_{\xi}(\mathbf{r})$ is the periodic corrector potential
\cite{Milton2002}, in units of $\left(\mathrm{m}\right)$, whose gradient is
dimensionless; the bracket is therefore the dimensionless
normalised cell field $\mathbf{E}^{\mathrm{cell}}_{\xi}/E_0$,
and the amplitude $E_0$ carries the physical units.
Equivalently, for a general dimensionless imposed direction
$\mathbf{v}=v_x\hat{\mathbf{e}}_x+v_y\hat{\mathbf{e}}_y$, the
associated electrostatic potential is
\begin{equation}
\Phi_{\mathbf{v}}(\mathbf{r})
= -\mathrm{E_0} \left[\,
\mathbf{v}\cdot\mathbf{r}+w_{\mathbf{v}}(\mathbf{r})
\,\right],
\label{eq:cell_potential}
\end{equation}
so that
$\mathbf{E}^{\mathrm{cell}}_{\mathbf{v}}
= -\nabla\Phi_{\mathbf{v}}
= \mathrm{E_0}\left(\mathbf{v}+\nabla w_{\mathbf{v}}\right)$,
and by linearity of the electrostatic problem
$w_{\mathbf{v}} = v_x w_x + v_y w_y$, with
$w_x = w_{\hat{\mathbf{e}}_x}$ and
$w_y = w_{\hat{\mathbf{e}}_y}$ the two basis correctors of
Eq.~\eqref{eq:cellfield}.
The corrector is determined by a constrained electrostatic cell
problem. Let $\Sigma=\bigcup_{m=1}^{N_{\mathrm{c}}}\Sigma_m$
denote the decomposition of the metallic region into
$N_{\mathrm{c}}$ electrically disconnected
perfect-electric-conductor (PEC) components, where connectivity
is understood on the periodically identified cell. The corrector associated with the imposed direction
$\mathbf{v}$ is defined by the minimisation
\begin{equation}
w_{\mathbf{v}}
=
\underset{w\in\mathcal{A}_{\mathbf{v}}}{\operatorname{arg\,min}}
\;
\frac{1}{2|U|}
\int_{\Omega}
\left|\mathbf{v}+\nabla w\right|^{2}\mathrm{d}A ,
\label{eq:corrector_minimisation}
\end{equation}
where $\mathcal{A}_{\mathbf{v}}$ is the admissible space of periodic finite-energy potentials
\begin{equation}
\begin{split}
\mathcal{A}_{\mathbf{v}}
= \Bigl\{
w\in H_{\mathrm{per}}^{1}(\Omega)
:\;
&\mathbf{v}\cdot\mathbf{r}+w(\mathbf{r})=c_m
\;\text{on }\partial\Sigma_m,\\
&\int_{\Omega}w\,\mathrm{d}A=0
\Bigr\}.
\end{split}
\label{eq:corrector_admissible_space}
\end{equation}
 $c_m$ is the normalised equipotential level of the $m$th floating PEC component, $m=1,\ldots,N_{\mathrm{c}}$. The corresponding equivalent potential is $-E_0c_m$. Since the
normalised cell field is $\mathbf{v}+\nabla w_{\mathbf{v}}$, it
is the total affine potential
$\mathbf{v}\cdot\mathbf{r}+w_{\mathbf{v}}$, rather than the
corrector alone, that is constant on each PEC component. Free
variation of the $c_m$ imposes the componentwise
charge-neutrality conditions
\begin{equation}
\int_{\partial\Sigma_m}
\hat{\mathbf{n}}\cdot
\left(\mathbf{v}+\nabla w_{\mathbf{v}}\right)\mathrm{d}s=0,
\qquad m=1,\ldots,N_{\mathrm{c}},
\label{eq:component_neutrality}
\end{equation}
so that each floating component polarises without acquiring net
charge. The Euler--Lagrange equations,
affine-periodic boundary conditions, discrete potential basis,
gauge fixing, and triangular-graph discretisation used in the
numerical solution are detailed in
Sec.~\ref{sec:supp_corrector} of the Supplemental Material~\cite{SupplementalMaterial}.

For the free-standing patch arrays considered here, the gap
region \(\Omega\) is filled with air, of permittivity
\(\varepsilon_0\). Multiplying the minimum value of the
functional in Eq.~\eqref{eq:corrector_minimisation} by
\(\varepsilon_0 E_0^2\) gives the cell-averaged electrostatic
energy density associated with the auxiliary two-dimensional
cell problem,
\begin{equation}
W_e(\mathbf{v})
= \frac{\varepsilon_{0}E_0^{2}}{2|U|}
\int_{\Omega}
\bigl|\mathbf{v} + \nabla w_{\mathbf{v}}\bigr|^2 \mathrm{d}A
= \tfrac{1}{2}\,E_0^{2}\,
\mathbf{v}\cdot\overline{\overline{C}}\cdot\mathbf{v}\,,
\label{eq:energy}
\end{equation}
in units of $\left(\mathrm{J/m^3}\right)$. Identifying the bilinear form, and writing the entries in terms of the dimensionless normalised cell fields $\mathbf{E}^{\mathrm{cell}}_{\xi}/\mathrm{E_0} = \hat{\mathbf{e}}_{\xi}+\nabla w_{\xi}$ of Eq.~\eqref{eq:cellfield} yields the capacitance integrals
\begin{equation}
C_{\xi\eta}[\Sigma]
=
\frac{\varepsilon_{0}}{|U|}
\int_{\Omega}
\bigl(\hat{\mathbf{e}}_{\xi}+\nabla w_{\xi}\bigr)
\cdot
\bigl(\hat{\mathbf{e}}_{\eta}+\nabla w_{\eta}\bigr)
\,\mathrm{d}A,
\;
\xi,\eta\in\{x,y\},
\label{eq:cap_integral}
\end{equation}
in which the impressed amplitude $\mathrm{E_0}$ has canceled. For a patch array printed between air and a semi-infinite dielectric substrate, $\varepsilon_0$ in Eqs.~\eqref{eq:energy} and~\eqref{eq:cap_integral} may be replaced by the quasi-static effective permittivity

\begin{equation}
\varepsilon_{\mathrm{eff}} = \frac{\varepsilon^{+}+\varepsilon^{-}}{2} = \frac{\varepsilon_0+\varepsilon_s}{2}=\varepsilon_{0}\frac{1+\varepsilon_{r,s}}{2}
\label{eq:eps_eff}
\end{equation}
where $\varepsilon_s=\varepsilon_0\varepsilon_{r,s}$ is the substrate permittivity.
\begin{figure}[t]
  \centering
  \includegraphics[width=\columnwidth]{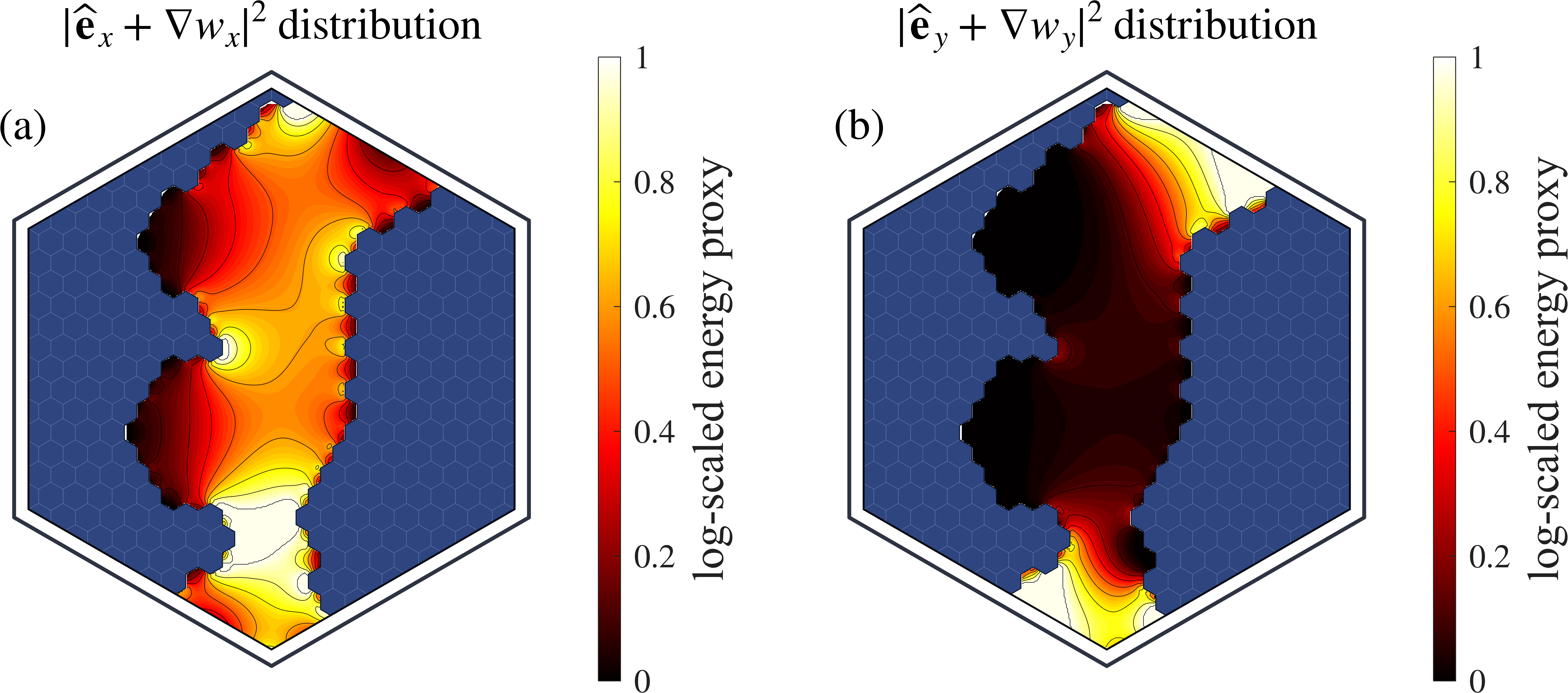}
  \caption{Normalised energy-density distributions for a
  representative irregular patch geometry $\Sigma$, computed from
  Eqs.~(\ref{eq:cellfield})--(\ref{eq:energy}).
The gap $\Omega$ is shown; the deep-blue
tessellated region is the metal patch $\Sigma$.
(a)~Distribution of $\vert \widehat{\mathbf e}_x+\nabla w_x \vert^2$ for the $x$-directed loading.
(b)~Distribution of $\vert \widehat{\mathbf e}_y+\nabla w_y \vert^2$ for the $y$-directed loading.
Both distributions are normalised and displayed on logarithmic colour scales. Field concentration near narrow opening and irregular patch edges produces enhanced energy density, increasing the local contributions to the corresponding diagonal capacitance response. The different spatial distributions under the two loading directions illustrate the direction-dependent response of the irregular geometry, while $C_{xy}=C_{yx}$ is determined by the
cross-field overlap in Eq.~\eqref{eq:cap_integral}}
  \label{fig:corrector}
\end{figure}
The dyadic defined by Eq.~\eqref{eq:cap_integral} is symmetric and positive definite, with units of $\left(\mathrm{F/m}\right)$. It is positive definite for the gap-connected geometries of  Sec.~\ref{subsec:topology}, and it reproduces the tensor form of
Table~\ref{tab:tensor_forms} from the symmetry of $\Sigma$. Figures~\ref{fig:corrector}(a) and (b) show the resulting energy-density distributions $|\hat{\mathbf{e}}_x + \nabla w_x|^2$ and $|\hat{\mathbf{e}}_y + \nabla w_y|^2$. The capacitance integrals over the gap region $\Omega$ in Eq.~\ref{eq:cap_integral}, give the entries of the capacitance dyad $\overline{\overline{C}}$. The diagonal entries $C_{xx}$ and $C_{yy}$ are the self-energy integrals of the $x$- and $y$-directed cell fields with themselves, while the off-diagonal entries $C_{xy}=C_{yx}$ is determined from the cross-field overlap.
Equation~\eqref{eq:cap_integral} is evaluated numerically for arbitrary
patch geometries in Sec.~\ref{sec:validation}.

\subsection{From geometric integral to equivalent sheet capacitance}
\label{subsec:calibration}
                 
\begin{figure}[t]
  \centering
  \includegraphics[width=\columnwidth]{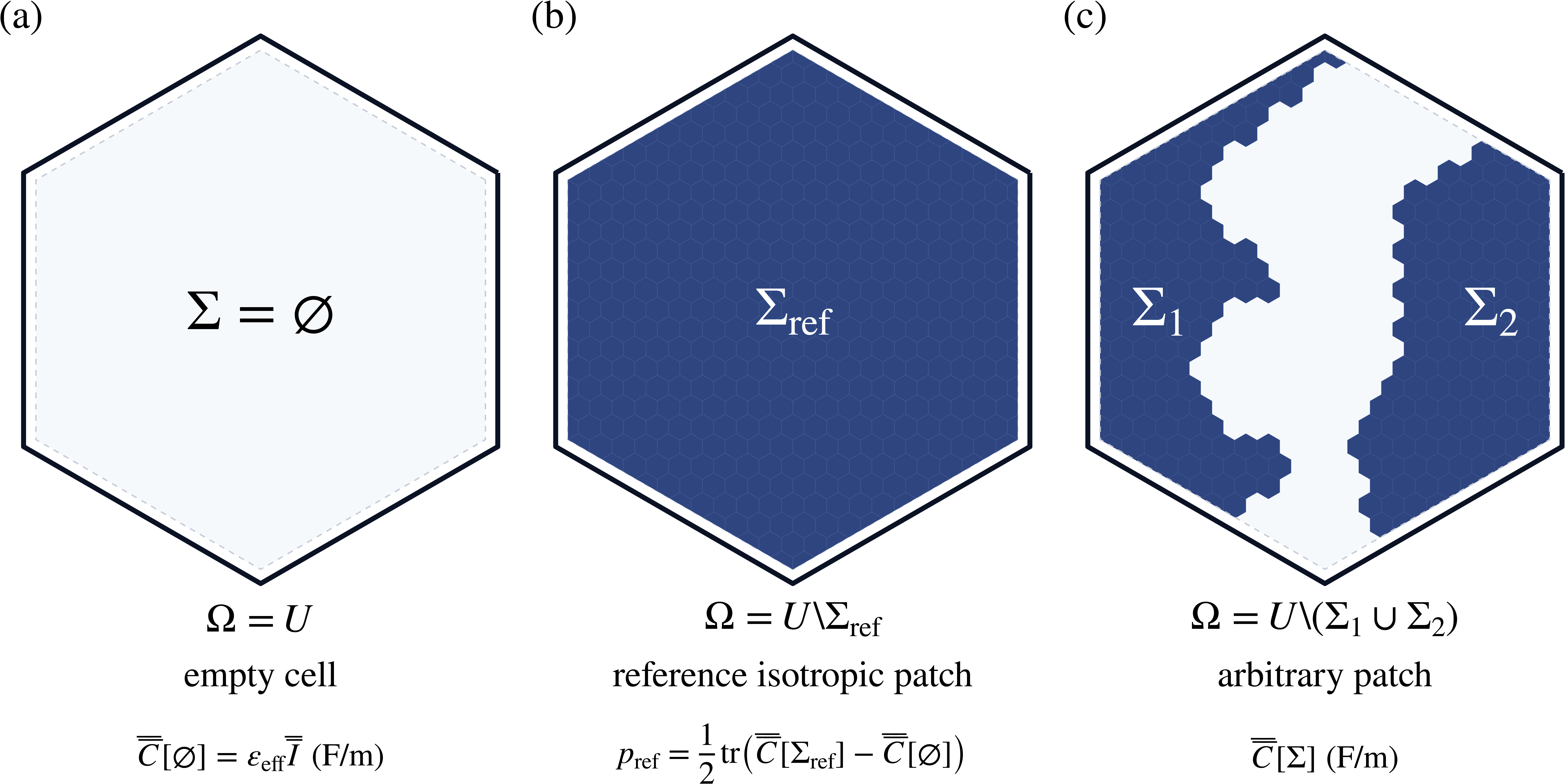}
  \caption{Cell and reference geometries used in the conversion from the geometric electrostatic response to the equivalent sheet capacitance.
  (a)~Empty cell: the cell integral gives the isotropic background
  contribution $\overline{\overline{C}}[\emptyset]=\varepsilon_0
  \overline{\overline{I}}$.
  (b)~Regular-hexagonal reference patch $\Sigma_{\mathrm{ref}}$, used to
  define the scalar reference excess $p_{\mathrm{ref}}$ and the equivalent reference scale.
  (c)~Arbitrary gap-connected patch geometry $\Sigma$, here comprising two electrically disconnected PEC components $\Sigma_1\cup\Sigma_2$. Its auxiliary cell response is converted into a model sheet capacitance using the same reference scale.}
  \label{fig:calibration_geometries}
\end{figure}

Equation~\eqref{eq:cap_integral} defines the symmetric cell-response
dyadic $\overline{\overline{C}}[\Sigma]$, in units of
$\left(\mathrm{F/m}\right)$, for the auxiliary two-dimensional electrostatic
problem. Its construction uses the periodic-corrector framework
of homogenization~\cite{Allaire1992} and the variational-energy
description of effective tensors~\cite{Milton2002},
with the PEC constraints specified in the preceding subsection.
The cell response includes the isotropic background of the empty
cell as well as the metal-induced contribution. It is distinct
from the equivalent sheet capacitance dyadic
$\overline{\overline{C}}_{\mathrm{eqv}}$, in units of
$\left(\mathrm{F}\right)$, which enters the transmission-line model: equivalent sheet
coefficients belong to an interface homogenization
description~\cite{HollowayKuester2016}. We first subtract the
background and normalise the remaining cell response, then
introduce an explicit reference-energy matching assumption to
set its equivalent sheet scale.

Figure~\ref{fig:calibration_geometries} shows the empty cell,
the reference patch $\Sigma_{\mathrm{ref}}$, and the target
geometry $\Sigma$. In the homogeneous free-space background,
the empty-cell correctors vanish, giving
\begin{equation}
\overline{\overline{C}}[\emptyset]
=\varepsilon_0\overline{\overline{I}}.
\label{eq:empty_cell_response}
\end{equation}
We define the metal-induced excess by
\begin{equation}
\overline{\overline{C}}_{\mathrm{exc}}[\Sigma]
=\overline{\overline{C}}[\Sigma]
-\overline{\overline{C}}[\emptyset].
\label{eq:PC_def}
\end{equation}
Thus $\overline{\overline{C}}_{\mathrm{exc}}[\emptyset]
=\overline{\overline{0}}$. Subtracting an isotropic dyadic
preserves the principal directions of the raw cell response,
but generally changes its eigenvalue ratio.

For the isolated PEC components considered here, the equipotential
boundary condition permits each corrector to be extended
continuously into the metal, with
\[
\nabla w_\xi=-\hat{\mathbf e}_\xi
\qquad \text{in }\Sigma.
\]
The normalised total field
$\hat{\mathbf e}_\xi+\nabla w_\xi$ therefore vanishes there,
so the integration domain in Eq.~\eqref{eq:cap_integral}
can be extended from $\Omega$ to the complete cell $U$.
Periodicity of the extended corrector gives
\[
\int_U\nabla w_\xi\,\mathrm{d}A=\mathbf0.
\]
On expanding the integrand, both cross terms consequently vanish,
while the constant-field term gives the empty-cell contribution.
Subtracting it according to Eq.~\eqref{eq:PC_def} yields
\begin{equation}
\left(
\overline{\overline{C}}_{\mathrm{exc}}[\Sigma]
\right)_{\xi\eta}
=\frac{\varepsilon_0}{|U|}
\int_U\nabla w_\xi\cdot\nabla w_\eta\,\mathrm{d}A,
\qquad \xi,\eta\in\{x,y\}.
\label{eq:calibration_excess_gram}
\end{equation}
This integral includes the corrector extension inside the metal;
its domain is $U$, not only the gap $\Omega$.
For every real in-plane vector
$\mathbf q=q_x\hat{\mathbf x}+q_y\hat{\mathbf y}$,
\begin{equation}
\mathbf q\cdot\overline{\overline{C}}_{\mathrm{exc}}\cdot\mathbf q
=\frac{\varepsilon_0}{|U|}
\int_U\left|q_x\nabla w_x+q_y\nabla w_y\right|^2
\mathrm{d}A\geq0.
\label{eq:calibration_excess_positive}
\end{equation}
The excess is therefore symmetric and positive semidefinite
within the stated two-dimensional problem.

A charge-free gap cavity enclosed entirely by one connected
PEC component has constant boundary potential and hence zero field
by electrostatic uniqueness~\cite{Jackson1998}.
Filling only such cavities $H_k$ therefore leaves the auxiliary
cell response unchanged:
\begin{equation}
\Sigma_{\mathrm{eff}}=\Sigma\cup\bigcup_k H_k,
\qquad
\overline{\overline{C}}[\Sigma_{\mathrm{eff}}]
=\overline{\overline{C}}[\Sigma].
\label{eq:calibration_closed_holes}
\end{equation}
We retain $\Sigma$ below. This equivalence is restricted to the
auxiliary two-dimensional problem: open notches and
boundary-connected slots are not filled, and apertures in a
physically thin sheets are not thereby equivalent to solid metal.

\begin{figure}[t]
  \centering
  \includegraphics[width=\columnwidth]{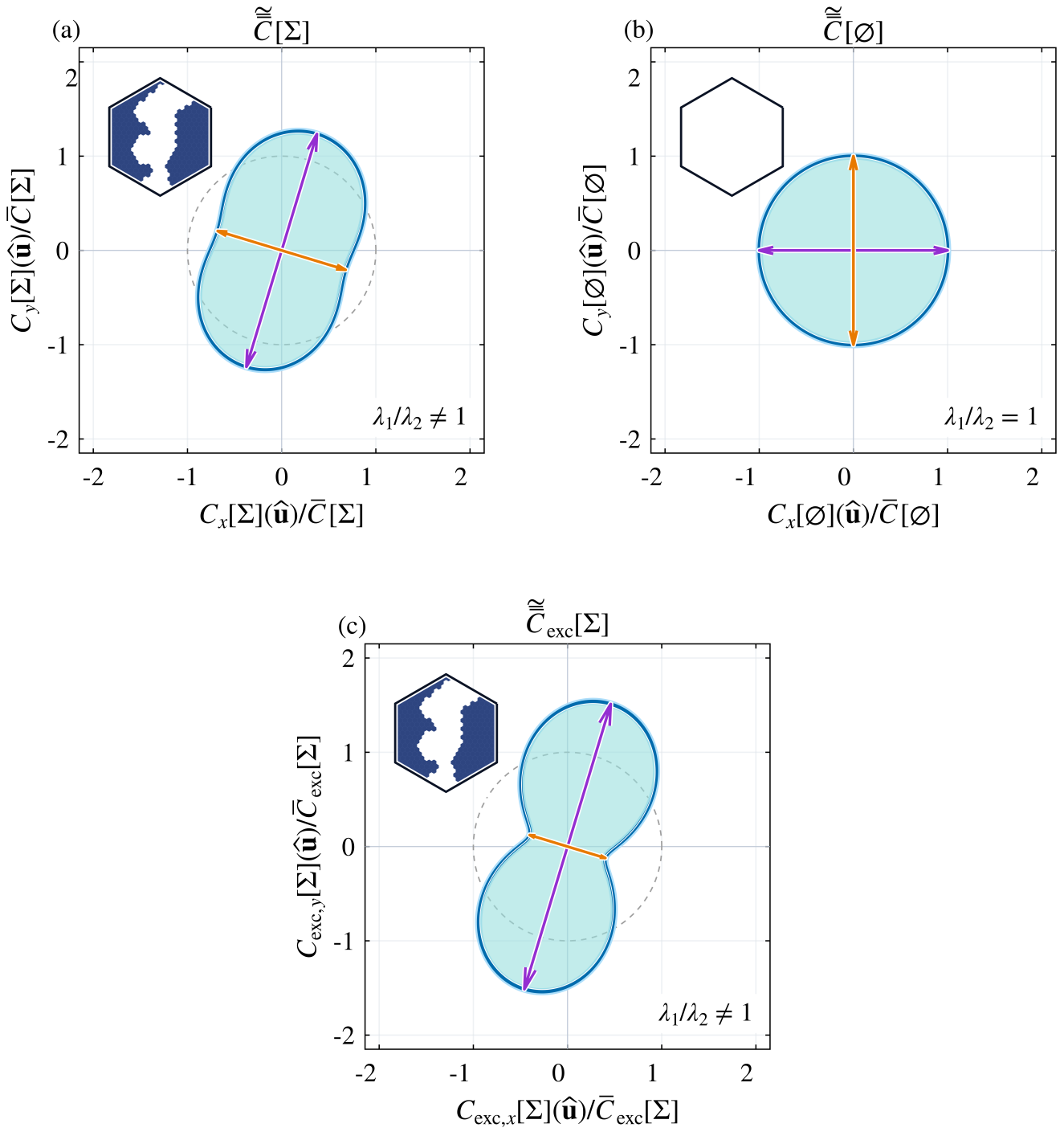}
  \caption{Empty-cell subtraction at the dyadic level
  [Eq.~\eqref{eq:PC_def}]. Each dyadic is represented by its
  normalised directional-capacitance indicator and compared with
  the dashed isotropic reference circle. Normalisation by one half
  of the trace removes the absolute scale and highlights the
  anisotropy and principal axes.
  (a)~Raw cell-response dyadic
  $\widetilde{\overline{\overline C}}[\Sigma]$.
  (b)~Empty-cell response
  $\widetilde{\overline{\overline C}}[\emptyset]
  =\overline{\overline I}$, which coincides with the reference circle.
  (c)~Excess dyadic
  $\widetilde{\overline{\overline C}}_{\mathrm{exc}}[\Sigma]$,
  obtained by subtracting the isotropic background before
  normalisation. Its principal directions coincide with those of
  the raw response, whereas its eigenvalue ratio generally differs.}
  \label{fig:empty_cell_subtraction}
\end{figure}

To display the anisotropy, each nonzero dyadic
\mbox{$\overline{\overline M}\in
\{\overline{\overline{C}}[\Sigma],
\overline{\overline{C}}[\emptyset],
\overline{\overline{C}}_{\mathrm{exc}}[\Sigma]\}$}
is represented by the normalized directional-capacitance indicator
\begin{equation}
\widetilde M(\hat{\mathbf u})
=\hat{\mathbf u}\cdot
\widetilde{\overline{\overline M}}\cdot\hat{\mathbf u},
\qquad
\widetilde{\overline{\overline M}}
=\frac{\overline{\overline M}}
{\tfrac12\operatorname{tr}(\overline{\overline M})},
\label{eq:calibration_directional_indicator}
\end{equation}
plotted as the radial coordinate for
\begin{equation}
\hat{\mathbf u}(\theta)
=\cos\theta\,\hat{\mathbf x}+\sin\theta\,\hat{\mathbf y}.
\label{eq:calibration_indicator_direction}
\end{equation}
Each nonzero dyadic above is positive semi-definite and therefore
has positive trace. This trace normalisation removes the absolute
scale while preserving the eigenvalue ratio and principal axes.
Figure~\ref{fig:empty_cell_subtraction} compares the raw response,
the analytically isotropic empty-cell background, and the excess.

The target and reference cell problems use the same lattice,
background permittivity, and boundary conditions.
The regular-hexagonal reference has an isotropic excess by the
symmetry of the complete cell, consistent with the tensor
constraints in Table~\ref{tab:tensor_forms}. Define
\begin{equation}
\begin{aligned}
\overline{\overline S}[\Sigma]
&=\frac{\overline{\overline C}_{\mathrm{exc}}[\Sigma]}
{p_{\mathrm{ref}}},\\
p_{\mathrm{ref}}
&=\tfrac12\operatorname{tr}\!\left(
\overline{\overline C}_{\mathrm{exc}}[\Sigma_{\mathrm{ref}}]
\right)>0.
\end{aligned}
\label{eq:S_def}
\end{equation}
Here $p_{\mathrm{ref}}$ has units of $\left(\mathrm{F/m}\right)$, whereas
$\overline{\overline S}$ is dimensionless. Consequently,
\begin{equation}
\overline{\overline S}[\Sigma_{\mathrm{ref}}]
=\overline{\overline I},
\qquad
\overline{\overline S}[\emptyset]
=\overline{\overline0}.
\label{eq:S_anchors}
\end{equation}
These are exact anchors of the continuum normalisation; their
numerical recovery requires consistent target and reference
discretisations.

The conversion to equivalent sheet units requires an additional
modeling assumption. For a real tangential loading $E_0\mathbf v$,
where $E_0$ has units of $\left(\mathrm{V/m}\right)$ and $\mathbf v$ is a
dimensionless in-plane vector, the auxiliary excess energy density
is
\begin{equation}
\Delta W_{\mathrm e}(\mathbf v)
=\frac{E_0^2}{2}\,
\mathbf v\cdot\overline{\overline C}_{\mathrm{exc}}
\cdot\mathbf v,
\label{eq:calibration_excess_energy}
\end{equation}
in units of $\left(\mathrm{J/m^3}\right)$. We approximate the sheet energy
per unit area by a common conversion length
$\ell_{\mathrm{ref}}>0$ multiplying this excess energy density:
\begin{equation}
\begin{aligned}
\mathcal U_{\mathrm s}^{\mathrm{model}}(\mathbf v)
&=\frac{E_0^2}{2}\,
\mathbf v\cdot\overline{\overline C}_{\mathrm{eqv}}
\cdot\mathbf v\\
&=\ell_{\mathrm{ref}}\,\Delta W_{\mathrm e}(\mathbf v).
\end{aligned}
\label{eq:calibration_sheet_energy_matching}
\end{equation}
The same macroscopic tangential-field convention is used on both
sides. This reference-energy ansatz assumes one scalar conversion length for both in plane directions and all target geometries in the selected family. It does not follow from the
two-dimensional cell problem or the normalisation anchors alone,
and is not a substitute for an equivalent interface-cell
derivation~\cite{HollowayKuester2016}.

Since Eq.~\eqref{eq:calibration_sheet_energy_matching} holds for
every $\mathbf v$, equality of the quadratic forms gives
$\overline{\overline C}_{\mathrm{eqv}}
=\ell_{\mathrm{ref}}\overline{\overline C}_{\mathrm{exc}}$.
Matching the isotropic reference to its prescribed equivalent sheet
capacitance $C_{\mathrm{ref}}^{\mathrm{eqv}}$ fixes
\begin{equation}
\ell_{\mathrm{ref}}p_{\mathrm{ref}}
=C_{\mathrm{ref}}^{\mathrm{eqv}},
\qquad
\ell_{\mathrm{ref}}
=\frac{C_{\mathrm{ref}}^{\mathrm{eqv}}}{p_{\mathrm{ref}}}.
\label{eq:calibration_conversion_length}
\end{equation}
The resulting equivalent sheet-capacitance model is
\begin{equation}
\begin{aligned}
\overline{\overline C}_{\mathrm{eqv}}[\Sigma]
&=C_{\mathrm{ref}}^{\mathrm{eqv}}\,
\overline{\overline S}[\Sigma]\\
&=\frac{C_{\mathrm{ref}}^{\mathrm{eqv}}}{p_{\mathrm{ref}}}
\left(\overline{\overline C}[\Sigma]
-\varepsilon_0\overline{\overline I}\right).
\end{aligned}
\label{eq:Cp_disp}
\end{equation}
The ratio in Eq.~\eqref{eq:calibration_conversion_length} has units
of length $\left(\mathrm{m}\right)$, so Eq.~\eqref{eq:Cp_disp} has units of $\left(\mathrm{F}\right)$. We provide the detailed derivation of the excess dyadic and
reference normalization in Sec.~\ref{sec:supp_calibration_proof} of the Supplemental Material~\cite{SupplementalMaterial},
where we also establish the uniqueness of the sheet calibration
within the assumed scalar-affine class and clarify the scope
of the common conversion-length assumption. No nonlinear eigenvalue transformation is introduced. In the orthonormal principal basis of the symmetric excess
dyadic~\cite{HighamMatrices},
\begin{equation}
\overline{\overline S}[\Sigma]
=\sum_{i=1}^{2}s_i\,
\hat{\mathbf v}_i\otimes\hat{\mathbf v}_i,
\qquad s_i\geq0,
\label{eq:spectral_decomposition}
\end{equation}
the equivalent principal capacitances are
$C_i^{\mathrm{eqv}}=C_{\mathrm{ref}}^{\mathrm{eqv}}s_i$.
Positive reference scaling therefore preserves the excess dyadic's
principal directions, its eigenvalue ratio when both eigenvalues
are positive, its nonnegative energy, and the symmetry constraints
of Table~\ref{tab:tensor_forms}.

The reference value is an independent input. For the
free-standing regular-hexagonal array, we retain the reference
\begin{equation}
\begin{aligned}
C_{\mathrm{ref}}^{\mathrm{eqv}}&=2\varepsilon_0 A_{\mathrm{iso}},\\
A_{\mathrm{iso}}
&=\frac{\sqrt3\,D_h}{2\pi}
\ln\!\left[\csc\!\left(\frac{\pi g}{2D_h}\right)\right],
\end{aligned}
\label{eq:isotropic_reference_capacitance}
\end{equation}
where $D_h$ and $g$ denote the period parameter and uniform
interpatch gap of the reference geometry, respectively.
The logarithmic gap dependence is motivated by the canonical
electrically dense grid models of Luukkonen
\emph{et al.}~\cite{Luukkonen2008}, which treat strips and square
patches. The hexagonal prefactor in
Eq.~\eqref{eq:isotropic_reference_capacitance} is part of the
specified reference approximation, not an exact consequence of
that square-patch result. Its accuracy must be established
independently for the reference array. The relation
$C_{\mathrm{ref}}^{\mathrm{eqv}}=2\varepsilon_0 A_{\mathrm{iso}}$
sets the free-space grid-length convention used in the dispersion
model. A separately computed equivalent reference value can replace
Eq.~\eqref{eq:isotropic_reference_capacitance} without changing the
normalisation in Eq.~\eqref{eq:S_def}.

For any fixed positive reference value, the complete conversion
preserves the two anchors:
\begin{equation}
\overline{\overline C}_{\mathrm{eqv}}[\emptyset]
=\overline{\overline0},
\qquad
\overline{\overline C}_{\mathrm{eqv}}[\Sigma_{\mathrm{ref}}]
=C_{\mathrm{ref}}^{\mathrm{eqv}}\overline{\overline I}.
\label{eq:physical_capacitance_anchors}
\end{equation}
The construction combines the energy-derived excess dyadic with
one common reference scale, without eigenvalue compression or
geometry-dependent conversion factors. Symmetry, nonnegative
excess, and closed-cavity equivalence are properties of the
auxiliary cell problem. The accuracy of the reference
scale and of the common conversion length remains separate from
those exact two-dimensional results.

\section{Quasi-static surface-wave dispersion}
\label{sec:dispersion}

The equivalent sheet-capacitance dyadic
$\overline{\overline{C}}_{\mathrm{eqv}}[\Sigma]$ from Eq.~\eqref{eq:Cp_disp} enters the electromagnetic model through the averaged boundary condition at the MTS plane at $z=0$.
In the quasi-static limit, the grid admittance for a free-standing
patch array is purely capacitive~\cite{Luukkonen2008}. Hence, the homogenised sheet-admittance dyadic is
\begin{equation}
\overline{\overline{Y}}_{\mathrm{s}}(\omega) = j\omega\ \overline{\overline{C}}_{\mathrm{eqv}}[\Sigma].
\label{eq:sheet_admittance_dyad}
\end{equation}
For a transverse-electric (TE) surface wave propagating along $\mathbf{k}_t = k_x\hat{\mathbf{x}}+k_y\hat{\mathbf{y}}$, with $\hat{\mathbf{k}}_t=\mathbf{k}_t/|\mathbf{k}_t|$, the in-plane electric field is directed along $\hat{\mathbf{e}}_{\mathrm{TE}}=(\hat{z}\times\mathbf{k}_{t})/k_t$. The sheet-capacitance response sampled by the modal electric field is therefore the quadratic projection
$\hat{\mathbf{e}}_{\mathrm{TE}}\cdot\overline{\overline{C}}_{\mathrm{eqv}}\cdot\hat{\mathbf{e}}_{\mathrm{TE}}$ of the $\overline{\overline{C}}_{\mathrm{eqv}}[\Sigma]$ dyadic.

\subsection{Transverse resonance conditions}
\label{subsec:TRC}

For an isotropic sheet, or for propagation along a principal axis of an anisotropic sheet, the TE and TM polarisations are uncoupled. In this limit, a bound surface wave with transverse decay constant
$\gamma = \sqrt{k_t^2-k_0^2}$ \mbox{($k_t = |\mathbf{k}_t|; k_t > k_0$)} satisfies the two-sided scalar transverse-resonance condition on a free-standing sheet~\cite{Tretyakov2003,Luukkonen2008}
\begin{equation}
  2Y_0^{p} + Y_g^{p} = 0,
  \qquad
  p \in \{\mathrm{TE},\,\mathrm{TM}\},
  \label{eq:TRC}
\end{equation}
where $Y_0^p$ is the one-sided free-space line admittance and
$Y_g^p$ is the grid admittance for the transverse-electric (TE) or transverse-magnetic (TM) polarisation. Equation ~\eqref{eq:TRC} is the factorised scalar form of the more general anisotropic matrix condition introduced in Sec.~\ref{subsec:aniso_disp}.

For an uncoupled TM mode supported by a scalar capacitive patch array, both admittance of a capacitive sheet in Eq.~\eqref{eq:TRC} are positive imaginary%
\[
Y_0^{\mathrm{TM}}
=
j\frac{\omega\varepsilon_0}{\gamma},
\qquad
Y_g^{\mathrm{TM}}
=
j\frac{2\alpha_g}{\eta_0}.
\]
Consequently the TM condition cannot be equated to 0, and hence no bound TM solution exists. The uncoupled isotropic or principal-axis capacitive sheet model therefore supports a bound TE mode but not a bound TM mode~\cite{Sievenpiper1999}.

For the TE polarisation, whose electric field is perpendicular to
$\hat{\mathbf{k}}_t$), the free-space line admittance of a bound wave is
\[
Y_0^{\mathrm{TE}}
=
-j\frac{\gamma}{\omega\mu_0}.
\]
The Luukkonen grid admittance for a capacitive patch array
in free space is~\cite{Luukkonen2008} 
\begin{equation}
  Y_g^{\mathrm{TE}}
  = j\frac{2\alpha_g}{\eta_0}{F_L(k_t)},
  \label{eq:Yg_TE}
\end{equation}
where $F_L(k_t)= \left[1 - k_t^2/(2k_0^2)\right]$ is the $k_t$-dependent correction factor, and
$\alpha_g = k_0 A_g$ is the
dimensionless grid parameter, with $A_g$~(m) the Luukkonen grid length
of the patch array.
Substituting~\eqref{eq:Yg_TE} into~\eqref{eq:TRC} and using $\eta_0 = \omega\mu_0/k_0$
gives the scalar TE dispersion condition
\begin{equation}
  \gamma = k_0\,\alpha_g\,F_L(k_t).
  \label{eq:TE_disp_scalar}
\end{equation}

Equation~\eqref{eq:TE_disp_scalar} is valid in the window $k_0 < k_t < \sqrt{2}\,k_0$ where $F_L(k_t)>0$. Substituting $k_t^2 = k_0^2 + \gamma^2$ and solving the resulting quadratic in $\gamma$ gives the closed-form root
\begin{equation}
  \gamma
  =
  \frac{k_0}{\alpha_g}
  \left(
  \sqrt{1+\alpha_g^2}-1
  \right),
  \qquad
  k_t
  =
  k_0\sqrt{1+\frac{\gamma^2}{k_0^2}}.
  \label{eq:linear_disp}
\end{equation}
The solution of Eq.~\eqref{eq:linear_disp} automatically satisfies $\gamma < k_0$ and hence remains inside the validity window. Written in terms of the geometry-fixed grid length, 
\[
\gamma
=
\frac{
\sqrt{1+(k_0A_g)^2}-1
}{
A_g
}.
\]
The branch remains confined to the light line at low frequency and breaks-off quadratically as $k_0A_g$ grows, with the geometry entering through the single length $A_g$.

\subsection{General local anisotropic condition and quasi-TE reduction}
\label{subsec:aniso_disp}

The transverse-resonance conditions of
Sec.~\ref{subsec:TRC} are parameterised by a single
geometry-fixed scalar $A_g$. For a patch geometry whose point
group does not enforce in-plane isotropy, however, the sheet
capacitance is dyadic and the effective response depends on the
direction of propagation relative to its principal axes.
Accordingly, the scalar grid length must be replaced by a
direction-dependent projection of the capacitance dyadic.

The equivalent sheet-capacitance dyadic defines the symmetric
grid-length dyadic
\begin{equation}
\overline{\overline{A}}
=
\frac{
\overline{\overline{C}}_{\mathrm{eqv}}[\Sigma]
}{
2\varepsilon_0
},
\label{eq:Adyad}
\end{equation}
in units of ($\mathrm{m}$). For the isotropic reference patch,
\[
\overline{\overline{A}}[\Sigma_{\mathrm{ref}}]
=
A_{\mathrm{iso}}\overline{\overline{I}},
\]
recovering the scalar grid length introduced in
Sec.~\ref{subsec:TRC}. The symmetry constraints on the entries
and principal axes of
$\overline{\overline{A}}$ for the different point groups are
summarised in Table~\ref{tab:tensor_forms}.

For a nonzero in-plane wavevector
\[
\mathbf{k}_t
=
k_x\hat{\mathbf{x}}
+
k_y\hat{\mathbf{y}},
\qquad
k_t=|\mathbf{k}_t|,
\]
we introduce the propagation-fixed directions
\begin{equation}
\hat{\mathbf e}_{\mathrm{TM}}=\frac{\mathbf k_t}{k_t},
\qquad
\hat{\mathbf e}_{\mathrm{TE}}
=\frac{\hat{\mathbf z}\times\mathbf k_t}{k_t}.
\label{eq:eTE}
\end{equation}
Here $\hat{\mathbf e}_{\mathrm{TM}}$ and
$\hat{\mathbf e}_{\mathrm{TE}}$ denote the tangential
electric-field directions of the TM and TE waves,
respectively. At $\Gamma$, where $\mathbf k_t=\mathbf 0$,
these directions are understood in the one-sided limit
along the selected propagation path.

For a general anisotropic sheet, rotation into this
propagation-fixed basis produces longitudinal, transverse, and
cross projections of $\overline{\overline A}$. Consequently,
the local eigenmode is in general hybrid TE--TM rather than
strictly TE. The corresponding local transverse-resonance
condition may be written compactly as
\begin{equation}
\det
\left(
\overline{\overline{\mathcal T}}_0
\right)
=
0,
\label{eq:hybrid_local_dispersion}
\end{equation}
where $\overline{\overline{\mathcal T}}_0$ is the $2\times2$
transverse-resonance dyadic in the
$\{
\hat{\mathbf e}_{\mathrm{TM}},
\hat{\mathbf e}_{\mathrm{TE}}
\}$ basis. Its explicit construction, the resulting TE--TM
cross-coupling term, and the corresponding Schur-complement
reduction are given in Sec.~\ref{sec:supp_anisotropic_trc} of the Supplemental Material~\cite{SupplementalMaterial}.

The reduced model used here follows the dominant transverse
component of the TE-dominant branch. The effective grid length
sampled by this component is therefore the quadratic projection
\begin{equation}
A_0(\mathbf{k}_t)
\equiv
\hat{\mathbf e}_{\mathrm{TE}}
\cdot
\overline{\overline A}
\cdot
\hat{\mathbf e}_{\mathrm{TE}}
=
\frac{
A_{xx}k_y^2
-
2A_{xy}k_xk_y
+
A_{yy}k_x^2
}{
k_t^2
}.
\label{eq:alpha0_proj}
\end{equation}
For an isotropic sheet,
$A_0=A_g$ for every propagation direction. For an anisotropic
sheet, $A_0$ varies continuously as $\mathbf{k}_t$ rotates
relative to the principal axes of
$\overline{\overline A}$.

Using Eq.~\eqref{eq:Adyad}, the local sheet-admittance dyadic is
\begin{equation}
\overline{\overline Y}_s
=
j\omega
\overline{\overline C}_{\mathrm{eqv}}
=
j\frac{2k_0}{\eta_0}
\overline{\overline A}.
\label{eq:localconstitutivelaw}
\end{equation}
Neglecting the explicit TE--TM hybridisation correction while
retaining the dominant transverse response reduces the local
eigencondition to the quasi-TE closure
\begin{equation}
2Y_0^{\mathrm{TE}}
+
j\frac{2k_0}{\eta_0}
A_0(\mathbf{k}_t)F_L(k_t)
=
0,
\label{eq:qTE_TRC}
\end{equation}
where $F_L(k_t)$ is the $k_t$-dependent correction factor
introduced in Sec.~\ref{subsec:TRC}. Equation~\eqref{eq:qTE_TRC}
is the anisotropic extension of the scalar patch-array condition:
the complete geometric anisotropy enters through
$A_0(\mathbf{k}_t)$, while the scalar $k_t$-dependent correction
is retained through $F_L(k_t)$. No tensorial extension of
$F_L$ to the TE--TM cross-coupling term is assumed.

The reduction is exact when the propagation direction is aligned
with a principal axis of
$\overline{\overline A}$, for which the TE--TM cross projection
vanishes. Away from a principal axis, Eq.~\eqref{eq:qTE_TRC}
approximates the hybrid condition Eq.~\eqref{eq:hybrid_local_dispersion}. By Schur reduction in Sec.~\ref{sec:supp_anisotropic_trc} of the Supplemental Material~\cite{SupplementalMaterial}, the hybrid eigencondition is equivalent to
\begin{equation}
    T_{\mathrm{TE}}-\frac{T_{\times}^{2}}{T_{\mathrm{TM}}} = 0,
    \label{eq:hybrid_schur}
\end{equation}
whereas Eq.~\eqref{eq:qTE_TRC} retains only $T_{\mathrm{TE}}=0$. We quantify the neglected hybridisation by the relative frequency shift between the quasi-TE root and the hybrid root at fixed $\mathbf{k}_t$,
\begin{equation}
\delta_{\mathrm{hyb}}(\mathbf{k}_t)
=
\frac{
\bigl|f_{\mathrm{TE}}(\mathbf{k}_t)
-
f_{\mathrm{hybrid}}(\mathbf{k}_t)\bigr|
}{
f_{\mathrm{hybrid}}(\mathbf{k}_t)
},
\label{eq:delta_hyb}
\end{equation}
which vanishes on principal axes ($T_{\times}=0$) and is largest for low-symmetry cells. The full hybrid construction is given in Sec.~\ref{sec:supp_anisotropic_trc} of the Supplemental Material~\cite{SupplementalMaterial}.

Substituting
$\alpha_g\rightarrow k_0A_0(\mathbf{k}_t)$ into
Eq.~\eqref{eq:TE_disp_scalar} and retaining
$F_L(k_t)$ gives the quasi-TE evanescent decay constant
\begin{equation}
\Gamma_{\mathrm{TL}}(k_0,\mathbf{k}_t)
=
\frac{
\sqrt{
1+
\left[
k_0A_0(\mathbf{k}_t)
\right]^2
}
-1
}{
A_0(\mathbf{k}_t)
},
\label{eq:GammaTL}
\end{equation}
in units of $\mathrm{m}^{-1}$. The $k_t$-dependent correction
factor is already contained in this root and must therefore not
be applied a second time.

The corresponding anisotropic quasi-TE residual is
\begin{equation}
R_0(k_0,\mathbf{k}_t)
=
\sqrt{k_t^2-k_0^2}
-
\Gamma_{\mathrm{TL}}(k_0,\mathbf{k}_t),
\label{eq:R0_aniso}
\end{equation}
and the dispersion condition is
\[
R_0(k_0,\mathbf{k}_t)=0.
\]
Setting $A_0=A_g$ recovers the scalar result of
Sec.~\ref{subsec:TRC}.

Since $A_0$ depends on the propagation direction, the resulting
quasi-TE isofrequency contours are generally anisotropic, with
principal directions inherited from
$\overline{\overline A}$. In the weak-anisotropy limit they are
approximately elliptical. For the positive-definite capacitive
responses considered here the contours remain closed, in
contrast to the hyperbolic contours produced by sign-indefinite
anisotropic sheet responses~\cite{GomezDiaz2015,High2015,
GomezDiazOME2015}.

\section{Geometry-derived first-shell Floquet correction}
\label{sec:sideband}
The condition $R_0=0$ provides the local dispersion law while the
higher Floquet harmonics remain strongly evanescent.  As
$\mathbf{k}_t$ approaches the Brillouin-zone boundary, however, the
nearest nonzero reciprocal-lattice harmonics can become progressively
less confined and provide the leading spatial-dispersion correction
\cite{Belov2003,CorreasSerrano15,Brillouin1953,Munk2005}.
We therefore retain only the complete first reciprocal-shell as we will see in Sec.~\ref{sec:validation},
\begin{equation}
\mathbf G_p
\in
\left\{
\pm\mathbf b_1,\,
\pm\mathbf b_2,\,
\pm(\mathbf b_1+\mathbf b_2)
\right\},
\qquad
p=1,\ldots,6,
\label{eq:first_shell}
\end{equation}
where the reciprocal basis is defined by
$\mathbf a_i\!\cdot\!\mathbf b_j=2\pi\delta_{ij}$.
The six vectors in Eq.~\eqref{eq:first_shell} have the same reciprocal
magnitude and constitute the complete nearest nonzero shell.  The
particular member, or pair of members, that becomes least evanescent
depends on the propagation direction; retaining all six orders
explicitly therefore avoids introducing a path-dependent spectral
truncation \cite{RodriguezBerral2015,Mesa2016}.  The construction of
the first three reciprocal shells, together with their Floquet
propagation thresholds and evanescent decay lengths at $\Gamma$, $\mathrm{M}$,
and $\mathrm{K}$, is given in
Sec.~\ref{sec:supp_reciprocal_shells} of the Supplemental Material~\cite{SupplementalMaterial}.

\begin{figure}[t]
  \centering
  \includegraphics[width=.25\textwidth]{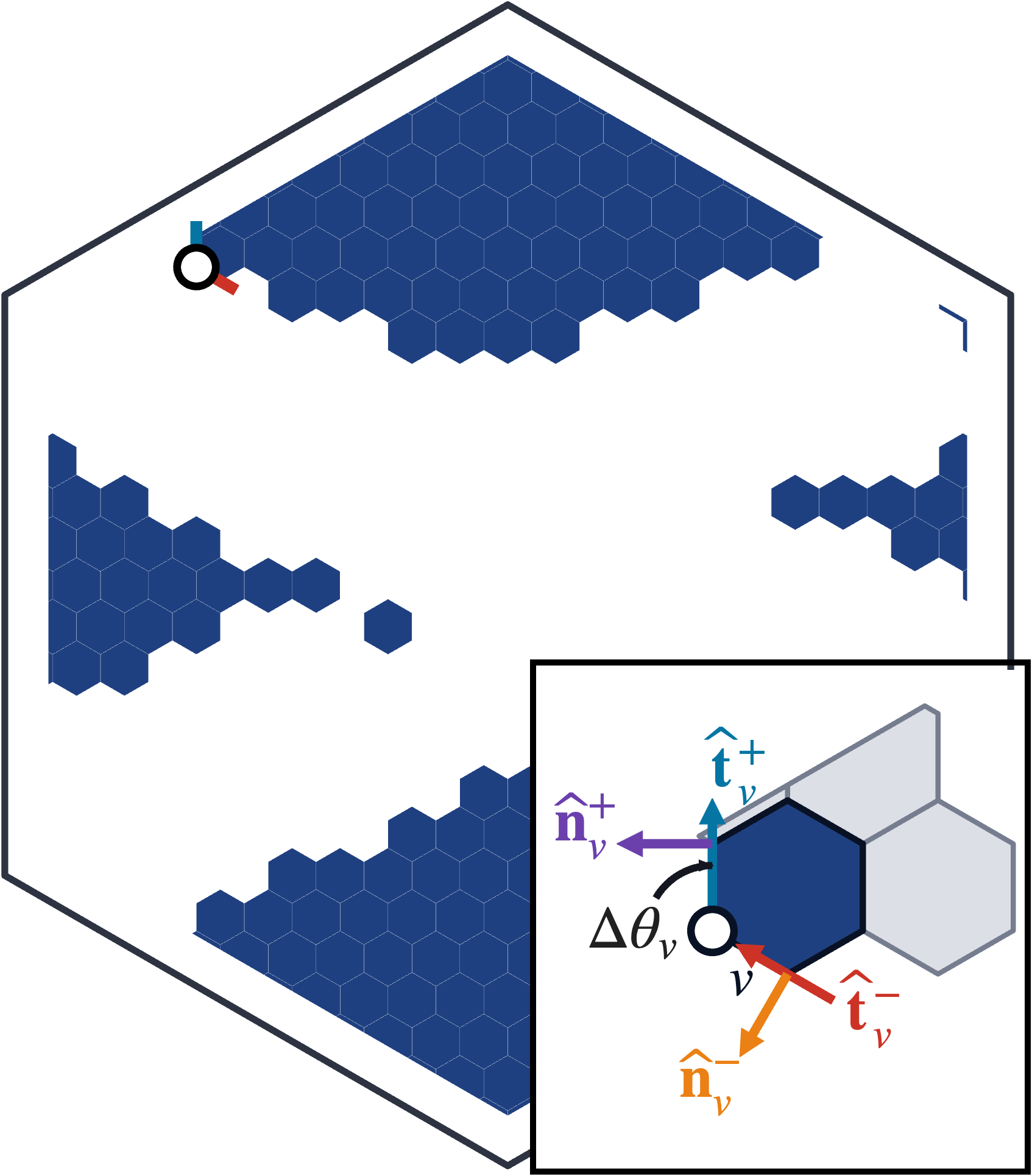}
  \caption{Random metallic patch inside the hexagonal unit cell and local
boundary geometry at a representative vertex \(v\). The one-sided
tangents \(\hat{\mathbf t}_v^\pm\), outward normals
\(\hat{\mathbf n}_v^\pm\), and exterior turning angle
\(\Delta\theta_v\) define the discrete-curvature contribution used in
Eqs.~\eqref{eq:Traw}--\eqref{eq:Tp_mp}. The same reconstructed boundary
\(\partial\Sigma\) is used in the Fourier overlap of
Eq.~\eqref{eq:Bmom}.}
  \label{fig:boundary_local_frame}
\end{figure}

For the $p$th retained sideband, the shifted transverse wavevector and
free-space decay constant are
\begin{equation}
\mathbf q_p
=
\mathbf k_t+\mathbf G_p,
\qquad
\gamma_{h_p}
=
\sqrt{|\mathbf q_p|^2-k_0^2},
\label{eq:qp_gammap}
\end{equation}
for an evanescent order.  Each sideband samples the same local sheet
dyadic through its own TE polarization,
\begin{equation}
\hat{\mathbf e}_p
=
\frac{\hat{\mathbf z}\times\mathbf q_p}{|\mathbf q_p|},
\qquad
A_p
=
\hat{\mathbf e}_p
\!\cdot\!
\overline{\overline A}
\!\cdot\!
\hat{\mathbf e}_p,
\label{eq:sideband_local}
\end{equation}
with the corresponding sheet-loaded decay $\Gamma_p$ obtained from
Eq.~\eqref{eq:GammaTL} evaluated at $A_p$.  Each retained order also
carries its own modal admittance \cite{Collin1990}.  Two quantities,
both computed directly from the reconstructed boundary
$\partial\Sigma$, then determine its coupling to the fundamental
channel.  Figure~\ref{fig:boundary_local_frame} shows the reconstructed
polygonal boundary and the one-sided tangent--normal frame used to
define these geometric coupling measures.

The complete six-order construction, including the reciprocal-shell
geometry, polygon-edge evaluation of the complex overlap, point-wise normalization of the directional-turning multipliers, is
detailed in Sec.~\ref{sec:supp_first_shell_correction} of the Supplemental Material~\cite{SupplementalMaterial}. The reciprocal-shell
geometry and its mirror pairing are illustrated in Supplemental
Fig.~\ref{fig:supp_first_shell_mirror}~\cite{SupplementalMaterial}.

\subsection{Boundary Fourier overlap of edge-charge profiles}
\label{subsec:boundary_fourier_overlap}

We describe coupling between the fundamental Floquet order
and the six retained first-shell orders through the
normalized overlap of model edge-charge profiles.
Charge-based modal descriptions motivate representing
interactions through spatial charge
distributions~\cite{HohenesterKrenn2005,Silveiro2015},
while Floquet expansions of surface-current profiles
underlie related analytical models of
metasurfaces~\cite{BaladiHum2021}.

We parameterize the reconstructed boundary by arclength
$\ell\in[0,L]$, where $L$ is its total length, and write
$\mathbf r(\ell)\in\partial\Sigma$.
For several boundary loops, their arclength
parameterizations are concatenated in a fixed order.
The in-plane outward normal $\hat{\mathbf n}(\ell)$
is piecewise constant on each open polygon edge.

The field behaviour and charge concentration near
conducting edges motivate a boundary-based description
of the capacitive response~\cite{Jackson1998,Meixner1972}.
Within the present reduced model, we adopt the
dimensionless edge-charge profiles
\begin{equation}
    \widetilde{\rho}_a(\ell;\mathbf k_t)
    =
    \bigl[
        \hat{\mathbf e}_a\cdot\hat{\mathbf n}(\ell)
    \bigr]
    e^{-i(\mathbf k_t+\mathbf G_a)\cdot\mathbf r(\ell)},
    \qquad a=0,\ldots,6.
    \label{eq:model_edge_charge_profiles}
\end{equation}
Here $\mathbf G_0=\mathbf0$, while
$\mathbf G_p$, $p=1,\ldots,6$, are the retained
first reciprocal-shell vectors.
We take
$\hat{\mathbf e}_0\equiv\hat{\mathbf e}_{\mathrm{TE}}$,
and define each sideband TE polarization
$\hat{\mathbf e}_p$ relative to its own tangential
wavevector $\mathbf k_t+\mathbf G_p$.
The polarization vectors are real.
The modal amplitudes are factored out, and the normal
projections provide geometric weights for the assumed
charge-profile shapes.

At fixed real $\mathbf k_t$, the cumulative cross moment is
\begin{align}
    B_{0p}(\ell)
    &=
    \int_0^\ell
    \widetilde{\rho}_0^{\,*}(\ell';\mathbf k_t)
    \widetilde{\rho}_p(\ell';\mathbf k_t)
    \,\mathrm d\ell'
    \nonumber\\
    &=
    \int_0^\ell
    \bigl[
        \hat{\mathbf e}_0\cdot\hat{\mathbf n}(\ell')
    \bigr]
    \bigl[
        \hat{\mathbf e}_p\cdot\hat{\mathbf n}(\ell')
    \bigr]
    e^{-i\mathbf G_p\cdot\mathbf r(\ell')}
    \,\mathrm d\ell',
    \label{eq:Bmom}
\end{align}
and the squared  profile norms are
\begin{align}
    N_a(\ell)
    &=
    \int_0^\ell
    \left|
        \widetilde{\rho}_a(\ell';\mathbf k_t)
    \right|^2
    \,\mathrm d\ell'
    \nonumber\\
    &=
    \int_0^\ell
    \bigl[
        \hat{\mathbf e}_a\cdot\hat{\mathbf n}(\ell')
    \bigr]^2
    \,\mathrm d\ell',
    \qquad a\in\{0,p\}.
    \label{eq:Nnorm}
\end{align}
The common Bloch phase cancels in the cross moment,
leaving the reciprocal phase
$e^{-i\mathbf G_p\cdot\mathbf r}$.

The full-boundary quantities are
$B_{0p}=B_{0p}(L)$ and $N_a=N_a(L)$.
For nonzero profile norms, their normalized overlap is
\begin{equation}
    \eta_{0p}(\mathbf k_t)
    =
    \frac{B_{0p}(L)}
    {\sqrt{N_0(L)\,N_p(L)}},
    \qquad |\eta_{0p}(\mathbf{k}_t)|\leq1.
    \label{eq:eta0p}
\end{equation}
Since $|e^{-i\mathbf G_p\cdot\mathbf r}|=1$,
the Cauchy--Schwarz inequality applied to the two
boundary profiles gives
\mbox{$|B_{0p}(L)|^2\leq N_0(L)N_p(L)$},
and hence $|\eta_{0p}(\mathbf{k}_t)|\leq1$.
The Floquet phase enters the cross moment but not
the norms.

The overlap measures the complex correlation between
the fundamental and sideband model charge profiles:
contributions from different edges reinforce or cancel
according to their normal projections and reciprocal
phases.
It provides a geometric analogue of a coupled-mode
overlap~\cite{Kogelnik1972,Yariv1973,Haus1984}.
Its use as an electromagnetic coupling measure is
part of the adopted edge-projection approximation. The full-boundary quantities are independent of the
starting points and ordering of the boundary loops,
although the cumulative functions depend on that choice.

The geometry dependence can also be expressed through
the boundary Fourier tensor
\begin{equation}
    \overline{\overline F}_{\partial\Sigma}(\mathbf G_p)
    =
    \oint_{\partial\Sigma}
    \hat{\mathbf n}\hat{\mathbf n}^{\mathsf T}
    e^{-i\mathbf G_p\cdot\mathbf r}
    \,\mathrm d\ell.
    \label{eq:orthogonality_boundary_tensor}
\end{equation}
The full-boundary cross moment is then
\begin{equation}
    B_{0p}(L)
    =
    \hat{\mathbf e}_0^{\mathsf T}
    \overline{\overline F}_{\partial\Sigma}(\mathbf G_p)
    \hat{\mathbf e}_p.
    \label{eq:boundary_tensor_contraction}
\end{equation}
This tensor contains the spatial distribution and
orientation of the boundary edges.
Its contraction need not vanish for
$\mathbf G_p\neq\mathbf0$:
complete-cell Floquet orthogonality concerns an
unweighted area inner product, whereas the present
overlap uses normal projections integrated over
the boundary.
The detailed derivation is given in Sec.~\ref{sec:supp_first_shell_correction}
of the Supplemental Material ~\cite{SupplementalMaterial}.

For the polygonal boundaries of
Sec.~\ref{sec:geometry_cp},
Eq.~\eqref{eq:Bmom} is evaluated exactly edge by edge.
The complex phase of $\eta_{0p}$ is retained throughout.

\begin{figure}[t]
    \centering
    \includegraphics[width=\columnwidth]
    {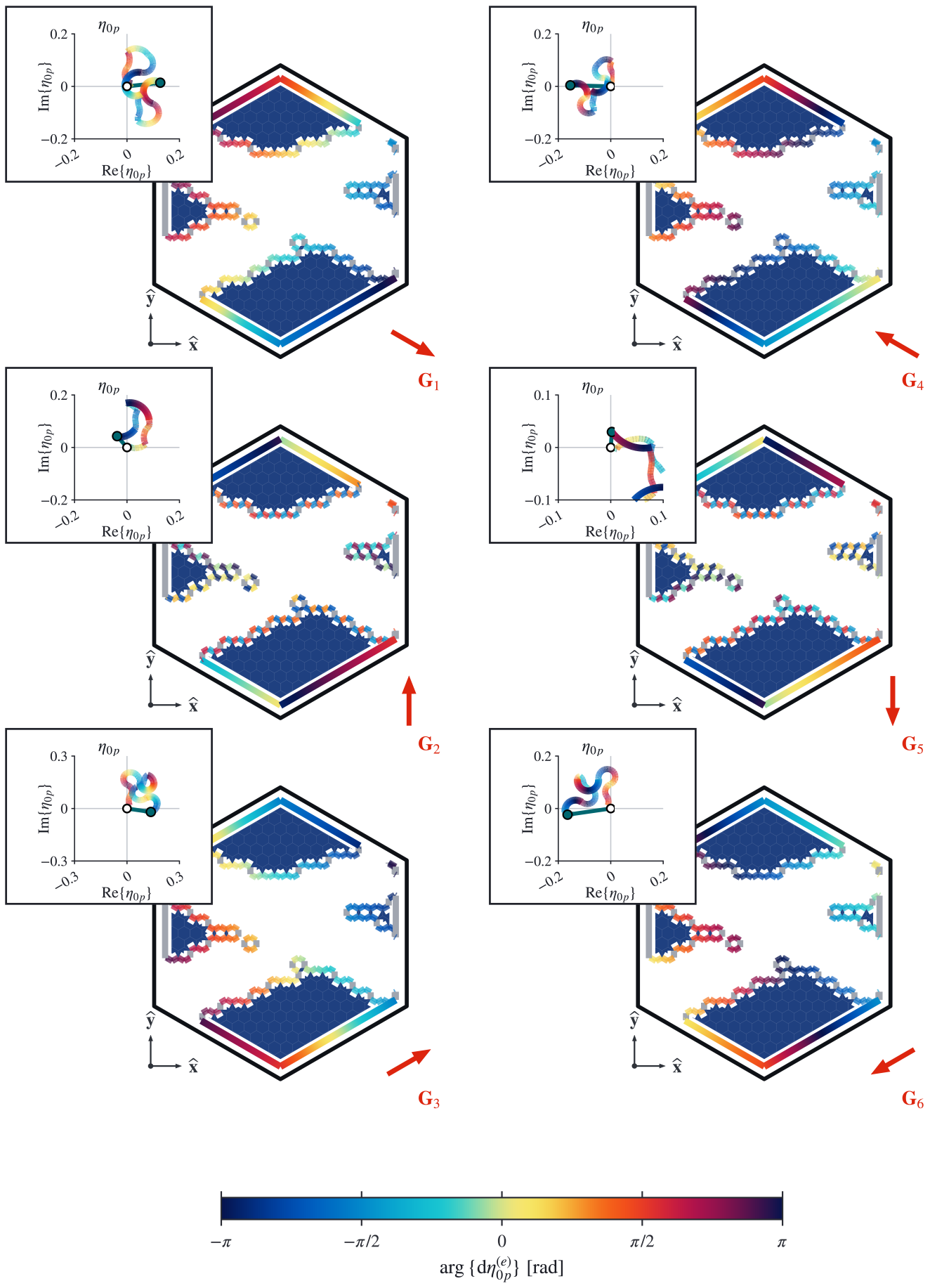}
    \caption{
    Complex boundary Fourier overlap for the six
    retained first-shell orders at the representative
    point \mbox{$\mathbf k_t=0.82\,\mathbf K$}
    on the $\Gamma$--$\mathrm K$ path.
    Boundary colour denotes the phase of the edgewise
    contribution to $B_{0p}$.
    Each inset shows the cumulative complex-plane
    trajectory of the boundary contributions,
    whose endpoint determines the normalized
    overlap $\eta_{0p}$.
    The complex phase is retained in the coherent
    sideband combination of
    Eq.~\eqref{eq:sigma_proj}.
    }
    \label{fig:six_eta_overlap}
\end{figure}

For the visualizations in
Figs.~\ref{fig:six_eta_overlap}
and~\ref{fig:six_turning_maps},
we select the representative interior point
$\mathbf k_t=0.82\,\mathbf K$.
The dispersion calculation evaluates the overlaps and turning multipliers pointwise over the complete sampled $\Gamma$--$\mathrm K$ segment. 

The edge-resolved contributions for the complete
first reciprocal-shell are illustrated in
Fig.~\ref{fig:six_eta_overlap}.
The boundary colour records the phase of the local
complex contribution, while each inset traces its
cumulative sum in the complex plane.

\subsection{Directional multipliers from boundary turning}
\label{subsec:directional_turning}

The directional multiplier $\widetilde m_p(\mathbf k_t)$ weights
fundamental-to-sideband coupling according to the
participation of the model edge-charge profiles at
changes in the boundary orientation.
The field behaviour near conducting corners motivates
including boundary turning in this
description~\cite{VanBladel1996}.
We adopt the following dimensionless geometric
construction within the reduced coupling model.

At each vertex $v$, let
$\hat{\mathbf t}_v^{\pm}$ and
$\hat{\mathbf n}_v^{\pm}$ denote the one-sided
unit tangents and normals.
The signed exterior turning angle is
\begin{equation}
    \Delta\theta_v
    =
    \operatorname{atan2}
    \left(
        \hat{\mathbf z}\cdot
        [\hat{\mathbf t}_v^{-}\times
         \hat{\mathbf t}_v^{+}],
        \hat{\mathbf t}_v^{-}\cdot
        \hat{\mathbf t}_v^{+}
    \right),
    \label{eq:directional_turning_angle}
\end{equation}
and the normalized total absolute turning is
\begin{equation}
    T_{\mathrm{abs}}
    =
    \frac{1}{2\pi}
    \sum_v|\Delta\theta_v|.
    \label{eq:directional_total_turning}
\end{equation}
This quantity measures the accumulated absolute rotation
of the boundary tangent, or equivalently of its normal.
For a single simple closed polygonal boundary,
the total-curvature theorem gives
$T_{\mathrm{abs}}\geq1$, with equality for a convex boundary~\cite{Milnor1950}.
Reversals of the turning direction increase this quantity. For several boundary loops, their absolute-turning contributions are added.

The participation of order $a$ at a vertex is measured
by the mean squared charge-profile amplitude on its
two adjacent edges:
\begin{equation}
    S_a(v;\mathbf k_t)
    =
    \frac12\sum_{\pm}
    \left[
        \hat{\mathbf e}_a(\mathbf k_t)\cdot
        \hat{\mathbf n}_v^{\pm}
    \right]^2,
    \qquad a\in\{0,p\}.
    \label{eq:Sdef}
\end{equation}
This follows from the model profiles in
Eq.~\eqref{eq:model_edge_charge_profiles}.
Since the polarization vectors and boundary normals
are unit vectors, $0\leq S_a\leq1$.
The raw turning participation is
\begin{equation}
    T_p^{\mathrm{raw}}(\mathbf k_t)
    =
    \frac{1}{2\pi}
    \sum_v|\Delta\theta_v|\,
    2\sqrt{
        S_0(v;\mathbf k_t)
        S_p(v;\mathbf k_t)
    }.
    \label{eq:Traw}
\end{equation}
Thus a vertex contributes strongly when its turn is
pronounced and both profiles participate on the
adjacent edges.
This nonnegative weight measures jointly sampled
boundary turning; cancellation from charge signs and
reciprocal phases is retained in the complex overlap
$\eta_{0p}$.

At each Bloch wavevector, we normalize the six raw
participations and define the amplitude multipliers as
\begin{align}
    T_p(\mathbf k_t)
    &=
    \frac{
        T_{\mathrm{abs}}\,
        T_p^{\mathrm{raw}}(\mathbf k_t)
    }{
        \dfrac16\sum_{r=1}^{6}
        T_r^{\mathrm{raw}}(\mathbf k_t)
    },
    \nonumber\\
    \widetilde m_p(\mathbf k_t)
    &=
    \sqrt{2T_p(\mathbf k_t)}.
    \label{eq:Tp_mp}
\end{align}
For a nonzero normalization denominator,
\begin{equation}
    \begin{aligned}
        \frac16\sum_{p=1}^{6}T_p(\mathbf k_t)
        &=T_{\mathrm{abs}},
        \\
        \frac16\sum_{p=1}^{6}
        \widetilde m_p^{\,2}(\mathbf k_t)
        &=2T_{\mathrm{abs}}.
    \end{aligned}
    \label{eq:directional_turning_mean}
\end{equation}
The normalization preserves the shell mean while
allowing its distribution among the orders to vary
with their polarization projections.
The square root converts participation into an
amplitude weight.
The factor of two specifies the adopted amplitude
normalization; no additional fitted scale multiplies
Eq.~\eqref{eq:Tp_mp}.
The shell root-mean-square multiplier is therefore
$\sqrt{2T_{\mathrm{abs}}}$ at every Bloch point.

Along the straight $\Gamma$--$\mathrm K$ segment,
\begin{equation}
    \begin{aligned}
        \mathbf k_t(s)
        &=s\,\mathbf K,
        \qquad 0\leq s\leq1,
        \\
        \mathbf K
        &=\frac{2\mathbf b_1+\mathbf b_2}{3},
    \end{aligned}
    \label{eq:GammaK_parameter}
\end{equation}
where $\mathbf b_1,\mathbf b_2$ are the primitive
reciprocal vectors and $\mathbf K$ locates the
selected Brillouin-zone corner.
At every sampled point, the coupling uses
$\widetilde m_p[\mathbf k_t(s)]$ and
$\eta_{0p}[\mathbf k_t(s)]$ evaluated at that same
wavevector.
The geometry remains fixed; the turning weights vary
through the polarization projections in
Eq.~\eqref{eq:Sdef}.
At $\Gamma$, the fundamental TE polarization and the
multipliers are defined by the limit $s\to0^+$ along
this path, with
$\hat{\mathbf e}_0=
\hat{\mathbf z}\times\mathbf K/|\mathbf K|$.

\begin{figure}[t]
    \centering
    \includegraphics[width=.35\textwidth]
    {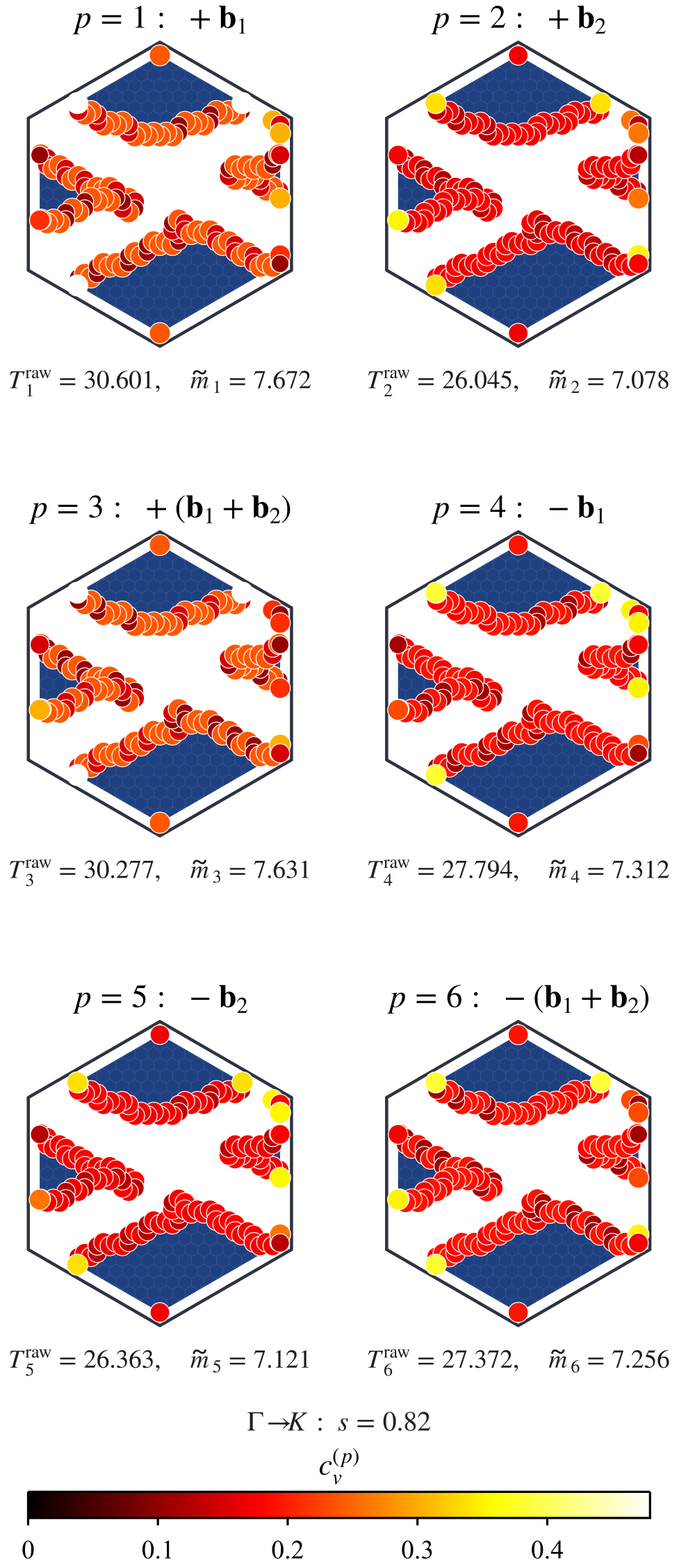}
    \caption{
    Directional turning participation for the six
    retained first-shell orders at
    $\mathbf k_t=0.82\,\mathbf K$.
    Marker colour and size indicate the vertex
    summands of Eq.~\eqref{eq:Traw}; all boundary
    vertices are shown.
    The values below each unit cell give
    $T_p^{\mathrm{raw}}$ and the pointwise amplitude
    multiplier $\widetilde m_p(\mathbf k_t)$.
    }
    \label{fig:six_turning_maps}
\end{figure}

Figure~\ref{fig:six_turning_maps} illustrates the
vertex contributions at one representative point.

Uniform scaling of the boundary and lattice preserves
these dimensionless turning quantities at corresponding
normalized Bloch wavevectors.
The prescription therefore introduces no additional
length scale.
The six multiplier traces along $\Gamma$--$\mathrm K$
are shown in Supplemental Fig.~\ref{fig:supp_multiplier_freezing}; the normalization
and scaling properties are detailed in Sec.~\ref{sec:supp_first_shell_correction} of the
Supplemental Material~\cite{SupplementalMaterial}.

\subsection{Sideband self-energy and corrected dispersion}
\label{subsec:sideband_self_energy}

At each Bloch wavevector, the fundamental order couples
to the six retained sidebands through
\begin{equation}
    \begin{aligned}
        v_p
        &=
        \widetilde m_p(\mathbf k_t)
        \sqrt{\Gamma_{\mathrm{TL}}\Gamma_p}\,
        \eta_{0p}(\mathbf k_t),
        \\
        D_p
        &=\gamma_{h_p}-\Gamma_p.
    \end{aligned}
    \label{eq:vp_Dp}
\end{equation}
Both quantities have units of $\mathrm{m}^{-1}$.
The overlap and turning multiplier are evaluated at
the current $\mathbf k_t$, while the decay quantities
retain their dependence on $(k_0,\mathbf k_t)$.
Each sideband detuning is determined by its own
tangential wavevector and sheet loading.

Define
$\mathbf v=(v_1,\ldots,v_6)^{\mathsf T}$,
$\mathbf D=\operatorname{diag}(D_1,\ldots,D_6)$,
and $\gamma_0=\sqrt{k_t^2-k_0^2}$.
The adopted $(1+6)$-mode system is
\begin{equation}
    \begin{pmatrix}
        \gamma_0-\Gamma_{\mathrm{TL}}
        &-\mathbf v^\dagger
        \\
        -\mathbf v&\mathbf D
    \end{pmatrix}
    \begin{pmatrix}
        a_0\\
        \mathbf a_{\mathrm{sb}}
    \end{pmatrix}
    =\mathbf0,
    \label{eq:sideband_block_schematic}
\end{equation}
where $a_0$ and $\mathbf a_{\mathrm{sb}}$ are the
fundamental and sideband amplitudes.
The diagonal sideband block omits direct coupling
between distinct sidebands.

For invertible $\mathbf D$, eliminating the sideband
amplitudes by a Schur
complement~\cite{Zhang2005,SupplementalMaterial}
gives
\begin{equation}
    R(k_0,\mathbf k_t)
    =
    \gamma_0-\Gamma_{\mathrm{TL}}
    -\Sigma_{\mathrm{sb}}
    =0.
    \label{eq:six_order_disp}
\end{equation}
Setting $\Sigma_{\mathrm{sb}}=0$ recovers the local
condition in Eq.~\eqref{eq:R0_aniso}.
Without a symmetry projection, the self-energy is
\begin{equation}
    \begin{aligned}
        \Sigma_{\mathrm{sb}}
        &=
        \mathbf v^\dagger\mathbf D^{-1}\mathbf v
        \\
        &=
        \Gamma_{\mathrm{TL}}
        \sum_{p=1}^{6}
        \frac{
            \widetilde m_p^{\,2}(\mathbf k_t)\,
            \Gamma_p\,|\eta_{0p}(\mathbf k_t)|^2
        }{
            \gamma_{h_p}-\Gamma_p
        }.
    \end{aligned}
    \label{eq:sigma_ind}
\end{equation}
All quantities in this expression are evaluated at
the same $(k_0,\mathbf k_t)$, with the full frequency
dependence of the detunings retained.
The turning factors and overlaps follow from the
reconstructed boundary, and the sheet-loaded decays
from the TE projections of the grid-length dyadic.
No sideband coupling or detuning is fitted separately.
At a singular sideband block, the eigencondition must
be obtained from the coupled system
\eqref{eq:sideband_block_schematic}.

When the propagation axis is a mirror of the complete
unit-cell geometry, the sidebands can be organized
into symmetry-adapted combinations.
Reflection about
$\hat{\mathbf u}=\hat{\mathbf k}_t$ is represented by
$M=2\hat{\mathbf u}\hat{\mathbf u}^{\mathsf T}
-\overline{\overline I}$ and induces the permutation
$M\mathbf G_p=\mathbf G_{\pi(p)}$.
In the adopted TE basis,
$M\hat{\mathbf e}_p=-\hat{\mathbf e}_{\pi(p)}$.
The basis carries the odd parity of the fundamental
TE field, so the allowed scalar amplitudes are equal
within each mirror pair~\cite{Bouckaert1936,
BradleyCracknell2009,Sakoda2004,Ochiai2001}.
Their subspace is specified by
\begin{equation}
    \begin{aligned}
        P_{\mathrm{TE}}
        &=
        \tfrac12
        \bigl(\overline{\overline I}_6+\Pi_M\bigr),
        \\
        \operatorname{range}(U)
        &=
        \operatorname{range}(P_{\mathrm{TE}}),
    \end{aligned}
    \label{eq:PTE}
\end{equation}
where $\Pi_M$ implements the permutation and
$U^\dagger U=\overline{\overline I}_r$.
Along $\Gamma$--$\mathrm K$, the six orders form
three mirror pairs, giving $r=3$.
The projected self-energy is
\begin{equation}
    \Sigma_{\mathrm{sb}}^{\mathrm{proj}}
    =
    (U^\dagger\mathbf v)^\dagger
    \bigl(U^\dagger\mathbf D U\bigr)^{-1}
    (U^\dagger\mathbf v).
    \label{eq:sigma_proj}
\end{equation}
When mirror-symmetry reduction applies, $\Sigma_{\mathrm{sb}}^{\mathrm{proj}}$ is used for $\Sigma_{\mathrm{sb}}$ in Eq.~\eqref{eq:six_order_disp}.
Taking $U=\overline{\overline I}_6$ recovers the
unprojected expression.

The projection retains the complex coupling amplitudes,
including the phases of $\eta_{0p}$.
For an exactly mirror-symmetric geometry represented
with its origin on the mirror axis,
$v_{\pi(p)}=v_p$ and $D_{\pi(p)}=D_p$.
Each allowed pair combination therefore retains the
full contribution of its two orders, while the
complementary combination decouples.
No additional multiplicity factor is introduced.
A displaced geometry requires the corresponding
Fourier phases in both the couplings and the mirror
representation.
The explicit permutation
$\pi=(1\,3)(2\,5)(4\,6)$ and construction of $U$ are
given in Sec.~\ref{sec:supp_first_shell_correction} of the Supplemental Material;
the pairs are illustrated in Supplemental Fig.~\ref{fig:supp_first_shell_mirror}(b).

The pointwise prescription also recovers the local
long-wavelength limit.
For fixed geometry with finite, nondispersive loading,
$\Gamma_{\mathrm{TL}},\Gamma_p=O(k_0^2)$.
As $k_0,\mathbf k_t\to0$ along the lowest branch,
$D_p\to|\mathbf G_p|>0$.
The bounded turning multipliers and normalized overlaps
then give $\Sigma_{\mathrm{sb}}=O(k_0^4)$ and
\begin{equation}
    \frac{\Sigma_{\mathrm{sb}}}
         {\Gamma_{\mathrm{TL}}}
    \longrightarrow0.
    \label{eq:sigma_qs_limit}
\end{equation}
The same limit holds after the mirror projection,
whose sideband block has a finite, invertible limit.
Local recovery therefore does not require the
individual turning multipliers to approach unity
or vanish.

The Schur elimination is exact within the adopted
finite coupled system.
The geometric coupling prescription and omission of
direct sideband interactions remain approximations
whose accuracy is assessed against the full-wave
benchmarks in Sec.~\ref{sec:validation}.
With the lattice fixed, the construction follows
\begin{equation}
    \begin{aligned}
        \{\overline{\overline C},\partial\Sigma\}
        &\longrightarrow
        \{\Gamma_{\mathrm{TL}},\Gamma_p,
          \widetilde m_p,\eta_{0p}\}
        \\
        &\longrightarrow
        \Sigma_{\mathrm{sb}}
        \longrightarrow R=0.
    \end{aligned}
    \label{eq:chain}
\end{equation}
The local branch depends on
$\overline{\overline C}$ through the projection
$A_0(\mathbf k_t)$.
The nonlocal correction additionally requires
$\partial\Sigma$, which supplies the Fourier overlap
and pointwise turning participation.

\section{Numerical Results}
\label{sec:validation}
The dispersion model of Sec.~\ref{sec:dispersion} is validated against full-wave eigenmode simulations for a set of geometries spanning the symmetry classes of Table~\ref{tab:tensor_forms}. Importantly, HFSS is not used to extract $\{\overline{\overline{C}}$ or to fit any sideband parameter. For each geometry, the dyadic is computed once from the cell problem of Sec.~\ref{sec:geometry_cp}, the Floquet correction is evaluated from $\partial\Sigma$, and the HFSS eigenmode sweep serves solely as an independent benchmark of the resulting bands. Each structure is a hexagonal unit cell of circumradius $R_{\mathrm{cell}} = 11.547$~mm with a minimum edge-to-edge gap of $1$~mm, with a random or canonical metal patch built from the honeycomb tessellation of Sec.~\ref{sec:geometry_cp}. The eigenfrequencies of the first bound TE surface-wave mode are computed with the HFSS eigenmode solver under the periodic boundary conditions, sweeping the tangential Bloch wavevector $\mathbf{k}_t$ along the $\Gamma$--$\mathrm{K}$ path of the hexagonal Brillouin zone. For each geometry, the capacitance dyadic $\overline{\overline{C}}$ is computed once from the cell problem of Sec.~\ref{sec:geometry_cp}, projected onto the TE polarisation along the swept direction via~\eqref{eq:alpha0_proj}, and inserted into the local~\eqref{eq:R0_aniso}, and sideband-corrected~\eqref{eq:six_order_disp} dispersion relations. Since the sweep direction is fixed while the principal axes of $\overline{\overline{C}}$ vary from structure to structure, the set probes the directional projection over a wide range of orientations relative to the anisotropy frame. The model accuracy is quantified by the relative frequency error at fixed $k_t$, $\epsilon(k_t) = |f_{\mathrm{model}} - f_{\mathrm{HFSS}}|/f_{\mathrm{HFSS}}$, evaluated at the simulated points. 

\begin{figure}[htbp]
\centering
\includegraphics[width=\linewidth]{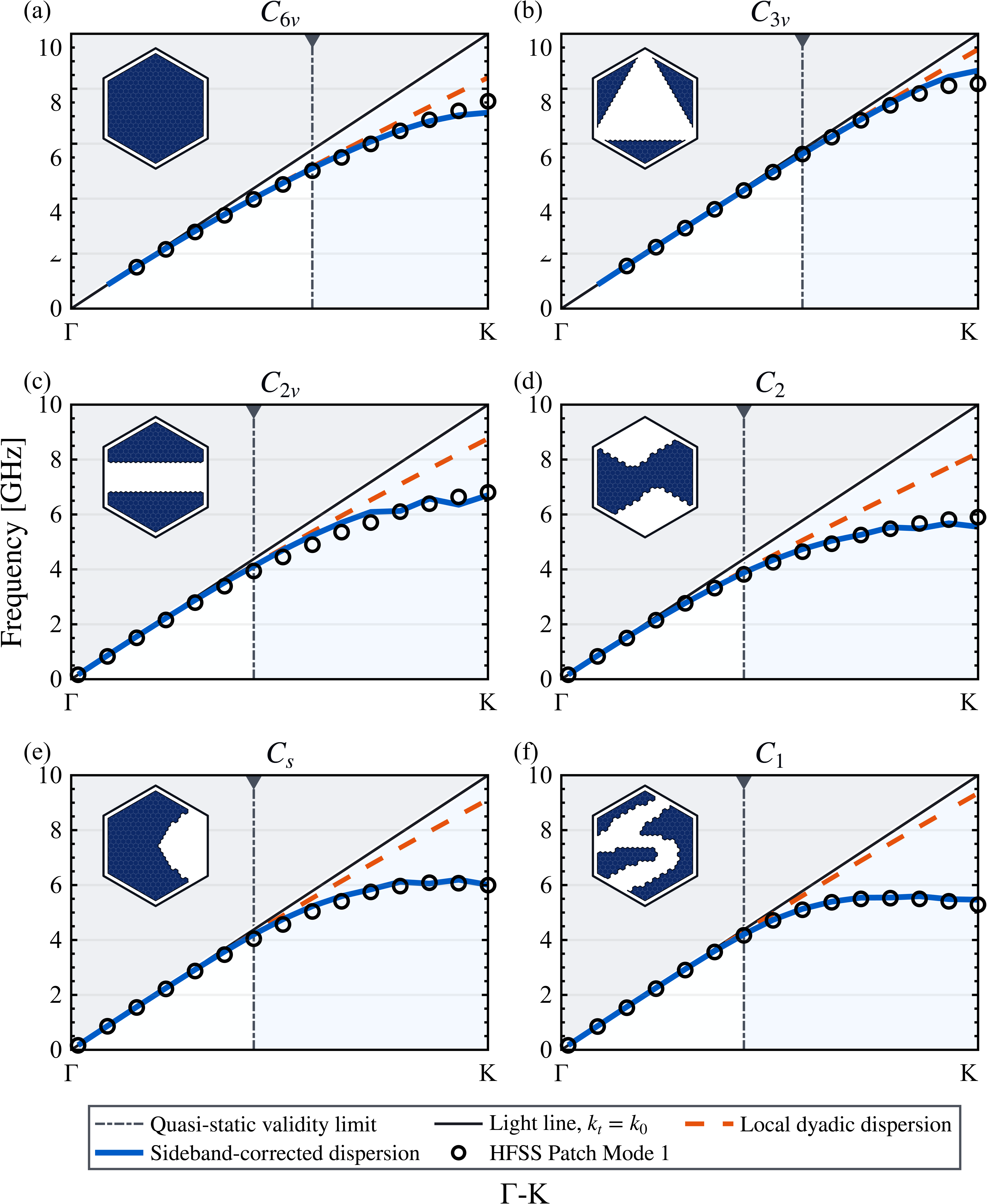}
\caption{\label{fig:dispersion_panels}Surface-wave dispersion along the $\Gamma$--$\mathrm{K}$ path for six representative structures, one per symmetry class of Table~\ref{tab:tensor_forms}. The vertical dash-dotted marker indicates the onset of spatial dispersion, beyond which the Floquet-sideband coupling becomes significant and the local and sideband-corrected models separate. Below the onset, both models coincide and follow the quasi-linear branch of Eq.~\eqref{eq:linear_disp}; beyond it the sideband-corrected model tracks the progressive flattening of the simulated band.}
\end{figure}

Figure~\ref{fig:dispersion_panels} shows the six representative structures, one per symmetry class. At small $k_t$ the local dyadic and sideband-corrected models are indistinguishable and both track the quasi-linear branch. Near the zone boundary they separate: the local model keeps following the projected-dyadic response, while the first-shell model bends with the HFSS band through coupling to the six retained Floquet orders. The detuning of the $p$th order,
\begin{equation}
D_p = \gamma_{h_p} - \Gamma_p,
\end{equation}
is set by the free-space decay $\gamma_{h_p}$ against the sheet-loaded decay $\Gamma_p$ of that order, both fixed by the geometry, not fitted. The coupling amplitude itself depends on the boundary overlap $\eta_{0p}$ and the directional multiplier $\widetilde{m}_p$, again both geometric and evaluated pointwise. No sideband detuning is fitted per geometry.

\begin{figure}[htbp]
\centering
\includegraphics[width=\columnwidth]{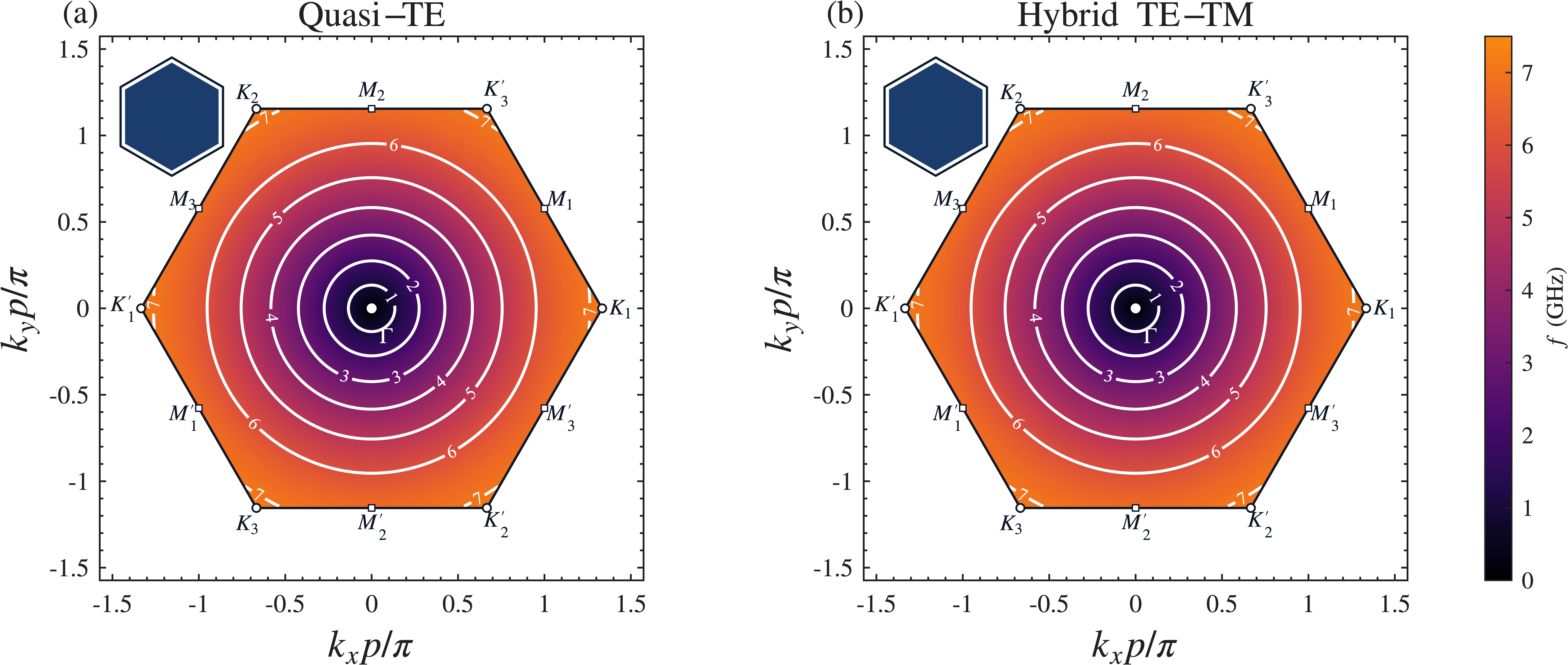}
\par\vspace{0.6em}
\includegraphics[width=\columnwidth]{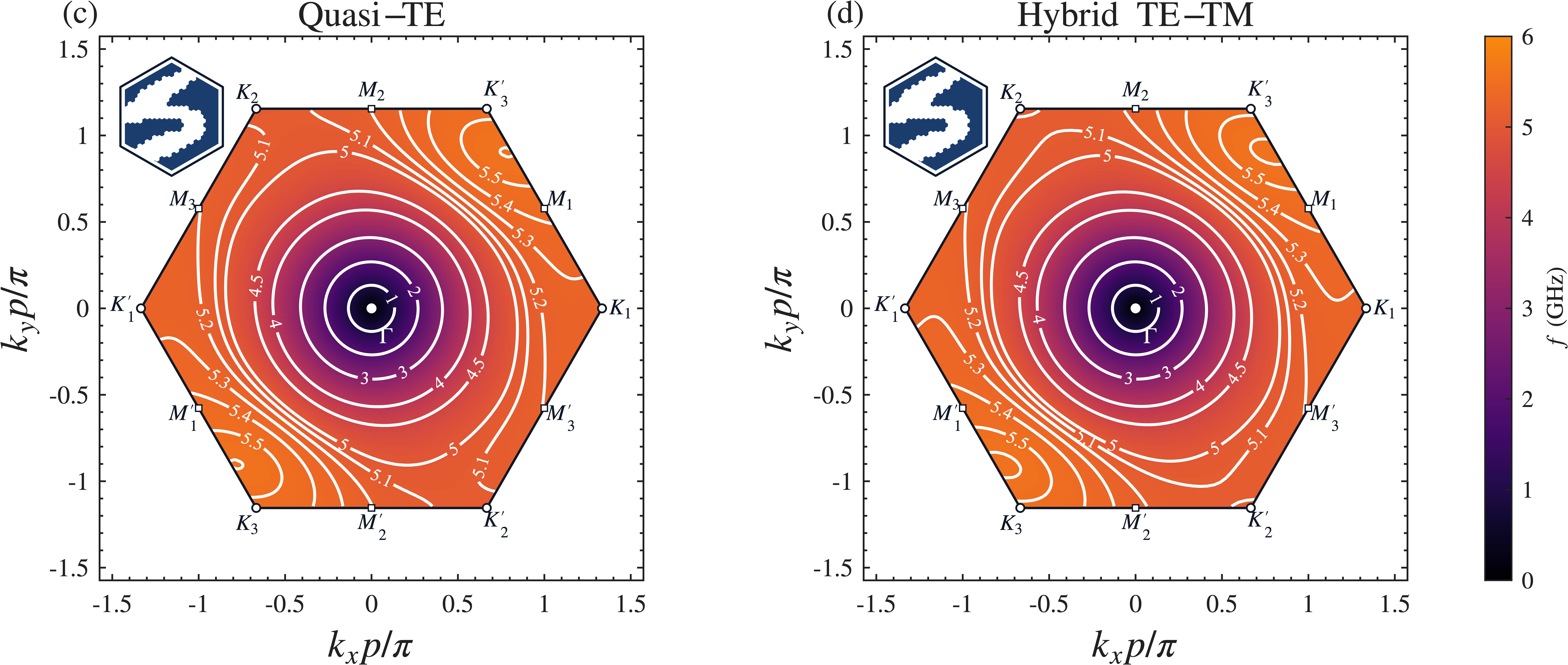}
\caption{\label{fig:fullbz_isofrequency}
Model-predicted isofrequency contours of the first TE-dominated surface-wave branch over the hexagonal first Brillouin zone for the $C_{6v}$ (top row) and $C_1$ (bottom row) patches. The left and right columns show the quasi-TE approximation and the hybrid TE--TM model, respectively; both retain the six Floquet orders of the first reciprocal-shell. The colour maps give the modal frequency in GHz, the white curves mark selected isofrequencies, and the insets show the corresponding patches. The $C_{6v}$ contours remain approximately circular, with weak deformation at larger wavevectors, whereas the $C_1$ contours develop a pronounced directional dependence. The two polarisation treatments give similar contours near $\Gamma$, with more visible differences in the outer $C_1$ contours.
}
\end{figure}

Figure~\ref{fig:fullbz_isofrequency} complements the
$\Gamma$--$\mathrm{K}$ cuts of Fig.~\ref{fig:dispersion_panels}
by showing the model-predicted angular response throughout
the first Brillouin zone.
The reduced hybrid TE--TM calculation restores the
longitudinal--transverse coupling within each retained
Floquet order, while retaining the same geometric TE--TE
coupling between orders as the quasi-TE calculation.
The corresponding closure and the numerical construction
of the full-Brillouin-zone isofrequency contours are
detailed in Sec.~\ref{sec:supp_anisotropic_trc} of the Supplemental Material ~\cite{SupplementalMaterial}.
The small contours surrounding $\Gamma$ are nearly circular
in both examples. At larger wavevectors, the Floquet correction depends on the propagation direction through the sideband decay constants $\gamma_{h_p}$, the detunings $D_p$, and the boundary coupling factors. Consequently, the same first-shell coupling that bends the $\Gamma$--$\mathrm{K}$ bands also reshapes the isofrequency contours in a direction-dependent manner. For the $C_{6v}$ patch, the local capacitance dyadic is isotropic, and the contours remain nearly circular over most of the plotted range; any finite-wavevector deformation must nevertheless preserve the sixfold symmetry and need not preserve exact circularity. For the $C_1$ patch, the anisotropic local dyadic combines with the direction-dependent Floquet correction to produce much stronger contour deformation towards the zone boundary. Thus, the low-frequency near-circular contours do not imply an isotropic $C_1$ dyadic. The quasi-TE and hybrid TE--TM calculations reproduce the same main directional features, although including polarisation coupling produces more visible shifts in the outer $C_1$ contours. Such shifts can become more apparent as the band flattens, because a smaller radial frequency slope makes the contour position more sensitive to changes in the modal frequency. These full-zone contours illustrate the model predictions; the independent HFSS validation reported here remains the $\Gamma$--$\mathrm{K}$ sweep.

\begin{table}[htbp]
\caption{\label{tab:validation_stats}
Validation statistics by symmetry class: sweep-mean relative frequency error for the local dyadic model $\bar{\epsilon}_{\mathrm{local\ dyadic}}$, and the sweep-mean relative frequency error for the sideband-corrected  model $\bar{\epsilon}_{\mathrm{sideband\text{-}corrected}}$ against HFSS eigenmode data along $\Gamma$--$\mathrm{K}.$
}
\begin{ruledtabular}
\begin{tabular}{lcc}
\textrm{Point group $G$} &
$\bar{\epsilon}_{\mathrm{local\ dyadic}}$
&
$\bar{\epsilon}_{\mathrm{sideband\text{-}corrected}}$\\
\colrule
$C_{6v}$ & $3.54\%$ &$1.48\%$  \\
$C_{3v}$ & $2.71\%$ &$0.73\%$  \\
$C_{2v}$ & $9.10\%$ &$4.04\%$  \\
$C_{2}$ & $12.92\%$ &$3.70\%$  \\
$C_{s}$ & $10.96\%$ &$2.19\%$  \\
$C_{1}$ & $13.51\%$ &$1.02\%$  \\

\end{tabular}
\end{ruledtabular}
\end{table}

Table~\ref{tab:validation_stats} has the validation statistics for each symmetry class. The largest deviations occur in the transition region immediately beyond the spatial-dispersion onset, where the model can overestimate the modal frequency. This is the region where the six-sideband truncation of Eq.~\eqref{eq:six_order_disp} is least accurate: the retained channels are no longer deeply evanescent.

\section{Conclusion}
\label{sec:conclusion}

We have developed a geometry-driven homogenisation framework that
maps the unit-cell geometry of patch-type MTSs onto an effective
surface-capacitance dyadic and its leading spatially dispersive
correction. The resulting model predicts anisotropic surface-wave
dispersion without per-geometry full-wave sheet retrieval, fitted
sideband parameters, or a full-wave eigenmode band sweep for each
new patch. The framework separates the three ingredients: Bravais periodicity defines the two-component
macroscopic tangential loading space, gap connectivity
ensures a non-degenerate capacitance response for the geometries
considered here, and the point-group symmetry of the complete
unit-cell geometry constrains the admissible form of the
capacitance dyadic. A quasi-static boundary-value problem on a
single periodic cell then determines its independent entries, scaled by the analytically calculated capacitance of
the isotropic hexagonal reference patch. This construction
provides the model's quasi-static loading directly, avoiding the
evaluation of frequency-dependent periodic Green's functions for
sheet-parameter retrieval and the extrapolation of finite-frequency
or finite-Bloch-phase simulation data toward the static,
long-wavelength limit. The resulting dyadic determines the local
anisotropic dispersion through its TE projection along the
propagation direction.

The leading spatially dispersive correction retains the six Floquet orders of the first reciprocal-shell. The coupling to the fundamental is determined by the complex boundary Fourier overlap and the directional turning multiplier, both evaluated at each sampled Bloch wavevector along the complete $\Gamma$--$K$ path. Each order enters through the detuning $D_p=\gamma_{h_p}-\Gamma_p$, determined by its free-space evanescence and sheet loading. For propagation along a mirror axis of the complete unit-cell geometry, the complex coupling amplitudes are projected onto symmetry-allowed combinations of mirror-related orders. This projection retains their overlap phases and the full contribution of each pair. The correction becomes negligible relative to the local loading in the long-wavelength limit and captures the leading spatial-dispersion toward the Brillouin-zone boundary. 

We validated against full-wave eigenmode simulations
for six representative symmetry classes spanning
$C_{6v}$ to $C_1$. The sideband-corrected model reduces
the sweep-mean frequency error of the local dyadic
model from 2.7--13.5\% to 0.7--4.0\%, with the largest
improvements occurring for the low-symmetry geometries.
A new gap-connected patch therefore yields a predicted
band diagram without repeating a full-wave eigenmode
campaign, while the same geometry-to-dispersion map
provides a route toward inverse design of anisotropic
surface-wave propagation.
The framework also provides a basis for extensions
to other propagation directions, including
$\Gamma$--$\mathrm M$, and to additional reciprocal
shells.
Substrate loading may be modeled using the effective permittivity of Eq.~\eqref{eq:eps_eff}, subject to separate validation.


\clearpage
\onecolumngrid

\setcounter{section}{0}
\setcounter{subsection}{0}
\setcounter{subsubsection}{0}
\setcounter{equation}{0}
\setcounter{figure}{0}
\setcounter{table}{0}
\renewcommand{\thesection}{S\arabic{section}}
\renewcommand{\theequation}{S\arabic{equation}}
\renewcommand{\thefigure}{S\arabic{figure}}
\renewcommand{\thetable}{S\arabic{table}}
\renewcommand{\theHsection}{supp.\arabic{section}}
\renewcommand{\theHsubsection}{\theHsection.\arabic{subsection}}
\renewcommand{\theHsubsubsection}{\theHsubsection.\arabic{subsubsection}}
\renewcommand{\theHequation}{supp.\arabic{equation}}
\renewcommand{\theHfigure}{supp.\arabic{figure}}
\renewcommand{\theHtable}{supp.\arabic{table}}

\begin{center}
{\large\textbf{Supplemental material for \\
Geometry-Derived Capacitance Dyadics with Floquet Sideband Correction for
Anisotropic Periodic Patch Metasurfaces}\par}
\medskip
Arghyadeep Pal, Federico Giusti, and Stefano Maci\par
\medskip
Department of Information Engineering and Mathematics,\par
University of Siena,\par
53100 Siena, Italy
\end{center}
\medskip

\section{Group-theoretic constraints on the surface-capacitance dyadic}
\label{sec:S1_symmetry}

This section provides the explicit derivation of the
group-theoretic tensor forms summarized in Table~\ref{tab:tensor_forms} of Sec.~\ref{subsec:group_theory} in the
main text.  For a reciprocal lossless sheet, the in-plane
surface-capacitance dyadic is real and symmetric, and may
therefore be written in an arbitrary Cartesian basis as
\begin{equation}
\overline{\overline C}
=
\begin{pmatrix}
C_{xx} & C_{xy}\\
C_{xy} & C_{yy}
\end{pmatrix}.
\label{eq:S_C_general}
\end{equation}
Let $G$ denote the in-plane point group of the complete
unit-cell geometry.  If $g\in G$ is represented in the
tangential plane by the orthogonal matrix
$\overline{\overline D}(g)$, invariance of the equivalent
response requires
\begin{equation}
\overline{\overline D}(g)
\cdot
\overline{\overline C}
\cdot
\overline{\overline D}^{\,T}(g)
=
\overline{\overline C},
\qquad
\forall g\in G .
\label{eq:S_invariance}
\end{equation}
The forms derived below are the symmetry-allowed tensor forms
obtained from Eq.~\eqref{eq:S_invariance}. Lower point symmetry
permits anisotropy but does not require it; a particular geometry
may still exhibit accidental isotropy.

\subsection{Rotational symmetry}

A counterclockwise in-plane rotation by an angle $\theta$ is
represented by
\begin{equation}
\overline{\overline R}(\theta)
=
\begin{pmatrix}
\cos\theta & -\sin\theta\\
\sin\theta & \cos\theta
\end{pmatrix}.
\label{eq:S_rotation_matrix}
\end{equation}
The capacitance dyadic is separated into its isotropic and
traceless parts by defining
\begin{equation}
 \overline C
=
\frac{C_{xx}+C_{yy}}{2},
\qquad
\Delta C
=
C_{xx}-C_{yy}.
\label{eq:S_mean_delta}
\end{equation}
Equation~\eqref{eq:S_C_general} can then be written as
\begin{equation}
\overline{\overline C}
=
\overline C\,\overline{\overline I}
+
\frac{1}{2}
\begin{pmatrix}
\Delta C & 2C_{xy}\\
2C_{xy} & -\Delta C
\end{pmatrix}.
\label{eq:S_iso_aniso_decomp}
\end{equation}
For convenience, we define the traceless anisotropy matrix
\begin{equation}
\overline{\overline {C}}_{\mathrm{an}}
=
\begin{pmatrix}
\Delta C & 2C_{xy}\\
2C_{xy} & -\Delta C
\end{pmatrix},
\label{eq:S_A_def}
\end{equation}
so that
\begin{equation}
\overline{\overline C}
=
\overline C\,\overline{\overline I}
+
\frac{1}{2}\overline{\overline {C}}_{\mathrm{an}}.
\label{eq:S_C_A}
\end{equation}

Under the rotation of Eq.~\eqref{eq:S_rotation_matrix}, the
transformed capacitance dyadic is
\begin{equation}
\overline{\overline C}'
=
\overline{\overline R}(\theta)
\cdot
\overline{\overline C}
\cdot
\overline{\overline R}^{\,T}(\theta).
\label{eq:S_C_transform}
\end{equation}
Substituting Eq.~\eqref{eq:S_C_A} gives
\begin{align}
\overline{\overline C}'
&=
\overline C\,
\overline{\overline R}(\theta)
\cdot
\overline{\overline I}
\cdot
\overline{\overline R}^{\,T}(\theta)
+
\frac{1}{2}
\overline{\overline R}(\theta)
\cdot
\overline{\overline {C}}_{\mathrm{an}}
\cdot
\overline{\overline R}^{\,T}(\theta)
\nonumber\\
&=
\overline C\,\overline{\overline I}
+
\frac{1}{2}
\overline{\overline R}(\theta)
\cdot
\overline{\overline {C}}_{\mathrm{an}}
\cdot
\overline{\overline R}^{\,T}(\theta),
\label{eq:S_C_transform_split}
\end{align}
where
\[
\overline{\overline R}(\theta)
\cdot
\overline{\overline R}^{\,T}(\theta)
=
\overline{\overline I}.
\]
Thus, only the traceless part needs to be transformed.

The first matrix multiplication gives
\begin{align}
\overline{\overline R}(\theta)
\cdot
\overline{\overline {C}}_{\mathrm{an}}
&=
\begin{pmatrix}
\cos\theta & -\sin\theta\\
\sin\theta & \cos\theta
\end{pmatrix}
\begin{pmatrix}
\Delta C & 2C_{xy}\\
2C_{xy} & -\Delta C
\end{pmatrix}
\nonumber\\[1mm]
&=
\begin{pmatrix}
\Delta C\cos\theta
-
2C_{xy}\sin\theta
&
2C_{xy}\cos\theta
+
\Delta C\sin\theta
\\
\Delta C\sin\theta
+
2C_{xy}\cos\theta
&
2C_{xy}\sin\theta
-
\Delta C\cos\theta
\end{pmatrix}.
\label{eq:S_RA}
\end{align}
Multiplication by
$\overline{\overline R}^{\,T}(\theta)$ then yields
\begin{align}
&
\overline{\overline R}(\theta)
\cdot
\overline{\overline {C}}_{\mathrm{an}}
\cdot
\overline{\overline R}^{\,T}(\theta)
\nonumber\\
&=
\begin{pmatrix}
\Delta C(\cos^2\theta-\sin^2\theta)
-
4C_{xy}\sin\theta\cos\theta
&
2\Delta C\sin\theta\cos\theta
+
2C_{xy}(\cos^2\theta-\sin^2\theta)
\\[1mm]
2\Delta C\sin\theta\cos\theta
+
2C_{xy}(\cos^2\theta-\sin^2\theta)
&
-\Delta C(\cos^2\theta-\sin^2\theta)
+
4C_{xy}\sin\theta\cos\theta
\end{pmatrix}.
\label{eq:S_RART_expanded}
\end{align}
Using
\begin{equation}
\cos 2\theta
=
\cos^2\theta-\sin^2\theta,
\qquad
\sin 2\theta
=
2\sin\theta\cos\theta,
\label{eq:S_double_angle_identities}
\end{equation}
this becomes
\begin{equation}
\overline{\overline R}(\theta)
\cdot
\overline{\overline {C}}_{\mathrm{an}}
\cdot
\overline{\overline R}^{\,T}(\theta)
=
\begin{pmatrix}
\Delta C\cos2\theta
-
2C_{xy}\sin2\theta
&
\Delta C\sin2\theta
+
2C_{xy}\cos2\theta
\\
\Delta C\sin2\theta
+
2C_{xy}\cos2\theta
&
-\Delta C\cos2\theta
+
2C_{xy}\sin2\theta
\end{pmatrix}.
\label{eq:S_RART_double}
\end{equation}
The transformed dyadic can also be written in the same
isotropic--traceless form,
\begin{equation}
\overline{\overline C}'
=
\overline C\,\overline{\overline I}
+
\frac{1}{2}
\begin{pmatrix}
\Delta C' & 2C_{xy}'\\
2C_{xy}' & -\Delta C'
\end{pmatrix}.
\label{eq:S_Cprime_decomp}
\end{equation}
Comparison with Eq.~\eqref{eq:S_RART_double} gives
\begin{align}
\Delta C'
&=
\Delta C\cos2\theta
-
2C_{xy}\sin2\theta,
\label{eq:S_Delta_transform}
\\
2C_{xy}'
&=
\Delta C\sin2\theta
+
2C_{xy}\cos2\theta.
\label{eq:S_Cxy_transform}
\end{align}
Hence,
\begin{equation}
\begin{pmatrix}
\Delta C'\\
2C_{xy}'
\end{pmatrix}
=
\begin{pmatrix}
\cos2\theta & -\sin2\theta\\
\sin2\theta & \cos2\theta
\end{pmatrix}
\begin{pmatrix}
\Delta C\\
2C_{xy}
\end{pmatrix}.
\label{eq:S_double_angle}
\end{equation}
Equation~\eqref{eq:S_double_angle} shows that the two independent
components of the traceless part of the symmetric in-plane
capacitance dyadic transform as
\begin{equation}
\begin{pmatrix}
\Delta C'\\
2C_{xy}'
\end{pmatrix}
=
\overline{\overline R}(2\theta)
\cdot
\begin{pmatrix}
\Delta C\\
2C_{xy}
\end{pmatrix}.
\label{eq:S_anisotropy_transform}
\end{equation}
Thus, the traceless part transforms in the same way as a
two-dimensional vector under a rotation through the doubled
angle $2\theta$.

For a rotation that belongs to the point group of the complete
unit-cell geometry, the capacitance dyadic must be invariant,
so that
\begin{equation}
\overline{\overline C}'
=
\overline{\overline C}.
\label{eq:S_rotational_invariance}
\end{equation}
Equivalently,
\begin{equation}
\Delta C'=\Delta C,
\qquad
C_{xy}'=C_{xy}.
\label{eq:S_component_invariance}
\end{equation}
Substituting these relations into
Eq.~\eqref{eq:S_anisotropy_transform}, the rotational
invariance condition becomes
\begin{equation}
\begin{pmatrix}
\Delta C\\
2C_{xy}
\end{pmatrix}
=
\overline{\overline R}(2\theta)
\cdot
\begin{pmatrix}
\Delta C\\
2C_{xy}
\end{pmatrix}.
\label{eq:S_anisotropy_invariance}
\end{equation}
Equation~\eqref{eq:S_anisotropy_invariance} gives the
rotational invariance condition for the anisotropic part of
the surface-capacitance dyadic.

\subsubsection{Three- and six-fold rotational symmetry}

The three-fold and six-fold cyclic groups are written explicitly as
\begin{align}
C_3
&=
\left\{
\overline{\overline I},
\overline{\overline R}\!\left(\frac{2\pi}{3}\right),
\overline{\overline R}\!\left(\frac{4\pi}{3}\right)
\right\},
\label{eq:S_C3_group}
\\
C_6
&=
\left\{
\overline{\overline I},
\overline{\overline R}\!\left(\frac{\pi}{3}\right),
\overline{\overline R}\!\left(\frac{2\pi}{3}\right),
\overline{\overline R}(\pi),
\overline{\overline R}\!\left(\frac{4\pi}{3}\right),
\overline{\overline R}\!\left(\frac{5\pi}{3}\right)
\right\}.
\label{eq:S_C6_group}
\end{align}
For a rotation belonging to the point group of the unit-cell
geometry, Eq.~\eqref{eq:S_anisotropy_invariance} can be
rearranged as
\begin{equation}
\left[
\overline{\overline I}
-
\overline{\overline R}(2\theta)
\right]
\begin{pmatrix}
\Delta C\\
2C_{xy}
\end{pmatrix}
=
\begin{pmatrix}
0\\
0
\end{pmatrix}.
\label{eq:S_rot_invariance_homogeneous}
\end{equation}
A nonzero anisotropic part can exist only if the matrix in
Eq.~\eqref{eq:S_rot_invariance_homogeneous} is singular, i.e.,
\begin{equation}
\det
\left[
\overline{\overline I}
-
\overline{\overline R}(2\theta)
\right]
=
0.
\label{eq:S_rot_det_condition}
\end{equation}
Using
\begin{equation}
\overline{\overline I}
-
\overline{\overline R}(2\theta)
=
\begin{pmatrix}
1-\cos2\theta & \sin2\theta\\
-\sin2\theta & 1-\cos2\theta
\end{pmatrix},
\label{eq:S_I_minus_R}
\end{equation}
the determinant becomes
\begin{align}
\det
\left[
\overline{\overline I}
-
\overline{\overline R}(2\theta)
\right]
&=
(1-\cos2\theta)^2+\sin^22\theta
\nonumber\\
&=
1-2\cos2\theta+\cos^22\theta+\sin^22\theta
\nonumber\\
&=
2-2\cos2\theta
\nonumber\\
&=
4\sin^2\theta .
\label{eq:S_rot_det}
\end{align}

For $C_3$, the physical rotation angle is
\begin{equation}
\theta=\frac{2\pi}{3}.
\label{eq:S_C3_theta}
\end{equation}
Equation~\eqref{eq:S_rot_invariance_homogeneous} therefore
takes the explicit form
\begin{equation}
\left[
\overline{\overline I}
-
\overline{\overline R}\!\left(\frac{4\pi}{3}\right)
\right]
\begin{pmatrix}
\Delta C\\
2C_{xy}
\end{pmatrix}
=
\begin{pmatrix}
0\\
0
\end{pmatrix},
\label{eq:S_C3_system}
\end{equation}
where
\begin{equation}
\overline{\overline I}
-
\overline{\overline R}\!\left(\frac{4\pi}{3}\right)
=
\begin{pmatrix}
\frac{3}{2} & -\frac{\sqrt{3}}{2}\\[1mm]
\frac{\sqrt{3}}{2} & \frac{3}{2}
\end{pmatrix}.
\label{eq:S_C3_matrix}
\end{equation}
Its determinant is
\begin{equation}
\det
\left[
\overline{\overline I}
-
\overline{\overline R}\!\left(\frac{4\pi}{3}\right)
\right]
=
3\neq0.
\label{eq:S_C3_det}
\end{equation}
Hence Eq.~\eqref{eq:S_C3_system} has only the trivial
solution,
\begin{equation}
\Delta C=0,
\qquad
C_{xy}=0.
\label{eq:S_C3_constraints}
\end{equation}
Since
\begin{equation}
\Delta C=C_{xx}-C_{yy},
\label{eq:S_delta_definition_C3}
\end{equation}
Eq.~\eqref{eq:S_C3_constraints} gives
\begin{equation}
C_{xx}=C_{yy}\equiv C,
\qquad
C_{xy}=0,
\label{eq:S_C3_components}
\end{equation}
and therefore
\begin{equation}
\overline{\overline C}
=
C\,\overline{\overline I}.
\label{eq:S_C3_isotropic}
\end{equation}

For $C_6$, the physical rotation angle is
\begin{equation}
\theta=\frac{\pi}{3}.
\label{eq:S_C6_theta}
\end{equation}
Equation~\eqref{eq:S_rot_invariance_homogeneous} becomes
\begin{equation}
\left[
\overline{\overline I}
-
\overline{\overline R}\!\left(\frac{2\pi}{3}\right)
\right]
\begin{pmatrix}
\Delta C\\
2C_{xy}
\end{pmatrix}
=
\begin{pmatrix}
0\\
0
\end{pmatrix},
\label{eq:S_C6_system}
\end{equation}
with
\begin{equation}
\overline{\overline I}
-
\overline{\overline R}\!\left(\frac{2\pi}{3}\right)
=
\begin{pmatrix}
\frac{3}{2} & \frac{\sqrt{3}}{2}\\[1mm]
-\frac{\sqrt{3}}{2} & \frac{3}{2}
\end{pmatrix}.
\label{eq:S_C6_matrix}
\end{equation}
Again,
\begin{equation}
\det
\left[
\overline{\overline I}
-
\overline{\overline R}\!\left(\frac{2\pi}{3}\right)
\right]
=
3\neq0,
\label{eq:S_C6_det}
\end{equation}
and hence
\begin{equation}
\Delta C=0,
\qquad
C_{xy}=0.
\label{eq:S_C6_constraints}
\end{equation}
Thus,
\begin{equation}
C_3,\ C_6:
\qquad
\overline{\overline C}
=
C\,\overline{\overline I}.
\label{eq:S_C3_C6_isotropy}
\end{equation}
The mirror-containing groups additionally contain reflections.
A mirror whose in-plane axis makes an angle $\phi$ with the
$x$-axis is represented by
\begin{equation}
\overline{\overline M}(\phi)
=
\overline{\overline R}(\phi)
\cdot
\begin{pmatrix}
1 & 0\\
0 & -1
\end{pmatrix}
\cdot
\overline{\overline R}^{\,T}(\phi).
\label{eq:S_general_mirror}
\end{equation}
Carrying out the multiplication gives
\begin{equation}
\overline{\overline M}(\phi)
=
\begin{pmatrix}
\cos 2\phi & \sin 2\phi\\
\sin 2\phi & -\cos 2\phi
\end{pmatrix}.
\label{eq:S_general_mirror_explicit}
\end{equation}
The three mirror operations of $C_{3v}$ are therefore
\begin{equation}
C_{3v}
=
C_3
\cup
\left\{
\overline{\overline M}(0),
\overline{\overline M}\!\left(\frac{\pi}{3}\right),
\overline{\overline M}\!\left(\frac{2\pi}{3}\right)
\right\},
\label{eq:S_C3v_group}
\end{equation}
whereas the six mirror operations of $C_{6v}$ give
\begin{align}
C_{6v}
=
C_6
\cup
\bigg\{
&
\overline{\overline M}(0),
\overline{\overline M}\!\left(\frac{\pi}{6}\right),
\overline{\overline M}\!\left(\frac{\pi}{3}\right),
\overline{\overline M}\!\left(\frac{\pi}{2}\right),
\nonumber\\
&
\overline{\overline M}\!\left(\frac{2\pi}{3}\right),
\overline{\overline M}\!\left(\frac{5\pi}{6}\right)
\bigg\}.
\label{eq:S_C6v_group}
\end{align}
Since
\begin{equation}
C_3\subset C_{3v},
\qquad
C_6\subset C_{6v},
\label{eq:S_Cnv_subgroups}
\end{equation}
invariance under $C_{3v}$ or $C_{6v}$ necessarily includes
invariance under the corresponding rotational subgroup.
Therefore, the rotational constraints derived above already give
\begin{equation}
C_3,\ C_{3v},\ C_6,\ C_{6v}:
\qquad
\overline{\overline C}
=
C\,\overline{\overline I}.
\label{eq:S_highrot_summary}
\end{equation}
\begin{table}[htbp]
\caption{
Rotational constraints on the anisotropic part of the
surface-capacitance dyadic. The angle $\theta$ denotes the
physical rotation angle contained in the corresponding point
group.
}
\label{tab:S_rotational_constraints}
\begin{ruledtabular}
\begin{tabular}{cccc}
Point group
&
$\theta$
&
$\det[\overline{\overline I}-\overline{\overline R}(2\theta)]$
&
Constraint
\\
\hline
$C_3$
&
$2\pi/3$
&
$3$
&
$\Delta C=C_{xy}=0$
\\
$C_6$
&
$\pi/3$
&
$3$
&
$\Delta C=C_{xy}=0$
\\
$C_{3v}$
&
$2\pi/3$
&
$3$
&
$\Delta C=C_{xy}=0$
\\
$C_{6v}$
&
$\pi/3$
&
$3$
&
$\Delta C=C_{xy}=0$
\end{tabular}
\end{ruledtabular}
\end{table}

\subsubsection{Two-fold rotational symmetry}

The two-fold cyclic group is
\begin{equation}
C_2
=
\left\{
\overline{\overline I},
\overline{\overline R}(\pi)
\right\}.
\label{eq:S_C2_group}
\end{equation}
For a rotation by $\pi$,
\begin{equation}
\overline{\overline R}(\pi)
=
\begin{pmatrix}
-1 & 0\\
0 & -1
\end{pmatrix}
=
-\overline{\overline I}.
\label{eq:S_C2_rotation}
\end{equation}
The invariance condition therefore gives
\begin{align}
\overline{\overline R}(\pi)
\cdot
\overline{\overline C}
\cdot
\overline{\overline R}^{\,T}(\pi)
&=
\left(-\overline{\overline I}\right)
\cdot
\overline{\overline C}
\cdot
\left(-\overline{\overline I}\right)
\nonumber\\
&=
\overline{\overline C}.
\label{eq:S_C2_invariance}
\end{align}
Thus the invariance condition is satisfied identically,
independently of the values of $C_{xx}$, $C_{yy}$, and
$C_{xy}$. Consequently, $C_2$ symmetry imposes no additional
constraint beyond the symmetry of the capacitance dyadic:
\begin{equation}
C_2:
\qquad
\overline{\overline C}
=
\begin{pmatrix}
C_{xx} & C_{xy}\\
C_{xy} & C_{yy}
\end{pmatrix}.
\label{eq:S_C2_general}
\end{equation}
The two eigenvalues and the orientation of the principal axes
therefore remain unrestricted by $C_2$ symmetry.

\subsection{Mirror symmetry}

We now determine the constraint imposed directly by a mirror operation on the capacitance dyadic. The general in-plane reflection $\overline{\overline M}(\phi)$ was introduced in Eq.~\eqref{eq:S_general_mirror_explicit}. Choosing a coordinate frame
$(\widehat{\mathbf e}_{\parallel},\widehat{\mathbf e}_{\perp})$ aligned with the mirror axis corresponds to $\phi=0$, in which case
\begin{equation}
    \overline{\overline M}(0)
    =
    \begin{pmatrix}
    1 & 0\\
    0 & -1
    \end{pmatrix}
    \label{eq:S_mirror_aligned}
\end{equation}
In this basis, the most general symmetric capacitance dyadic is 
\begin{equation}
\overline{\overline C}_{\parallel\perp}
=
\begin{pmatrix}
C_{\parallel\parallel} & C_{\parallel\perp}\\
C_{\parallel\perp} & C_{\perp\perp}
\end{pmatrix}.
\label{eq:S_C_mirror_frame}
\end{equation}
Mirror invariance requires
\begin{align}
\overline{\overline M}(0)
\cdot
\overline{\overline C}_{\parallel\perp}
\cdot
\overline{\overline M}^{\,T}(0)
&=
\begin{pmatrix}
1 & 0\\
0 & -1
\end{pmatrix}
\begin{pmatrix}
C_{\parallel\parallel} & C_{\parallel\perp}\\
C_{\parallel\perp} & C_{\perp\perp}
\end{pmatrix}
\begin{pmatrix}
1 & 0\\
0 & -1
\end{pmatrix}
\nonumber\\
&=
\begin{pmatrix}
C_{\parallel\parallel} & -C_{\parallel\perp}\\
-C_{\parallel\perp} & C_{\perp\perp}
\end{pmatrix}
=
\overline{\overline C}_{\parallel\perp}.
\label{eq:S_mirror_invariance}
\end{align}
Therefore,
\begin{equation}
C_{\parallel\perp}=0,
\label{eq:S_mirror_offdiag}
\end{equation}
and the allowed dyadic form in the mirror-aligned basis is
\begin{equation}
\overline{\overline C}_{\parallel\perp}
=
\begin{pmatrix}
C_{\parallel} & 0\\
0 & C_{\perp}
\end{pmatrix}.
\label{eq:S_mirror_diagonal}
\end{equation}
Thus the directions parallel and perpendicular to the mirror
are principal axes of the capacitance dyadic.

\subsubsection{$C_s$ symmetry}

The group $C_s$ contains the identity and a single mirror
operation,
\begin{equation}
C_s
=
\left\{
\overline{\overline I},
\overline{\overline M}(\phi)
\right\},
\label{eq:S_Cs_group}
\end{equation}
where the mirror axis makes an angle $\phi$ with the fixed
Cartesian $x$-axis.  In the mirror-aligned basis,
Eq.~\eqref{eq:S_mirror_diagonal} gives
\begin{equation}
\overline{\overline C}_{\parallel\perp}
=
\begin{pmatrix}
C_{\parallel} & 0\\
0 & C_{\perp}
\end{pmatrix}.
\label{eq:S_Cs_mirror_frame}
\end{equation}
Transforming this dyadic to the fixed Cartesian frame gives
\begin{equation}
\overline{\overline C}
=
\overline{\overline R}(\phi)
\cdot
\begin{pmatrix}
C_{\parallel} & 0\\
0 & C_{\perp}
\end{pmatrix}
\cdot
\overline{\overline R}^{\,T}(\phi).
\label{eq:S_Cs_rotation}
\end{equation}
Carrying out the multiplication,
\begin{equation}
\overline{\overline C}
=
\begin{pmatrix}
C_{\parallel}\cos^2\phi+C_{\perp}\sin^2\phi
&
(C_{\parallel}-C_{\perp})\sin\phi\cos\phi
\\[1mm]
(C_{\parallel}-C_{\perp})\sin\phi\cos\phi
&
C_{\parallel}\sin^2\phi+C_{\perp}\cos^2\phi
\end{pmatrix}.
\label{eq:S_Cs_explicit}
\end{equation}
Therefore,
\begin{align}
C_{xx}
&=
C_{\parallel}\cos^2\phi
+
C_{\perp}\sin^2\phi,
\label{eq:S_Cs_xx}
\\
C_{yy}
&=
C_{\parallel}\sin^2\phi
+
C_{\perp}\cos^2\phi,
\label{eq:S_Cs_yy}
\\
C_{xy}
&=
\left(C_{\parallel}-C_{\perp}\right)
\sin\phi\cos\phi .
\label{eq:S_Cs_xy}
\end{align}
Hence the allowed tensor form may be summarized as
\begin{equation}
C_s:
\qquad
\overline{\overline C}
=
\overline{\overline R}(\phi)
\cdot
\begin{pmatrix}
C_{\parallel} & 0\\
0 & C_{\perp}
\end{pmatrix}
\cdot
\overline{\overline R}^{\,T}(\phi).
\label{eq:S_Cs_summary}
\end{equation}
Thus the two principal capacitances remain independent, while
their principal axes are fixed parallel and perpendicular to
the mirror.

\subsubsection{$C_{2v}$ symmetry}

Choosing the Cartesian axes along the two mutually
perpendicular mirror lines, the $C_{2v}$ group can be written as
\begin{equation}
C_{2v}
=
\left\{
\overline{\overline I},
\overline{\overline R}(\pi),
\overline{\overline M}(0),
\overline{\overline M}\!\left(\frac{\pi}{2}\right)
\right\}.
\label{eq:S_C2v_group}
\end{equation}
The two mirror operations are
\begin{equation}
\overline{\overline M}(0)
=
\begin{pmatrix}
1 & 0\\
0 & -1
\end{pmatrix},
\qquad
\overline{\overline M}\!\left(\frac{\pi}{2}\right)
=
\begin{pmatrix}
-1 & 0\\
0 & 1
\end{pmatrix}.
\label{eq:S_C2v_mirrors}
\end{equation}
Starting from the general symmetric dyadic
\begin{equation}
\overline{\overline C}
=
\begin{pmatrix}
C_{xx} & C_{xy}\\
C_{xy} & C_{yy}
\end{pmatrix},
\label{eq:S_C2v_general}
\end{equation}
invariance under the first mirror gives
\begin{align}
\overline{\overline M}(0)
\cdot
\overline{\overline C}
\cdot
\overline{\overline M}^{\,T}(0)
&=
\begin{pmatrix}
C_{xx} & -C_{xy}\\
-C_{xy} & C_{yy}
\end{pmatrix}
\nonumber\\
&=
\overline{\overline C},
\label{eq:S_C2v_mirror_invariance}
\end{align}
and therefore
\begin{equation}
C_{xy}=0.
\label{eq:S_C2v_offdiag}
\end{equation}
The second mirror gives the same condition,
\begin{equation}
\overline{\overline M}\!\left(\frac{\pi}{2}\right)
\cdot
\overline{\overline C}
\cdot
\overline{\overline M}^{\,T}\!\left(\frac{\pi}{2}\right)
=
\overline{\overline C},
\qquad
C_{xy}=0.
\label{eq:S_C2v_second_mirror}
\end{equation}
The two-fold rotation
$\overline{\overline R}(\pi)=-\overline{\overline I}$
introduces no additional constraint, as shown in
Eq.~\eqref{eq:S_C2_invariance}.  The allowed form is therefore
\begin{equation}
C_{2v}:
\qquad
\overline{\overline C}
=
\begin{pmatrix}
C_{xx} & 0\\
0 & C_{yy}
\end{pmatrix}.
\label{eq:S_C2v}
\end{equation}
Thus $C_{xx}$ and $C_{yy}$ remain independent, while the two
mirror axes fix the principal directions of the dyadic.

\subsection{Trivial symmetry $C_1$}

The trivial point group contains only the identity operation,
\begin{equation}
C_1
=
\left\{
\overline{\overline I}
\right\}.
\label{eq:S_C1_group}
\end{equation}
The invariance condition is therefore
\begin{equation}
\overline{\overline I}
\cdot
\overline{\overline C}
\cdot
\overline{\overline I}^{\,T}
=
\overline{\overline C},
\label{eq:S_C1_invariance}
\end{equation}
which is identically satisfied. Hence $C_1$ imposes no
point-symmetry constraint on the capacitance dyadic. For a
reciprocal lossless sheet, the most general allowed form is
\begin{equation}
C_1:
\qquad
\overline{\overline C}
=
\begin{pmatrix}
C_{xx} & C_{xy}\\
C_{xy} & C_{yy}
\end{pmatrix}.
\label{eq:S_C1}
\end{equation}
The response therefore contains three independent real
parameters, and the orientation of the principal axes is not
fixed by point symmetry.

\subsection{Directional-capacitance form}

Let $\lambda_1$ and $\lambda_2$ denote the two
principal values of the capacitance dyadic, and let the first
principal axis make an angle $\phi_C$ with the Cartesian
$x$-axis. The dyadic can then be written as
\begin{equation}
\overline{\overline C}
=
\overline{\overline R}(\phi_C)
\cdot
\begin{pmatrix}
\lambda_1 & 0\\
0 & \lambda_2
\end{pmatrix}
\cdot
\overline{\overline R}^{\,T}(\phi_C).
\label{eq:S_C_principal}
\end{equation}

The principal values and their orientation can be obtained
directly from the Cartesian components of the dyadic. Defining
the mean capacitance as
\begin{equation}
\overline C
=
\frac{C_{xx}+C_{yy}}{2}
=
\frac{1}{2}
\operatorname{tr}
\left(
\overline{\overline C}
\right),
\label{eq:S_mean_principal}
\end{equation}
the two principal values are
\begin{equation}
\lambda_{1,2}
=
\overline C
\pm
\frac{1}{2}
\sqrt{
\left(C_{xx}-C_{yy}\right)^2
+
4C_{xy}^{\,2}
},
\label{eq:S_principal_values}
\end{equation}
where the upper sign corresponds to $\lambda_1$. Their
orientation is
\begin{equation}
\phi_C
=
\frac{1}{2}
\operatorname{atan2}
\left(
2C_{xy},
C_{xx}-C_{yy}
\right)
\pmod{\pi}.
\label{eq:S_principal_angle}
\end{equation}
When $\lambda_1=\lambda_2$, the response is isotropic and
$\phi_C$ is arbitrary.

For an arbitrary in-plane direction
\begin{equation}
\widehat{\mathbf u}(\theta)
=
\cos\theta\,\widehat{\mathbf x}
+
\sin\theta\,\widehat{\mathbf y},
\label{eq:S_u_direction}
\end{equation}
the directional capacitance is
\begin{equation}
C(\widehat{\mathbf u})
=
\widehat{\mathbf u}
\cdot
\overline{\overline C}
\cdot
\widehat{\mathbf u}.
\label{eq:S_directional_definition}
\end{equation}
Substitution of Eq.~\eqref{eq:S_C_principal} gives
\begin{align}
C(\widehat{\mathbf u})
&=
\lambda_1
\cos^2(\theta-\phi_C)
+
\lambda_2
\sin^2(\theta-\phi_C)
\nonumber\\
&=
\frac{\lambda_1+\lambda_2}{2}
+
\frac{\lambda_1-\lambda_2}{2}
\cos\!\left[2(\theta-\phi_C)\right].
\label{eq:S_directional_capacitance}
\end{align}

The normalized directional-capacitance indicator plotted in
Fig.~\ref{fig:group_theory} of the main text becomes
\begin{equation}
\frac{C(\widehat{\mathbf u})}{\overline C}
=
1
+
\frac{\lambda_1-\lambda_2}
{\lambda_1+\lambda_2}
\cos\!\left[2(\theta-\phi_C)\right].
\label{eq:S_directional_normalized}
\end{equation}


\newpage
\section{Quasi-static corrector problem and numerical evaluation of the capacitance dyadic}
\label{sec:supp_corrector}

This section gives the mathematical and numerical details of the quasi-static
calculation introduced in Sec.~\ref{subsec:diff_geom} of the main text. We first define the
affine-periodic electrostatic cell problem and its two basis correctors. We
then derive the corresponding minimum-energy and Euler--Lagrange formulations,
describe the periodic triangular-graph approximation used for arbitrary patch
geometries, and finally identify the entries of the auxiliary capacitance
dyadic. The dyadic obtained here has units of \(\mathrm{F/m}\). Empty-cell
subtraction and conversion to the equivalent sheet-capacitance dyadic, with units
of \(\mathrm{F}\), are separate operations described in Sec.~\ref{subsec:calibration} of the main
text.

\subsection{Affine-periodic electrostatic cell problem}
\label{subsec:supp_cell_problem}

Let \(U\subset\mathbb R^2\) be a primitive cell with lattice translations
\(\mathbf a_1\) and \(\mathbf a_2\). The metallic region is the union of
\(N_c\) electrically disconnected perfect-electric-conductor (PEC)
components,
\begin{equation}
    \Sigma=\bigcup_{m=1}^{N_c}\Sigma_m,
    \qquad
    \Omega=U\setminus\Sigma,
    \label{eq:supp_geometry}
\end{equation}
where \(\Omega\) is the dielectric gap. Connectivity is understood on the
periodically identified cell; consequently, portions meeting through opposite
cell boundaries belong to the same physical component. We consider the
gap-connected geometries described in Sec.~\ref{subsec:topology} of the main text, for which
independent winding paths exist along both primitive translations.

For a dimensionless imposed in-plane direction
\begin{equation}
    \mathbf v
    =v_x\widehat{\mathbf e}_x+v_y\widehat{\mathbf e}_y,
    \label{eq:supp_loading}
\end{equation}
let \(E_0\) be the equivalent field amplitude, in \(\mathrm{V/m}\). Introduce
the normalized total potential-like function
\begin{equation}
    u_{\mathbf v}(\mathbf r)
    =\mathbf v\cdot\mathbf r+w_{\mathbf v}(\mathbf r),
    \qquad
    \Phi_{\mathbf v}(\mathbf r)=-E_0u_{\mathbf v}(\mathbf r).
    \label{eq:supp_total_potential}
\end{equation}
The associated microscopic electric field is
\begin{equation}
    \mathbf E_{\mathbf v}^{\mathrm{cell}}(\mathbf r)
    =-\nabla\Phi_{\mathbf v}
    =E_0\left(\mathbf v+\nabla w_{\mathbf v}\right).
    \label{eq:supp_cell_field}
\end{equation}
Thus \(u_{\mathbf v}\) and \(w_{\mathbf v}\) have units of length, whereas
\(\mathbf v+\nabla w_{\mathbf v}\) is dimensionless.

Only the microscopic correction is periodic:
\begin{subequations}
\label{eq:supp_affine_periodicity}
\begin{align}
    w_{\mathbf v}(\mathbf r+\mathbf a_i)
        &=w_{\mathbf v}(\mathbf r),
        \label{eq:supp_corrector_periodicity}\\
    u_{\mathbf v}(\mathbf r+\mathbf a_i)
        &=u_{\mathbf v}(\mathbf r)
          +\mathbf v\cdot\mathbf a_i,
        \qquad i=1,2.
        \label{eq:supp_total_quasiperiodicity}
\end{align}
\end{subequations}
Accordingly, the total potential is affine-periodic, while its gradient and
the microscopic cell field are periodic.

The tangential electric field vanishes on each PEC component. Therefore it is
the total potential, rather than the corrector alone, that is constant on a
conductor:
\begin{equation}
    u_{\mathbf v}(\mathbf r)
    =\mathbf v\cdot\mathbf r+w_{\mathbf v}(\mathbf r)
    =c_m(\mathbf v),
    \qquad \mathbf r\in\partial\Sigma_m.
    \label{eq:supp_pec_condition}
\end{equation}
The constants \(c_m\) are not prescribed because the conductors are floating;
their equivalent potentials are \(-E_0c_m\).

Let \(H^1_{\mathrm{per}}(\Omega)\) denote the Sobolev space whose traces agree
on paired cell boundaries. For a prescribed loading \(\mathbf v\), define the
affine admissible set
\begin{equation}
    \mathcal A_{\mathbf v}
    =\left\{
       w\in H^1_{\mathrm{per}}(\Omega):
       \mathbf v\cdot\mathbf r+w=c_m
       \ \text{on }\partial\Sigma_m
       \ \text{for some }c_m\in\mathbb R
      \right\}.
    \label{eq:supp_admissible_set}
\end{equation}
The field is unchanged under \(w_{\mathbf v}\mapsto w_{\mathbf v}+C\). A
convenient continuum gauge is
\begin{equation}
    \int_{\Omega}w_{\mathbf v}\,\mathrm dA=0.
    \label{eq:supp_continuum_gauge}
\end{equation}
This fixes only the arbitrary potential zero and has no effect on the field or
on the capacitance dyadic.

\subsection{Electrostatic energy and variational formulation}
\label{subsec:supp_variational_formulation}

For a trial corrector \(w\in\mathcal A_{\mathbf v}\), the cell-averaged
electrostatic energy density of the two-dimensional surrogate is
\begin{equation}
    \mathcal W_e[w;\mathbf v]
    =\frac{\epsilon_0E_0^2}{2|U|}
     \int_{\Omega}
     \left|\mathbf v+\nabla w\right|^2\,\mathrm dA.
    \label{eq:supp_trial_physical_energy}
\end{equation}
Here \(\epsilon_0E_0^2\) has units of \(\mathrm{J/m^3}\), while the area
integral divided by \(|U|\) is dimensionless. Equation
\eqref{eq:supp_trial_physical_energy} is therefore the raw energy-density
quantity associated with the auxiliary two-dimensional problem, not yet the
energy of the final sheet model. For a substrate-backed array,
\(\epsilon_0\) is replaced by the quasi-static effective permittivity defined
in Eq.~\eqref{eq:eps_eff} of the main text.

Since the positive factor \(\epsilon_0E_0^2\) does not affect the minimizer,
define the normalized Dirichlet functional
\begin{equation}
    \mathcal J_{\mathbf v}[w]
    =\frac{1}{2|U|}
     \int_{\Omega}
     \left|\mathbf v+\nabla w\right|^2\,\mathrm dA.
    \label{eq:supp_normalized_functional}
\end{equation}
The corrector is the minimum-energy solution
\begin{equation}
    w_{\mathbf v}
    \in
    \underset{w\in\mathcal A_{\mathbf v}/\mathbb R}
              {\operatorname{arg\,min}}
    \mathcal J_{\mathbf v}[w],
    \qquad
    W_e(\mathbf v)
    =\epsilon_0E_0^2\mathcal J_{\mathbf v}[w_{\mathbf v}].
    \label{eq:supp_minimization_problem}
\end{equation}

The admissible variations form the space
\begin{equation}
    \mathcal V
    =\left\{
       \varphi\in H^1_{\mathrm{per}}(\Omega):
       \varphi=d_m
       \ \text{on }\partial\Sigma_m
       \ \text{for some }d_m\in\mathbb R
      \right\}.
    \label{eq:supp_variation_space}
\end{equation}
The constants \(d_m\) may vary independently because the corresponding
conductor potentials are floating. For every \(\varphi\in\mathcal V\), the
first variation is
\begin{align}
    0
    &=\left.
      \frac{\mathrm d}{\mathrm dt}
      \mathcal J_{\mathbf v}[w_{\mathbf v}+t\varphi]
      \right|_{t=0}
      \nonumber\\
    &=\frac{1}{|U|}
      \int_{\Omega}
      \left(\mathbf v+\nabla w_{\mathbf v}\right)
      \cdot\nabla\varphi\,\mathrm dA.
    \label{eq:supp_first_variation}
\end{align}
Hence the weak cell problem is
\begin{equation}
    \int_{\Omega}
    \left(\mathbf v+\nabla w_{\mathbf v}\right)
    \cdot\nabla\varphi\,\mathrm dA=0,
    \qquad \forall\varphi\in\mathcal V.
    \label{eq:supp_weak_problem}
\end{equation}

Compactly supported variations give Laplace's equation in the gap. The
contributions from paired cell boundaries cancel because the field and the
test function are periodic. Independent variation of each \(d_m\) gives zero
total flux on every floating conductor. The equivalent strong problem is
\begin{subequations}
\label{eq:supp_strong_problem}
\begin{align}
    \nabla\cdot
    \left(\mathbf v+\nabla w_{\mathbf v}\right)&=0
        &&\text{in }\Omega,
        \label{eq:supp_gap_equation}\\
    \mathbf v\cdot\mathbf r+w_{\mathbf v}&=c_m(\mathbf v)
        &&\text{on }\partial\Sigma_m,
        \label{eq:supp_strong_pec}\\
    \int_{\partial\Sigma_m}
    \widehat{\mathbf n}_m\cdot
    \left(\mathbf v+\nabla w_{\mathbf v}\right)\,\mathrm d\ell&=0
        &&m=1,\ldots,N_c,
        \label{eq:supp_component_neutrality}\\
    w_{\mathbf v}(\mathbf r+\mathbf a_i)&=w_{\mathbf v}(\mathbf r)
        &&i=1,2,
        \label{eq:supp_strong_periodicity}
\end{align}
\end{subequations}
together with the gauge in Eq.~\eqref{eq:supp_continuum_gauge}. Here
\(\widehat{\mathbf n}_m\) points from the metal into the gap. Multiplication
of Eq.~\eqref{eq:supp_component_neutrality} by \(\epsilon_0E_0\) gives the
zero-net-charge condition on the \(m\)th floating conductor.

The cell problem is linear in the imposed loading. With the gauge fixed,
\begin{equation}
    w_{\mathbf v}=v_xw_x+v_yw_y,
    \qquad
    c_m(\mathbf v)=v_xc_{m,x}+v_yc_{m,y},
    \label{eq:supp_corrector_linearity}
\end{equation}
where \(w_x\) and \(w_y\) are generated by
\(\mathbf v=\widehat{\mathbf e}_x\) and
\(\mathbf v=\widehat{\mathbf e}_y\), respectively. Only these two basis
problems must therefore be solved.

\subsection{Periodic triangular-graph approximation}
\label{subsec:supp_graph_approximation}

The numerical implementation replaces the continuum problem by a periodic
triangular graph. The primitive parallelogram is sampled at
\begin{equation}
    \mathbf r_{ij}
    =s_i\mathbf a_1+t_j\mathbf a_2,
    \qquad
    s_i=\frac{i-\tfrac12}{N}-\frac12,
    \qquad
    t_j=\frac{j-\tfrac12}{N}-\frac12,
    \label{eq:supp_grid_points}
\end{equation}
for \(i,j=1,\ldots,N\). The value of \(N\) is chosen odd so that sampling
points do not lie systematically on symmetry or polygon-boundary lines. For
the metal-membership test, each point is folded into the Wigner--Seitz
representative of the same periodic cell; the associated lattice translation
is retained when affine edge offsets are formed.

Each sample is connected to its periodic neighbours in the three positive
triangular-grid directions
\begin{equation}
    \frac{\mathbf a_1}{N},
    \qquad
    \frac{\mathbf a_2}{N},
    \qquad
    \frac{-\mathbf a_1+\mathbf a_2}{N}.
    \label{eq:supp_bond_directions}
\end{equation}
Thus, each undirected physical bond is stored exactly once. Before contraction,
the periodic graph contains \(N^2\) samples and \(3N^2\) oriented bond
representatives.

Let
\begin{equation}
    \mathcal Q_N=\{\mathbf r_q\}_{q=1}^{N^2}
    =\mathcal G_N\mathbin{\dot\cup}
     \left(\mathbin{\dot\bigcup}_{m=1}^{N_c}\Sigma_{m,N}\right)
    \label{eq:supp_sample_partition}
\end{equation}
be the sampled points, partitioned into the gap samples \(\mathcal G_N\) and
the samples belonging to each periodically connected PEC component. Every gap
sample carries one independent degree of freedom, whereas all samples in the
same PEC component are contracted to one metal supernode. If \(N_d\) is the
number of gap samples, the number of graph unknowns is
\begin{equation}
    N_v=N_d+N_c.
    \label{eq:supp_number_unknowns}
\end{equation}

Introduce the point-to-unknown map
\begin{equation}
    \nu(q)=
    \begin{cases}
       m, & \mathbf r_q\in\Sigma_{m,N},\\
       N_c+j(q), & \mathbf r_q\text{ is the }j(q)\text{th gap sample},
    \end{cases}
    \label{eq:supp_variable_map}
\end{equation}
and the graph coordinate functions
\begin{equation}
    \psi_p(\mathbf r_q)=\delta_{p,\nu(q)},
    \qquad p=1,\ldots,N_v.
    \label{eq:supp_graph_coordinate_basis}
\end{equation}

The resulting discrete potential space is
\begin{equation}
    \mathcal X_N
    =
    \operatorname{span}
    \left\{
    \psi_1,\ldots,\psi_{N_v}
    \right\}.
    \label{eq:supp_discrete_space}
\end{equation}
For a gap degree of freedom, the corresponding basis function is nonzero at
only one sampled gap point. By contrast, the basis function associated with
the \(m\)th metal supernode is equal to unity at every sampled point belonging
to \(\Sigma_{m,N}\) and vanishes elsewhere. Thus, the contraction of a
connected PEC component to one basis function imposes its equipotential
condition directly.

These are coordinate functions on the sampled graph and are not being
identified with conforming \(H^1\) finite-element basis functions.

For loading direction \(\xi\in\{x,y\}\), let
\(\mathbf z_\xi\in\mathbb R^{N_v}\) contain the graph values of the normalized
total affine potential. At a representative sample,
\begin{equation}
    u_{\xi,N}(\mathbf r_q)
    =\sum_{p=1}^{N_v}z_{\xi,p}\psi_p(\mathbf r_q)
    =z_{\xi,\nu(q)},
    \label{eq:supp_discrete_total_potential}
\end{equation}
up to the known affine lattice translation attached to that representative.
In particular, \(z_{\xi,m}=c_{m,\xi}\) is the constant total potential of the
\(m\)th metal component; it is not a constant value of the periodic
corrector. In continuum notation the corresponding corrector trace is
\begin{equation}
    w_\xi(\mathbf r)
    =c_{m,\xi}-\widehat{\mathbf e}_\xi\cdot\mathbf r,
    \qquad \mathbf r\in\partial\Sigma_m.
    \label{eq:supp_corrector_metal_trace}
\end{equation}

Choose one orientation for each stored bond. An edge
\(e=(q,r;\mathbf R_e)\) joins sample \(q\) to the appropriate periodic image
of sample \(r\). The vector \(\mathbf R_e\) collects the lattice translation
produced by a periodic wrap and, when a metal sample has been folded to the
Wigner--Seitz representative, the corresponding change of representative.
After metal contraction, define the incidence matrix by
\begin{equation}
    B_{e,\nu(q)}=-1,
    \qquad
    B_{e,\nu(r)}=+1,
    \label{eq:supp_incidence_convention}
\end{equation}
with all other entries in row \(e\) equal to zero. A bond whose two endpoints
map to the same contracted unknown has a zero incidence row and is discarded
from the active solve-edge set
\begin{equation}
    \mathcal E_N^{\mathrm{act}}
    =\{e:\nu(q_e)\neq\nu(r_e)\}.
    \label{eq:supp_active_edges}
\end{equation}

For the \(\xi\)-directed basis problem, affine periodicity supplies the known
offset
\begin{equation}
    (\mathbf g_\xi)_e
    =\widehat{\mathbf e}_\xi\cdot\mathbf R_e.
    \label{eq:supp_periodic_edge_offset}
\end{equation}
The oriented total-potential drop is therefore
\begin{align}
    d_{\xi,e}
    &=z_{\xi,\nu(r_e)}-z_{\xi,\nu(q_e)}
      +\widehat{\mathbf e}_\xi\cdot\mathbf R_e
      \nonumber\\
    &=\left(\mathbf B\mathbf z_\xi+\mathbf g_\xi\right)_e.
    \label{eq:supp_edge_drop}
\end{align}
The entries of \(\mathbf z_\xi\), \(\mathbf g_\xi\), and
\(\mathbf d_\xi\) all have units of length.

For an equilateral triangular grid, the cotangent weight of every periodic
bond is
\begin{equation}
    \kappa_\triangle
    =\frac12\left(\cot\frac{\pi}{3}+\cot\frac{\pi}{3}\right)
    =\frac{1}{\sqrt3}.
    \label{eq:supp_triangular_weight}
\end{equation}
Because every undirected bond is stored once, the discrete energy used to
assemble the graph problem is
\begin{equation}
    \mathcal J_{\xi,N}^{\mathrm{act}}(\mathbf z_\xi)
    =\frac{\kappa_\triangle}{2|U|}
     \sum_{e\in\mathcal E_N^{\mathrm{act}}}d_{\xi,e}^{\,2}.
    \label{eq:supp_discrete_functional}
\end{equation}
There is no additional factor associated with duplicated edge orientations,
because the implementation does not store the reverse copy of each bond.

Writing
\begin{equation}
    \mathbf W=\kappa_\triangle\mathbf I,
    \qquad
    \mathbf L=\mathbf B^{T}\mathbf W\mathbf B,
    \qquad
    \mathbf b_\xi=-\mathbf B^{T}\mathbf W\mathbf g_\xi,
    \label{eq:supp_graph_matrices}
\end{equation}
stationarity of Eq.~\eqref{eq:supp_discrete_functional} gives
\begin{equation}
    \mathbf L\mathbf z_\xi=\mathbf b_\xi.
    \label{eq:supp_graph_normal_equations}
\end{equation}
The equation associated with a freely varying metal supernode is the discrete
counterpart of the component-neutrality condition in
Eq.~\eqref{eq:supp_component_neutrality}.

The Laplacian has one constant null vector on every connected component of the
active graph. The implementation removes these gauge freedoms by selecting
one index \(f_c\) in every graph component \(c\) and imposing
\begin{equation}
    z_{\xi,f_c}=0.
    \label{eq:supp_pinned_gauge}
\end{equation}
Equivalently, if \(\mathcal R\) denotes the unpinned indices, the computed
solution satisfies
\begin{equation}
    \mathbf L_{\mathcal R\mathcal R}
    \mathbf z_{\xi,\mathcal R}
    =\mathbf b_{\xi,\mathcal R},
    \qquad
    z_{\xi,f_c}=0.
    \label{eq:supp_graph_reduced_system}
\end{equation}
The full-size form used in the code may be written by defining a diagonal mask
\(\mathbf D\), whose entries are zero at the pinned indices and unity
elsewhere, and \(\mathbf P=\mathbf I-\mathbf D\):
\begin{equation}
    \left(\mathbf D\mathbf L\mathbf D+\mathbf P\right)
    \mathbf z_\xi
    =\mathbf D\mathbf b_\xi.
    \label{eq:supp_pinned_full_system}
\end{equation}
This pinned-node convention chooses a different representative from the
mean-zero gauge in Eq.~\eqref{eq:supp_continuum_gauge}, but both give the same
edge drops and hence the same capacitance dyadic.

\subsection{Identification and numerical evaluation of the capacitance dyadic}
\label{subsec:supp_dyadic_identification}

After solving the two basis problems, define
\begin{equation}
    \mathbf F_\xi
    =\widehat{\mathbf e}_\xi+\nabla w_\xi,
    \qquad \xi\in\{x,y\}.
    \label{eq:supp_basis_fields}
\end{equation}
Equation~\eqref{eq:supp_corrector_linearity} gives
\begin{equation}
    \mathbf v+\nabla w_{\mathbf v}
    =v_x\mathbf F_x+v_y\mathbf F_y.
    \label{eq:supp_field_linearity}
\end{equation}
Substitution into the minimum energy yields
\begin{align}
    W_e(\mathbf v)
    =\frac{\epsilon_0E_0^2}{2|U|}
      \int_\Omega
      \bigl(&v_x^2\mathbf F_x\cdot\mathbf F_x
      +2v_xv_y\mathbf F_x\cdot\mathbf F_y
      \nonumber\\[-0.2em]
      &+v_y^2\mathbf F_y\cdot\mathbf F_y\bigr)\,\mathrm dA.
    \label{eq:supp_expanded_minimum_energy}
\end{align}
The minimum energy is a quadratic form,
\begin{equation}
    W_e(\mathbf v)
    =\frac{E_0^2}{2}
     \mathbf v\cdot\overline{\overline C}[\Sigma]\cdot\mathbf v,
    \label{eq:supp_energy_quadratic_form}
\end{equation}
with entries
\begin{equation}
    C_{\xi\eta}[\Sigma]
    =\frac{\epsilon_0}{|U|}
     \int_\Omega
     \left(\widehat{\mathbf e}_\xi+\nabla w_\xi\right)
     \cdot
     \left(\widehat{\mathbf e}_\eta+\nabla w_\eta\right)
     \,\mathrm dA,
    \qquad \xi,\eta\in\{x,y\}.
    \label{eq:supp_capacitance_entries}
\end{equation}
The area integral divided by \(|U|\) is dimensionless, so
\([C_{\xi\eta}]=\mathrm{F/m}\). The off-diagonal entry is the signed overlap
of the two basis fields and cannot, in general, be reconstructed from the two
self-energies alone. Symmetry of the bilinear form gives
\begin{equation}
    C_{xy}=C_{yx}.
    \label{eq:supp_reciprocity}
\end{equation}

At finite resolution, the implementation distinguishes the edge set used to
solve the graph problem from the edge set used to approximate the integral
over the open dielectric region \(\Omega\). Equation
\eqref{eq:supp_discrete_functional} is assembled on all active bonds,
including gap--gap and gap--metal bonds. The latter couple the gap unknowns to
the equipotential metal supernodes. The dyadic quadrature, however, uses the
subset
\begin{equation}
    \mathcal E_N^{\mathrm{gg}}
    =\{e\in\mathcal E_N^{\mathrm{act}}:
      q_e\in\mathcal G_N,\ r_e\in\mathcal G_N\},
    \label{eq:supp_gapgap_edges}
\end{equation}
containing only bonds whose two endpoints lie in the sampled gap. This avoids
assigning a wholly dielectric energy contribution to bonds cut by the PEC
boundary.

With the current numerical option, the computed entries are therefore
\begin{equation}
    C_{\xi\eta}^{(N)}[\Sigma]
    =\frac{\epsilon_0\kappa_\triangle}{|U|}
     \sum_{e\in\mathcal E_N^{\mathrm{gg}}}
     d_{\xi,e}d_{\eta,e},
    \qquad \xi,\eta\in\{x,y\}.
    \label{eq:supp_discrete_capacitance}
\end{equation}
For a substrate-backed calculation, \(\epsilon_0\) in
Eq.~\eqref{eq:supp_discrete_capacitance} is replaced by the same effective
permittivity used in the continuum cell problem. The distinction between
\(\mathcal E_N^{\mathrm{act}}\) and \(\mathcal E_N^{\mathrm{gg}}\) is a
finite-resolution boundary treatment: interface bonds participate in the
solution but not in the final gap-domain quadrature. Its numerical influence
is consequently assessed by the resolution comparison described below.

Equation~\eqref{eq:supp_discrete_capacitance} is a Gram form. In particular,
its off-diagonal entries are algebraically symmetric, and for any
\(\mathbf q=q_x\widehat{\mathbf e}_x+q_y\widehat{\mathbf e}_y\),
\begin{equation}
    \mathbf q\cdot
    \overline{\overline C}^{(N)}\cdot\mathbf q
    =\frac{\epsilon_0\kappa_\triangle}{|U|}
     \sum_{e\in\mathcal E_N^{\mathrm{gg}}}
     \left(q_xd_{x,e}+q_yd_{y,e}\right)^2
    \geq0.
    \label{eq:supp_discrete_positive_semidefinite}
\end{equation}
The continuum counterpart is
\begin{equation}
    \mathbf q\cdot\overline{\overline C}\cdot\mathbf q
    =\frac{\epsilon_0}{|U|}
     \int_\Omega
     \left|\mathbf q+\nabla w_{\mathbf q}\right|^2\,\mathrm dA
    \geq0.
    \label{eq:supp_positive_semidefinite}
\end{equation}
If the right-hand side vanished, integration of the zero field along the two
independent gap-winding cycles \(\gamma_i\) would give
\begin{equation}
    0=\int_{\gamma_i}
      \left(\mathbf q+\nabla w_{\mathbf q}\right)
      \cdot\mathrm d\boldsymbol\ell
     =\mathbf q\cdot\mathbf a_i,
    \qquad i=1,2.
    \label{eq:supp_winding_argument}
\end{equation}
Because \(\mathbf a_1\) and \(\mathbf a_2\) are linearly independent, this
implies \(\mathbf q=\mathbf0\). Hence the continuum dyadic is positive
definite for the gap-connected geometries considered here. In the empty-cell
limit, \(w_x=w_y=0\), so the analytic background response is
\begin{equation}
    \overline{\overline C}[\varnothing]
    =\epsilon_0\overline{\overline I}.
    \label{eq:supp_empty_cell_limit}
\end{equation}

To monitor spatial resolution, the raw dyadic is recomputed on a second grid.
If the first calculation uses \(N\) samples along each primitive direction,
the second uses
\begin{equation}
    N'=\operatorname{odd}\!\left(\operatorname{round}(1.5N)\right),
    \label{eq:supp_refined_grid_size}
\end{equation}
where \(\operatorname{odd}(n)=n\) for odd \(n\) and
\(\operatorname{odd}(n)=n+1\) for even \(n\). The recorded relative change is
\begin{equation}
    \delta_{N,N'}
    =\frac{
       \left\|\overline{\overline C}^{(N)}
       -\overline{\overline C}^{(N')}\right\|_F}
       {\left\|\overline{\overline C}^{(N')}\right\|_F}.
    \label{eq:supp_resolution_change}
\end{equation}
The calculations reported in the present implementation use \(N=151\), for
which \(N'=227\). A value \(\delta_{N,N'}>10^{-2}\) produces a numerical
warning and indicates that \(N\) should be increased. This comparison is a
two-resolution consistency indicator rather than an a posteriori error bound.
No equivalent sheet scaling or dispersion information enters this auxiliary
corrector calculation.

\newpage
\section{Mathematical derivation of the reference-calibrated sheet capacitance}
\label{sec:supp_calibration_proof}
Starting from the auxiliary dyadic defined in
Eq.~\eqref{eq:supp_capacitance_entries}, we derive the excess response,
its reference normalization, and the equivalent sheet capacitance. We use
the energy interpretation and empty-cell limit established in
Sec.~\ref{sec:supp_corrector}. We distinguish the resulting continuum
identities from the additional assumption relating auxiliary volume
energy to sheet energy. Throughout this section, we consider a uniform
free-space background of permittivity $\epsilon_0$.

\subsection{Full-cell extension and the excess capacitance dyadic}
\label{subsec:supp_cal_excess}

We retain the correctors $w_x,w_y$, the normalized fields $\mathbf F_x,
\mathbf F_y$, and the gap-domain gauge of
Eq.~\eqref{eq:supp_continuum_gauge}. For an isolated PEC component,
we use Eq.~\eqref{eq:supp_pec_condition} to extend the correctors as
\begin{equation}
    w_\xi(\mathbf r)
    =c_{m,\xi}-\widehat{\mathbf e}_\xi\cdot\mathbf r
    \quad\text{in }\Sigma_m,
    \qquad
    \nabla w_\xi=-\widehat{\mathbf e}_\xi
    \quad\text{in }\Sigma,
    \label{eq:supp_cal_metal_extension}
\end{equation}
for $\xi\in\{x,y\}$. We interpret the constant on a continuous physical
representative of the conductor. If that conductor is cut by the chosen
cell boundary, its translated representatives follow the affine
periodicity of Eq.~\eqref{eq:supp_total_quasiperiodicity}. In particular,
for a lattice translation $\mathbf R\in\Lambda$,
\begin{equation}
    c_{m,\xi}^{(\mathbf R)}
    =c_{m,\xi}+\widehat{\mathbf e}_\xi\cdot\mathbf R,
    \qquad
    w_\xi(\mathbf r+\mathbf R)=w_\xi(\mathbf r).
    \label{eq:supp_cal_translated_extension}
\end{equation}
Thus the extension respects periodic identification even when a physical
patch appears in several pieces in the chosen primitive cell. The argument
requires isolated conductors for which both imposed basis loadings are
admissible; it does not extend unchanged to a conductor winding around
the periodic cell.

For Lipschitz PEC boundaries, the matching traces of the gap and metal
expressions give an extended function in $H^1_{\mathrm{per}}(U)$. We retain
the notation $w_\xi$ for this extension. Its mean over $U$ need not vanish:
the original gauge on $\Omega$ is sufficient. Using periodicity, we obtain
\begin{equation}
    \int_U\nabla w_\xi\,\mathrm dA
    =\int_{\partial U}w_\xi\widehat{\mathbf n}\,\mathrm d\ell
    =\mathbf0.
    \label{eq:supp_cal_mean_gradient}
\end{equation}
For smooth periodic functions the boundary integrals cancel between
paired faces because the traces agree and the normals are opposite.
We obtain the same identity for $H^1_{\mathrm{per}}(U)$ by approximation
with smooth periodic functions.

The normalized total field vanishes in the metal,
$\mathbf F_\xi=\widehat{\mathbf e}_\xi+\nabla w_\xi=\mathbf0$ in
$\Sigma$. We therefore extend the integral in
Eq.~\eqref{eq:supp_capacitance_entries} from $\Omega$ to $U$ and expand:
\begin{align}
    C_{\xi\eta}[\Sigma]
    ={}&\frac{\epsilon_0}{|U|}
    \int_U
    (\widehat{\mathbf e}_\xi+\nabla w_\xi)
    \cdot(\widehat{\mathbf e}_\eta+\nabla w_\eta)\,\mathrm dA
    \nonumber\\
    ={}&\epsilon_0\delta_{\xi\eta}
    +\frac{\epsilon_0}{|U|}
      \widehat{\mathbf e}_\xi\cdot\int_U\nabla w_\eta\,\mathrm dA
    +\frac{\epsilon_0}{|U|}
      \widehat{\mathbf e}_\eta\cdot\int_U\nabla w_\xi\,\mathrm dA
    \nonumber\\
    &+\frac{\epsilon_0}{|U|}
      \int_U\nabla w_\xi\cdot\nabla w_\eta\,\mathrm dA.
    \label{eq:supp_cal_component_expansion}
\end{align}
The mixed terms vanish by Eq.~\eqref{eq:supp_cal_mean_gradient}.
Using the empty-cell value already established in
Eq.~\eqref{eq:supp_empty_cell_limit}, we define
\begin{equation}
    \overline{\overline C}_{\mathrm{exc}}[\Sigma]
    =\overline{\overline C}[\Sigma]
     -\overline{\overline C}[\varnothing]
    =\overline{\overline C}[\Sigma]
     -\epsilon_0\overline{\overline I}.
    \label{eq:supp_cal_empty_excess}
\end{equation}
From Eq.~\eqref{eq:supp_cal_component_expansion}, we obtain
\begin{equation}
    \left(\overline{\overline C}_{\mathrm{exc}}[\Sigma]\right)_{\xi\eta}
    =\frac{\epsilon_0}{|U|}
     \int_U\nabla w_\xi\cdot\nabla w_\eta\,\mathrm dA.
    \label{eq:supp_cal_excess_gram}
\end{equation}
The full-cell integration is essential. Indeed, the metal extension
contributes
\begin{equation}
    \int_\Sigma\nabla w_\xi\cdot\nabla w_\eta\,\mathrm dA
    =|\Sigma|\delta_{\xi\eta},
    \label{eq:supp_cal_metal_gram_contribution}
\end{equation}
which would be omitted if $U$ were replaced by $\Omega$ in
Eq.~\eqref{eq:supp_cal_excess_gram}.

Using the loading linearity in Eq.~\eqref{eq:supp_corrector_linearity},
we obtain, for every real in-plane vector $\mathbf v$,
\begin{align}
    \mathbf v\cdot\overline{\overline C}_{\mathrm{exc}}\cdot\mathbf v
    &=\frac{\epsilon_0}{|U|}
      \int_U|v_x\nabla w_x+v_y\nabla w_y|^2\,\mathrm dA
      \nonumber\\
    &=\frac{\epsilon_0}{|U|}
      \int_U|\nabla w_{\mathbf v}|^2\,\mathrm dA
      \geq0.
    \label{eq:supp_cal_excess_psd}
\end{align}
Thus the excess is symmetric and positive semidefinite. For a nonempty
positive-area isolated patch, we obtain the stronger estimate
\begin{equation}
    \mathbf v\cdot\overline{\overline C}_{\mathrm{exc}}\cdot\mathbf v
    \geq\frac{\epsilon_0}{|U|}
          \int_\Sigma|\nabla w_{\mathbf v}|^2\,\mathrm dA
    =\epsilon_0\frac{|\Sigma|}{|U|}|\mathbf v|^2
    \label{eq:supp_cal_excess_pd}
\end{equation}
which proves positive definiteness, since $\nabla w_{\mathbf v}=-\mathbf v$
in the metal. These statements concern the continuum cell problem.
The empty geometry has zero excess.

We subtract the two energies at the same imposed macroscopic field.
Using Eq.~\eqref{eq:supp_energy_quadratic_form}, we write
\begin{equation}
    \Delta W_e[\Sigma;\mathbf v]
    =W_e[\Sigma;\mathbf v]-W_e[\varnothing;\mathbf v]
    =\frac{E_0^2}{2}
     \mathbf v\cdot\overline{\overline C}_{\mathrm{exc}}[\Sigma]
     \cdot\mathbf v.
    \label{eq:supp_cal_excess_energy}
\end{equation}
This nonnegative quantity has units of $\mathrm{J/m^3}$. Its
quadratic-form interpretation follows the variational energy framework
of Milton~\cite[Sec.~13.1]{Milton2002}.

We also use the extension to establish the equivalence of closed
cavities. Let $H\subset\Omega$ be a charge-free cavity whose entire boundary belongs
to one connected PEC component, with no additional conductor inside.
The total normalized potential $u_{\mathbf v}$ is harmonic in $H$ and
has the constant trace $c_m(\mathbf v)$ on $\partial H$. We test its
weak harmonic equation with
$u_{\mathbf v}-c_m(\mathbf v)\in H^1_0(H)$ to obtain
\begin{equation}
    \int_H
    \left|\nabla\bigl(u_{\mathbf v}-c_m(\mathbf v)\bigr)\right|^2
    \,\mathrm dA=0.
    \label{eq:supp_cal_cavity_zero}
\end{equation}
Hence the total field vanishes in $H$. Filling $H$ with metal adds a
constraint already satisfied by the old minimizer. The new admissible
set is a subset of the old one, so its minimum energy cannot be smaller;
the old minimizer remains admissible and attains the same energy.
From equality for every loading, we conclude that
\begin{equation}
    \overline{\overline C}[\Sigma\cup H]
    =\overline{\overline C}[\Sigma],
    \qquad
    \overline{\overline C}_{\mathrm{exc}}[\Sigma\cup H]
    =\overline{\overline C}_{\mathrm{exc}}[\Sigma].
    \label{eq:supp_cal_cavity_invariance}
\end{equation}
The conclusion extends to any finite collection of such cavities.
It applies only to this auxiliary two-dimensional problem: open slots
are not filled, and an aperture in a physical thin sheet is not thereby
equivalent to solid metal.

\subsection{Reference normalization and directional representation}
\label{subsec:supp_cal_reference}

We use the same lattice, permittivity, field convention, and boundary
conditions for the target and reference cell problems. For the centered regular
hexagonal reference, the complete problem is invariant under the rotation
$\mathbf R$ through $\pi/3$. Rotating an admissible trial field preserves
its energy and maps the admissible set for loading $\mathbf v$ bijectively
onto that for loading $\mathbf R\mathbf v$. Consequently,
$W_e[\Sigma_{\mathrm{ref}};\mathbf R\mathbf v]
=W_e[\Sigma_{\mathrm{ref}};\mathbf v]$, and
\begin{equation}
    \mathbf R^{T}
    \overline{\overline C}_{\mathrm{exc}}[\Sigma_{\mathrm{ref}}]
    \mathbf R
    =\overline{\overline C}_{\mathrm{exc}}[\Sigma_{\mathrm{ref}}].
    \label{eq:supp_cal_reference_symmetry}
\end{equation}
To establish isotropy explicitly, we write the symmetric tensor
as $\left(\begin{smallmatrix}a&b\\b&d\end{smallmatrix}\right)$ and use
$\mathbf R=\cos(\pi/3)\mathbf I+\sin(\pi/3)\mathbf J$, where
$\mathbf J=\left(\begin{smallmatrix}0&-1\\1&0\end{smallmatrix}\right)$.
Equation~\eqref{eq:supp_cal_reference_symmetry} implies commutation with
$\mathbf J$, which gives $b=0$ and $a=d$. Therefore,
\begin{equation}
    \overline{\overline C}_{\mathrm{exc}}[\Sigma_{\mathrm{ref}}]
    =p_{\mathrm{ref}}\overline{\overline I},
    \qquad
    p_{\mathrm{ref}}
    =\frac12\operatorname{tr}\!\left(
      \overline{\overline C}_{\mathrm{exc}}[\Sigma_{\mathrm{ref}}]
      \right)>0.
    \label{eq:supp_cal_ref_isotropic}
\end{equation}
Using Eq.~\eqref{eq:supp_cal_excess_pd}, we obtain
$p_{\mathrm{ref}}\geq\epsilon_0|\Sigma_{\mathrm{ref}}|/|U|>0$.
The symmetry argument concerns the complete cell, including its lattice.

We define the dimensionless normalized excess by
\begin{equation}
    \overline{\overline S}[\Sigma]
    =\frac{\overline{\overline C}_{\mathrm{exc}}[\Sigma]}
          {p_{\mathrm{ref}}},
    \qquad
    [p_{\mathrm{ref}}]=\mathrm{F/m},
    \qquad
    [\overline{\overline S}]=1.
    \label{eq:supp_cal_S_definition}
\end{equation}
The empty-cell subtraction and isotropic reference give
\begin{equation}
    \overline{\overline S}[\varnothing]=\overline{\overline0},
    \qquad
    \overline{\overline S}[\Sigma_{\mathrm{ref}}]
    =\overline{\overline I}.
    \label{eq:supp_cal_S_anchors}
\end{equation}

For the figures, we separately normalize each tensor by half its own
trace. For a nonzero symmetric positive-semidefinite tensor
$\overline{\overline M}$, we define
\begin{equation}
    \widetilde{\overline{\overline M}}
    =\frac{\overline{\overline M}}
           {\tfrac12\operatorname{tr}(\overline{\overline M})},
    \qquad
    r_M(\theta)
    =\widehat{\mathbf u}(\theta)\cdot
     \widetilde{\overline{\overline M}}
     \cdot\widehat{\mathbf u}(\theta),
    \label{eq:supp_cal_indicator_def}
\end{equation}
where $\widehat{\mathbf u}(\theta)
=\cos\theta\,\widehat{\mathbf e}_x+
 \sin\theta\,\widehat{\mathbf e}_y$.
Let $m_1\geq m_2\geq0$ be the eigenvalues, and let an eigenvector
for $m_1$ make angle $\varphi$ with the $x$ axis. Evaluating in the
orthonormal principal basis~\cite[App.~B.5]{HighamMatrices}, we obtain
\begin{align}
    r_M(\theta)
    &=\frac{2\left[m_1\cos^2(\theta-\varphi)
                   +m_2\sin^2(\theta-\varphi)\right]}{m_1+m_2}
      \nonumber\\
    &=1+\frac{m_1-m_2}{m_1+m_2}
          \cos\left(2(\theta-\varphi)\right).
    \label{eq:supp_cal_indicator_formula}
\end{align}
The denominator is positive because the tensor is nonzero and
positive semidefinite. Thus this normalization preserves the principal
basis and eigenvalue ratio when defined, and gives the unit circle for
an isotropic tensor. The radial curve is a directional quadratic-form
indicator, not generally an ellipse. The zero tensor has no such
trace-normalized representation.

For the excess, we perform the subtraction before the plotting
normalization:
\begin{equation}
    \widetilde{\overline{\overline C}}_{\mathrm{exc}}
    =\frac{\overline{\overline C}-\epsilon_0\overline{\overline I}}
           {\tfrac12\operatorname{tr}(\overline{\overline C})-\epsilon_0},
    \label{eq:supp_cal_subtract_then_normalize}
\end{equation}
provided the excess is nonzero. The trace normalization in the figures
does not enter the sheet-capacitance calculation.

\subsection{Reference-energy matching and the equivalent sheet capacitance}
\label{subsec:supp_cal_sheet}

We now relate the auxiliary excess energy to the equivalent sheet
energy. For the same real macroscopic field $E_0\mathbf v$, we write
\begin{align}
    \Delta W_e[\Sigma;\mathbf v]
    &=\frac{E_0^2}{2}
      \mathbf v\cdot\overline{\overline C}_{\mathrm{exc}}[\Sigma]
      \cdot\mathbf v,
      & [\Delta W_e]&=\mathrm{J/m^3},
      \nonumber\\
    \mathcal U_s^{\mathrm{model}}[\Sigma;\mathbf v]
    &=\frac{E_0^2}{2}
      \mathbf v\cdot\overline{\overline C}_{\mathrm{eqv}}[\Sigma]
      \cdot\mathbf v,
      & [\mathcal U_s^{\mathrm{model}}]&=\mathrm{J/m^2}.
    \label{eq:supp_cal_sheet_energy}
\end{align}
We use a reciprocal electric-sheet constitutive
model~\cite{HollowayKuester2016}. For a selected geometry family
$\mathfrak F$, we introduce the energy-matching assumption
\begin{equation}
    \exists\,\ell_{\mathrm{ref}}>0:
    \quad
    \mathcal U_s^{\mathrm{model}}[\Sigma;\mathbf v]
    =\ell_{\mathrm{ref}}\Delta W_e[\Sigma;\mathbf v],
    \qquad
    \forall\Sigma\in\mathfrak F,
    \quad\forall\mathbf v\in\mathbb R^2.
    \label{eq:supp_cal_matching_assumption}
\end{equation}
We therefore assume one conversion length for all geometries and both
in-plane polarizations, under the same macroscopic-field convention.

For $E_0\ne0$, substitution gives
\begin{equation}
    \mathbf v^{T}\mathbf D\mathbf v=0
    \quad\forall\mathbf v,
    \qquad
    \mathbf D
    =\overline{\overline C}_{\mathrm{eqv}}
     -\ell_{\mathrm{ref}}\overline{\overline C}_{\mathrm{exc}},
    \qquad
    \mathbf D^T=\mathbf D.
    \label{eq:supp_cal_quadratic_difference}
\end{equation}
We evaluate this quadratic form on three loadings:
\begin{align}
    \mathbf v=\widehat{\mathbf e}_x
    &\quad\Longrightarrow\quad D_{xx}=0,
    \nonumber\\
    \mathbf v=\widehat{\mathbf e}_y
    &\quad\Longrightarrow\quad D_{yy}=0,
    \nonumber\\
    \mathbf v=\widehat{\mathbf e}_x+\widehat{\mathbf e}_y
    &\quad\Longrightarrow\quad
    D_{xx}+2D_{xy}+D_{yy}=0.
    \label{eq:supp_cal_polarization}
\end{align}
Consequently,
\begin{equation}
    D_{xx}=D_{yy}=D_{xy}=D_{yx}=0
    \quad\Longrightarrow\quad
    \overline{\overline C}_{\mathrm{eqv}}[\Sigma]
    =\ell_{\mathrm{ref}}
     \overline{\overline C}_{\mathrm{exc}}[\Sigma].
    \label{eq:supp_cal_proportionality}
\end{equation}

For the isotropic reference, we impose the independently prescribed
sheet value $C_{\mathrm{ref}}^{\mathrm{eqv}}>0$:
\begin{align}
    \overline{\overline C}_{\mathrm{exc}}[\Sigma_{\mathrm{ref}}]
    &=p_{\mathrm{ref}}\overline{\overline I},
    \qquad p_{\mathrm{ref}}>0,
    \nonumber\\
    \overline{\overline C}_{\mathrm{eqv}}[\Sigma_{\mathrm{ref}}]
    &=C_{\mathrm{ref}}^{\mathrm{eqv}}\overline{\overline I}.
    \label{eq:supp_cal_reference_match}
\end{align}
Combining Eqs.~\eqref{eq:supp_cal_proportionality} and
\eqref{eq:supp_cal_reference_match}, we obtain
\begin{equation}
    \ell_{\mathrm{ref}}p_{\mathrm{ref}}\overline{\overline I}
    =C_{\mathrm{ref}}^{\mathrm{eqv}}\overline{\overline I}
    \quad\Longrightarrow\quad
    \ell_{\mathrm{ref}}
    =\frac{C_{\mathrm{ref}}^{\mathrm{eqv}}}{p_{\mathrm{ref}}}.
    \label{eq:supp_cal_length}
\end{equation}
Therefore,
\begin{align}
    \overline{\overline C}_{\mathrm{eqv}}[\Sigma]
    &=\frac{C_{\mathrm{ref}}^{\mathrm{eqv}}}{p_{\mathrm{ref}}}
      \overline{\overline C}_{\mathrm{exc}}[\Sigma]
      \nonumber\\
    &=C_{\mathrm{ref}}^{\mathrm{eqv}}
      \overline{\overline S}[\Sigma]
      \nonumber\\
    &=\frac{C_{\mathrm{ref}}^{\mathrm{eqv}}}{p_{\mathrm{ref}}}
      \left(\overline{\overline C}[\Sigma]
           -\epsilon_0\overline{\overline I}\right).
    \label{eq:supp_cal_final_map}
\end{align}
We verify the units and the two reference conditions directly:
\begin{equation}
    [\ell_{\mathrm{ref}}]
    =\frac{\mathrm F}{\mathrm{F/m}}=\mathrm m,
    \qquad
    [\overline{\overline C}_{\mathrm{eqv}}]
    =\mathrm m\,\frac{\mathrm F}{\mathrm m}=\mathrm F,
    \label{eq:supp_cal_units}
\end{equation}
\begin{equation}
    \overline{\overline C}_{\mathrm{eqv}}[\varnothing]
    =\overline{\overline0},
    \qquad
    \overline{\overline C}_{\mathrm{eqv}}[\Sigma_{\mathrm{ref}}]
    =C_{\mathrm{ref}}^{\mathrm{eqv}}\overline{\overline I}.
    \label{eq:supp_cal_sheet_anchors}
\end{equation}

We can also derive this normalization as the unique map within the
scalar-affine calibration class
\begin{equation}
    \overline{\overline C}_{\mathrm{eqv}}[\Sigma]
    =a\,\overline{\overline C}[\Sigma]
     +b\,\overline{\overline I},
    \qquad [a]=\mathrm m,
    \qquad [b]=\mathrm F,
    \label{eq:supp_cal_affine_class}
\end{equation}
where $a,b$ are independent of $\Sigma$. Indeed,
\begin{align}
    \overline{\overline C}[\varnothing]
    &=\epsilon_0\overline{\overline I},
    \nonumber\\
    \overline{\overline C}[\Sigma_{\mathrm{ref}}]
    &=(\epsilon_0+p_{\mathrm{ref}})\overline{\overline I},
    \label{eq:supp_cal_affine_inputs}
\end{align}
so we write the two matching conditions as
\begin{equation}
    \begin{pmatrix}
        \epsilon_0 & 1\\
        \epsilon_0+p_{\mathrm{ref}} & 1
    \end{pmatrix}
    \begin{pmatrix}a\\b\end{pmatrix}
    =\begin{pmatrix}0\\C_{\mathrm{ref}}^{\mathrm{eqv}}\end{pmatrix},
    \qquad
    \det\begin{pmatrix}
        \epsilon_0 & 1\\
        \epsilon_0+p_{\mathrm{ref}} & 1
    \end{pmatrix}
    =-p_{\mathrm{ref}}\ne0.
    \label{eq:supp_cal_affine_system}
\end{equation}
Solving this system, we obtain
\begin{equation}
    a=\frac{C_{\mathrm{ref}}^{\mathrm{eqv}}}{p_{\mathrm{ref}}}
     =\ell_{\mathrm{ref}},
    \qquad
    b=-\epsilon_0
       \frac{C_{\mathrm{ref}}^{\mathrm{eqv}}}{p_{\mathrm{ref}}}
     =-\epsilon_0\ell_{\mathrm{ref}}.
    \label{eq:supp_cal_affine_solution}
\end{equation}
Thus the empty-cell condition fixes the offset and the reference fixes
the slope. Uniqueness is conditional on the class
\eqref{eq:supp_cal_affine_class}.

To identify the spectral consequences, we write~\cite[App.~B.5]{HighamMatrices}
\begin{equation}
    \overline{\overline C}_{\mathrm{exc}}
    =\sum_{i=1}^{2}\lambda_i
       \widehat{\mathbf v}_i\otimes\widehat{\mathbf v}_i,
    \qquad
    \lambda_i\geq0,
    \qquad
    \widehat{\mathbf v}_i\cdot\widehat{\mathbf v}_j=\delta_{ij}.
    \label{eq:supp_cal_excess_spectrum}
\end{equation}
Then
\begin{align}
    \overline{\overline C}
    &=\sum_{i=1}^{2}(\epsilon_0+\lambda_i)
      \widehat{\mathbf v}_i\otimes\widehat{\mathbf v}_i,
      \nonumber\\
    \overline{\overline S}
    &=\sum_{i=1}^{2}\frac{\lambda_i}{p_{\mathrm{ref}}}
      \widehat{\mathbf v}_i\otimes\widehat{\mathbf v}_i,
      \nonumber\\
    \overline{\overline C}_{\mathrm{eqv}}
    &=\sum_{i=1}^{2}
      \frac{C_{\mathrm{ref}}^{\mathrm{eqv}}\lambda_i}{p_{\mathrm{ref}}}
      \widehat{\mathbf v}_i\otimes\widehat{\mathbf v}_i.
    \label{eq:supp_cal_sheet_spectral}
\end{align}
In particular,
\begin{equation}
    s_i=\frac{\lambda_i}{p_{\mathrm{ref}}},
    \qquad
    C_i^{\mathrm{eqv}}=C_{\mathrm{ref}}^{\mathrm{eqv}}s_i,
    \qquad
    \frac{C_1^{\mathrm{eqv}}}{C_2^{\mathrm{eqv}}}
    =\frac{\lambda_1}{\lambda_2}
    \quad(\lambda_2>0),
    \label{eq:supp_cal_principal_values}
\end{equation}
\begin{equation}
    \operatorname{ker}\overline{\overline C}_{\mathrm{eqv}}
    =\operatorname{ker}\overline{\overline C}_{\mathrm{exc}},
    \qquad
    \operatorname{rank}\overline{\overline C}_{\mathrm{eqv}}
    =\operatorname{rank}\overline{\overline C}_{\mathrm{exc}},
    \label{eq:supp_cal_rank_preservation}
\end{equation}
\begin{equation}
    \mathbf v\cdot\overline{\overline C}_{\mathrm{eqv}}\cdot\mathbf v
    =\ell_{\mathrm{ref}}\sum_{i=1}^{2}
      \lambda_i(\mathbf v\cdot\widehat{\mathbf v}_i)^2\geq0.
    \label{eq:supp_cal_sheet_positive}
\end{equation}
The principal basis is shared; it is not unique for an isotropic tensor.
For an orthogonal symmetry $\mathbf R$, we also obtain
\begin{equation}
    \mathbf R\overline{\overline C}\mathbf R^T
    =\overline{\overline C}
    \quad\Longrightarrow\quad
    \mathbf R\overline{\overline C}_{\mathrm{eqv}}\mathbf R^T
    =\ell_{\mathrm{ref}}
      \left(\mathbf R\overline{\overline C}\mathbf R^T
           -\epsilon_0\overline{\overline I}\right)
    =\overline{\overline C}_{\mathrm{eqv}}.
    \label{eq:supp_cal_sheet_symmetry}
\end{equation}

We prescribe the reference value independently of the auxiliary cell
problem. For the free-standing regular hexagonal reference, we use
\begin{align}
    C_{\mathrm{ref}}^{\mathrm{eqv}}
    &=2\epsilon_0A_{\mathrm{iso}},
      \nonumber\\
    A_{\mathrm{iso}}
    &=\frac{\sqrt3D_h}{2\pi}
      \ln\!\left[\csc\!\left(\frac{\pi g}{2D_h}\right)\right],
    \qquad 0<g<D_h.
    \label{eq:supp_cal_reference_approximation}
\end{align}
Here $[A_{\mathrm{iso}}]=\mathrm m$ and
$C_{\mathrm{ref}}^{\mathrm{eqv}}>0$. We motivate the logarithmic dependence
by the strip and square-patch models of Luukkonen
\emph{et al.}~\cite{Luukkonen2008} and treat the hexagonal prefactor as an
independently testable approximation. We may replace this expression
with a computed reference capacitance.

Finally, we identify what the isotropic reference leaves undetermined.
We decompose the excess into its isotropic and traceless parts:
\begin{equation}
    \overline{\overline C}_{\mathrm{exc}}
    =t\overline{\overline I}
     +\overline{\overline C}_{\mathrm{dev}},
    \qquad
    t=\frac12\operatorname{tr}
       \overline{\overline C}_{\mathrm{exc}},
    \qquad
    \operatorname{tr}\overline{\overline C}_{\mathrm{dev}}=0.
    \label{eq:supp_cal_iso_dev_split}
\end{equation}
For comparison, we consider the more general rotation-covariant linear
calibration
\begin{equation}
    \mathcal L\!\left(\overline{\overline C}_{\mathrm{exc}}\right)
    =\ell_{\mathrm{iso}}\,t\overline{\overline I}
     +\ell_{\mathrm{dev}}\,
       \overline{\overline C}_{\mathrm{dev}},
    \qquad
    [\ell_{\mathrm{iso}}]=[\ell_{\mathrm{dev}}]=\mathrm m.
    \label{eq:supp_cal_two_scales}
\end{equation}
For the reference,
\begin{equation}
    t_{\mathrm{ref}}=p_{\mathrm{ref}},
    \qquad
    \overline{\overline C}_{\mathrm{dev}}[\Sigma_{\mathrm{ref}}]
    =\overline{\overline0},
    \label{eq:supp_cal_ref_deviator_zero}
\end{equation}
and therefore
\begin{equation}
    \mathcal L(p_{\mathrm{ref}}\overline{\overline I})
    =\ell_{\mathrm{iso}}p_{\mathrm{ref}}\overline{\overline I}
    =C_{\mathrm{ref}}^{\mathrm{eqv}}\overline{\overline I}
    \quad\Longrightarrow\quad
    \ell_{\mathrm{iso}}
    =\frac{C_{\mathrm{ref}}^{\mathrm{eqv}}}{p_{\mathrm{ref}}},
    \label{eq:supp_cal_iso_scale_only}
\end{equation}
with no condition on $\ell_{\mathrm{dev}}$. Positivity alone does not
force equality: the eigenvalues of the output in
Eq.~\eqref{eq:supp_cal_two_scales} are
\begin{equation}
    \mu_i
    =\ell_{\mathrm{dev}}\lambda_i
     +(\ell_{\mathrm{iso}}-\ell_{\mathrm{dev}})t
    \geq0
    \qquad\text{if}\qquad
    0\leq\ell_{\mathrm{dev}}\leq\ell_{\mathrm{iso}}.
    \label{eq:supp_cal_two_scale_positive}
\end{equation}
In the present model, we choose
\begin{equation}
    \ell_{\mathrm{dev}}=\ell_{\mathrm{iso}}
    =\ell_{\mathrm{ref}}
    =\frac{C_{\mathrm{ref}}^{\mathrm{eqv}}}{p_{\mathrm{ref}}}.
    \label{eq:supp_cal_equal_scale_assumption}
\end{equation}
We thus use the reference to fix the overall scale and assume equal
scaling of the isotropic and anisotropic parts throughout the selected
geometry family. This is the physical hypothesis in
Eq.~\eqref{eq:supp_cal_matching_assumption}. Its validity requires
electromagnetic validation or an interface-cell reduction; the two
reference conditions alone do not establish it.

\subsection{Continuum identities and numerical consistency}
\label{subsec:supp_cal_numerical_scope}

We have derived the excess Gram identity and its positivity for the
continuum problem. At finite resolution, we solve on the active graph
bonds described in Sec.~\ref{sec:supp_corrector} and evaluate the raw
dyadic using the gap--gap subset in
Eq.~\eqref{eq:supp_discrete_capacitance}. The latter is a Gram form and
therefore gives a positive-semidefinite raw tensor. That fact alone does
not imply positivity after empty-cell subtraction:
\begin{equation}
    \overline{\overline C}^{(N)}\geq0
    \quad\not\!\Longrightarrow\quad
    \overline{\overline C}^{(N)}
    -\epsilon_0\overline{\overline I}\geq0.
    \label{eq:supp_cal_discrete_caveat}
\end{equation}
The same distinction applies when a numerically computed empty-cell tensor
is subtracted. The gap--gap quadrature does not, by itself, provide a
discrete counterpart of the full-cell identity
\eqref{eq:supp_cal_excess_gram}.

We therefore require a consistent background-subtraction convention and
discretization for the target and reference calculations. Numerical
checks should examine the excess eigenvalues, the positivity of the
computed $p_{\mathrm{ref}}$, and the stability of the normalized tensor
under refinement. Small relative changes of the raw tensor, as measured by
Eq.~\eqref{eq:supp_resolution_change}, need not imply equally small
relative changes of the excess when the background subtraction causes
cancellation. These checks assess numerical recovery of the continuum
identities; we do not assert a discrete convergence theorem here.

For the physical calibration, we obtain Eq.~\eqref{eq:supp_cal_final_map}
uniquely once we assume the common conversion length in
Eq.~\eqref{eq:supp_cal_matching_assumption} and prescribe a positive
independent reference capacitance. Applying that same length across
geometries and principal directions requires electromagnetic validation
or a further interface-cell derivation. The auxiliary normalization
alone does not establish this applicability.

\newpage
\section{Full local anisotropic transverse-resonance condition and quasi-TE reduction}
\label{sec:supp_anisotropic_trc}

This section derives the local anisotropic eigencondition underlying
the quasi-TE dispersion model used in the main text. We first express
the geometry-derived grid-length dyadic in a basis fixed to the
in-plane propagation direction, then derive the complete coupled
TE--TM transverse-resonance condition, and finally obtain the
TE-dominant reduction used for the dispersion calculations.

The equivalent sheet-capacitance dyadic is related to the grid-length
dyadic by
\begin{equation}
\overline{\overline A}
=
\frac{
\overline{\overline C}_{\mathrm{eqv}}
}{
2\varepsilon_0
},
\label{eq:supp_Adyad}
\end{equation}
so that, in a fixed Cartesian basis,
\begin{equation}
\overline{\overline A}
=
\begin{pmatrix}
A_{xx} & A_{xy}\\
A_{xy} & A_{yy}
\end{pmatrix}.
\label{eq:supp_A_cartesian}
\end{equation}
The symmetry of
$\overline{\overline A}$ follows from the reciprocity and symmetry
of the capacitance dyadic established in the main text.

\subsection{Propagation-fixed representation of the grid-length dyadic}
\label{subsec:supp_propagation_basis}

For a nonzero in-plane Bloch wavevector
\begin{equation}
\mathbf k_t
=
k_x\hat{\mathbf x}
+
k_y\hat{\mathbf y},
\qquad
k_t
=
\sqrt{k_x^2+k_y^2},
\label{eq:supp_kt}
\end{equation}
we define the longitudinal and transverse unit vectors
\begin{equation}
\hat{\mathbf e}_{\parallel}
=
\frac{\mathbf k_t}{k_t}
=
\frac{1}{k_t}
\begin{pmatrix}
k_x\\
k_y
\end{pmatrix},
\qquad
\hat{\mathbf e}_{\perp}
=
\hat{\mathbf e}_{\mathrm{TE}}
=
\frac{\hat{\mathbf z}\times\mathbf k_t}{k_t}
=
\frac{1}{k_t}
\begin{pmatrix}
-k_y\\
k_x
\end{pmatrix}.
\label{eq:supp_propagation_vectors}
\end{equation}
The pair
$\{
\hat{\mathbf e}_{\parallel},
\hat{\mathbf e}_{\perp}
\}$
forms an orthonormal basis of the tangential plane. The associated
orthogonal transformation is
\begin{equation}
Q(\mathbf k_t)
=
\begin{pmatrix}
\hat{\mathbf e}_{\parallel}
&
\hat{\mathbf e}_{\perp}
\end{pmatrix}
=
\frac{1}{k_t}
\begin{pmatrix}
k_x & -k_y\\
k_y &  k_x
\end{pmatrix},
\label{eq:supp_Q}
\end{equation}
for which
\begin{equation}
Q^{\mathsf T}Q
=
QQ^{\mathsf T}
=
I.
\label{eq:supp_Q_orthogonal}
\end{equation}

The representation of the grid-length dyadic in the
propagation-fixed basis is therefore
\begin{equation}
Q^{\mathsf T}
\overline{\overline A}
Q
=
\begin{pmatrix}
A_{\parallel} & A_{\times}\\
A_{\times}    & A_{\perp}
\end{pmatrix},
\label{eq:supp_A_rotated}
\end{equation}
where the diagonal entries are the longitudinal and transverse
quadratic projections and the off-diagonal entry is the
longitudinal--transverse cross projection.

Using Eqs.~\eqref{eq:supp_A_cartesian} and
\eqref{eq:supp_propagation_vectors},
\begin{align}
A_{\parallel}
&=
\hat{\mathbf e}_{\parallel}
\cdot
\overline{\overline A}
\cdot
\hat{\mathbf e}_{\parallel}
\nonumber\\
&=
\frac{
A_{xx}k_x^2
+
2A_{xy}k_xk_y
+
A_{yy}k_y^2
}{
k_t^2
},
\label{eq:supp_Aparallel}
\\[1ex]
A_{\perp}
&=
\hat{\mathbf e}_{\perp}
\cdot
\overline{\overline A}
\cdot
\hat{\mathbf e}_{\perp}
\nonumber\\
&=
\frac{
A_{xx}k_y^2
-
2A_{xy}k_xk_y
+
A_{yy}k_x^2
}{
k_t^2
},
\label{eq:supp_Aperp}
\\[1ex]
A_{\times}
&=
\hat{\mathbf e}_{\parallel}
\cdot
\overline{\overline A}
\cdot
\hat{\mathbf e}_{\perp}
\nonumber\\
&=
\frac{
A_{xy}(k_x^2-k_y^2)
+
(A_{yy}-A_{xx})k_xk_y
}{
k_t^2
}.
\label{eq:supp_Across}
\end{align}
Consequently,
\begin{equation}
\begin{split}
\overline{\overline A}
={}&
A_{\parallel}
\hat{\mathbf e}_{\parallel}
\hat{\mathbf e}_{\parallel}
+
A_{\perp}
\hat{\mathbf e}_{\perp}
\hat{\mathbf e}_{\perp}
\\
&+
A_{\times}
\left(
\hat{\mathbf e}_{\parallel}
\hat{\mathbf e}_{\perp}
+
\hat{\mathbf e}_{\perp}
\hat{\mathbf e}_{\parallel}
\right).
\end{split}
\label{eq:supp_A_dyadic}
\end{equation}
Juxtaposition of two unit vectors denotes their dyadic product.

The physical meaning of these three quantities is immediate:
$A_{\parallel}$ is the response sampled by a tangential electric
field parallel to $\mathbf k_t$, $A_{\perp}$ is that sampled by
the TE-polarised electric field, and $A_{\times}$ couples these
two field components.

The condition for exact TE--TM decoupling can also be seen
directly in the principal frame of
$\overline{\overline A}$. Let its eigenvalues be
$A_1$ and $A_2$, and let $\vartheta$ denote the angle between
$\mathbf k_t$ and the first principal axis. In this frame,
\begin{align}
A_{\parallel}
&=
A_1\cos^2\vartheta
+
A_2\sin^2\vartheta,
\label{eq:supp_Apar_principal}
\\
A_{\perp}
&=
A_1\sin^2\vartheta
+
A_2\cos^2\vartheta,
\label{eq:supp_Aperp_principal}
\\
A_{\times}
&=
\frac{A_2-A_1}{2}
\sin 2\vartheta.
\label{eq:supp_Across_principal}
\end{align}
Thus $A_{\times}=0$ identically for an isotropic sheet,
$A_1=A_2$, and also whenever the propagation direction is
aligned with either principal axis,
$\vartheta=0$ or $\pi/2$. Away from these directions,
an anisotropic sheet generally has
$A_{\times}\neq0$.

At the $\Gamma$ point,
$\mathbf k_t=\mathbf 0$, the propagation-fixed basis itself is
undefined. Direction-dependent quantities at $\Gamma$ are therefore
understood as one-sided limits along the selected path in reciprocal
space.

\subsection{Exact local TE--TM transverse-resonance condition}
\label{subsec:supp_exact_hybrid}

The tangential electric field of the most general local eigenmode is
expanded in the propagation-fixed basis as
\begin{equation}
\mathbf E_t
=
E_{\parallel}
\hat{\mathbf e}_{\parallel}
+
E_{\perp}
\hat{\mathbf e}_{\perp}.
\label{eq:supp_Et}
\end{equation}
The local constitutive relation of the capacitive sheet is
\begin{equation}
\overline{\overline Y}_s
=
j\omega
\overline{\overline C}_{\mathrm{eqv}}
=
j\frac{2k_0}{\eta_0}
\overline{\overline A}.
\label{eq:supp_Ys}
\end{equation}

The exterior free-space admittance dyadic is diagonal in the
propagation-fixed basis because the TE and TM plane-wave
polarisations are eigenpolarisations of homogeneous free space.
It may therefore be written as
\begin{equation}
\overline{\overline Y}_0
=
Y_0^{\mathrm{TM}}
\hat{\mathbf e}_{\parallel}
\hat{\mathbf e}_{\parallel}
+
Y_0^{\mathrm{TE}}
\hat{\mathbf e}_{\perp}
\hat{\mathbf e}_{\perp}.
\label{eq:supp_Y0_dyad}
\end{equation}
For a free-standing sheet in the same medium on both sides, the
two half-spaces contribute equally, so that the source-free
transverse-resonance condition is
\begin{equation}
\left(
2\overline{\overline Y}_0
+
\overline{\overline Y}_s
\right)
\cdot
\mathbf E_t
=
\mathbf 0_t.
\label{eq:supp_TRC_dyadic}
\end{equation}
Defining
\begin{equation}
\overline{\overline{\mathcal T}}_0
=
2\overline{\overline Y}_0
+
\overline{\overline Y}_s,
\label{eq:supp_T_def}
\end{equation}
Eq.~\eqref{eq:supp_TRC_dyadic} becomes
\begin{equation}
\overline{\overline{\mathcal T}}_0
\cdot
\mathbf E_t
=
\mathbf 0_t.
\label{eq:supp_TRC_compact}
\end{equation}

Using Eq.~\eqref{eq:supp_A_rotated}, the matrix representation of
the transverse-resonance dyadic is
\begin{equation}
\left[
\overline{\overline{\mathcal T}}_0
\right]_{\parallel\perp}
=
\begin{pmatrix}
\mathcal T_{\mathrm{TM}}
&
\mathcal T_{\times}
\\
\mathcal T_{\times}
&
\mathcal T_{\mathrm{TE}}
\end{pmatrix},
\label{eq:supp_T_matrix}
\end{equation}
with
\begin{align}
\mathcal T_{\mathrm{TM}}
&=
2Y_0^{\mathrm{TM}}
+
j\frac{2k_0}{\eta_0}
A_{\parallel},
\label{eq:supp_TTM}
\\
\mathcal T_{\mathrm{TE}}
&=
2Y_0^{\mathrm{TE}}
+
j\frac{2k_0}{\eta_0}
A_{\perp},
\label{eq:supp_TTE}
\\
\mathcal T_{\times}
&=
j\frac{2k_0}{\eta_0}
A_{\times}.
\label{eq:supp_Tcross}
\end{align}
Equation~\eqref{eq:supp_TRC_compact} is therefore equivalent to
the pair
\begin{align}
\mathcal T_{\mathrm{TM}}E_{\parallel}
+
\mathcal T_{\times}E_{\perp}
&=0,
\label{eq:supp_row1}
\\
\mathcal T_{\times}E_{\parallel}
+
\mathcal T_{\mathrm{TE}}E_{\perp}
&=0.
\label{eq:supp_row2}
\end{align}

A nontrivial field requires the determinant of
Eq.~\eqref{eq:supp_T_matrix} to vanish:
\begin{equation}
\det
\left(
\overline{\overline{\mathcal T}}_0
\right)
=
\mathcal T_{\mathrm{TM}}
\mathcal T_{\mathrm{TE}}
-
\mathcal T_{\times}^{\,2}
=
0.
\label{eq:supp_hybrid_det}
\end{equation}
Explicitly,
\begin{equation}
\left(
2Y_0^{\mathrm{TM}}
+
j\frac{2k_0}{\eta_0}A_{\parallel}
\right)
\left(
2Y_0^{\mathrm{TE}}
+
j\frac{2k_0}{\eta_0}A_{\perp}
\right)
-
\left(
j\frac{2k_0}{\eta_0}A_{\times}
\right)^2
=
0.
\label{eq:supp_hybrid_expanded}
\end{equation}
This is the complete local anisotropic TE--TM eigencondition.
The cross term is generated entirely by the representation of the
anisotropic sheet response in a propagation-fixed frame. It
vanishes for an isotropic sheet and along the principal directions
of an anisotropic sheet. Otherwise, the eigenpolarisation is
generally hybrid TE--TM, consistently with the behaviour of
anisotropic electromagnetic sheets~\cite{Yermakov2015}.

\subsection{Schur-complement reduction and hybridisation measure}
\label{subsec:supp_schur}

For the positive-definite capacitive patch arrays considered in
this work, the TM diagonal factor does not independently vanish
on the bound branch of interest. In the bound-wave convention
used in the main text,
\begin{equation}
Y_0^{\mathrm{TM}}
=
j\frac{\omega\varepsilon_0}{\gamma},
\qquad
\gamma
=
\sqrt{k_t^2-k_0^2}>0.
\label{eq:supp_YTM_bound}
\end{equation}
Since positive definiteness of
$\overline{\overline A}$ implies
$A_{\parallel}>0$ for every nonzero direction,
\begin{equation}
\mathcal T_{\mathrm{TM}}
=
j
\left(
\frac{2\omega\varepsilon_0}{\gamma}
+
\frac{2k_0}{\eta_0}A_{\parallel}
\right)
\neq0.
\label{eq:supp_TTM_nonzero}
\end{equation}
Thus no independent bound TM root is produced by the uncoupled
capacitive sheet.

Equation~\eqref{eq:supp_row1} can therefore be solved for the
longitudinal field component:
\begin{equation}
E_{\parallel}
=
-
\frac{
\mathcal T_{\times}
}{
\mathcal T_{\mathrm{TM}}
}
E_{\perp}.
\label{eq:supp_Eparallel}
\end{equation}
Substitution into Eq.~\eqref{eq:supp_row2} yields
\begin{equation}
\left[
\mathcal T_{\mathrm{TE}}
-
\frac{
\mathcal T_{\times}^{\,2}
}{
\mathcal T_{\mathrm{TM}}
}
\right]
E_{\perp}
=
0.
\label{eq:supp_schur_before}
\end{equation}
Hence the exact hybrid eigencondition may equivalently be written
in Schur-complement form as
\begin{equation}
\mathcal T_{\mathrm{TE}}
-
\frac{
\mathcal T_{\times}^{\,2}
}{
\mathcal T_{\mathrm{TM}}
}
=
0.
\label{eq:supp_schur}
\end{equation}
The second term is the correction to the uncoupled TE
transverse-resonance condition produced by anisotropic TE--TM
hybridisation.

Equation~\eqref{eq:supp_Eparallel} also provides a direct measure
of the hybrid character of the mode. We define
\begin{equation}
\zeta_{\mathrm{hyb}}
=
\left|
\frac{
E_{\parallel}
}{
E_{\perp}
}
\right|
=
\left|
\frac{
\mathcal T_{\times}
}{
\mathcal T_{\mathrm{TM}}
}
\right|.
\label{eq:supp_zeta}
\end{equation}
When $A_{\times}=0$,
$\mathcal T_{\times}=0$ and therefore
$\zeta_{\mathrm{hyb}}=0$; the TE and TM sectors decouple
exactly. When $\zeta_{\mathrm{hyb}}\ll1$, the longitudinal
component is perturbatively small and the eigenmode is
TE dominant.

For the positive-definite capacitive patch arrays considered
here, $\mathcal T_{\mathrm{TM}}\neq0$ on the bound branch of
interest, so the uncoupled TM sector does not support an
independent bound solution. The hybrid branch is therefore
continuously connected to the pure TE solution in the limit
$A_{\times}\rightarrow0$, while the longitudinal component
$E_{\parallel}$ is induced by the anisotropic cross coupling.
Indeed, from Eq.~\eqref{eq:supp_Eparallel},
\begin{equation}
\frac{E_{\parallel}}{E_{\perp}}
=
-
\frac{\mathcal T_{\times}}
{\mathcal T_{\mathrm{TM}}},
\end{equation}
so that the longitudinal field is first order in the coupling
ratio. By contrast, the correction to the transverse-resonance
condition enters through the Schur complement as
\begin{equation}
-
\frac{\mathcal T_{\times}^{\,2}}
{\mathcal T_{\mathrm{TM}}},
\end{equation}
and is therefore quadratic in the cross-coupling term. Hence,
when $\zeta_{\mathrm{hyb}}\ll1$, the leading-order dispersion
of the hybrid mode is governed by its transverse component,
with the longitudinal field providing only a perturbative
correction.

The reduced dispersion model used in the main text therefore
follows the dominant transverse component of this TE-dominant
branch. Its effective capacitive response is the transverse
quadratic projection
$A_0(\mathbf k_t)=A_{\perp}$, while the explicit
Schur-complement correction is omitted in the reduced closure.

\subsection{Quasi-TE reduction used in the dispersion model}
\label{subsec:supp_quasiTE}

The transverse response sampled by the TE-dominant field is
\begin{equation}
A_0(\mathbf k_t)
\equiv
A_{\perp}
=
\hat{\mathbf e}_{\mathrm{TE}}
\cdot
\overline{\overline A}
\cdot
\hat{\mathbf e}_{\mathrm{TE}},
\label{eq:supp_A0_definition}
\end{equation}
or, in Cartesian components,
\begin{equation}
A_0(\mathbf k_t)
=
\frac{
A_{xx}k_y^2
-
2A_{xy}k_xk_y
+
A_{yy}k_x^2
}{
k_t^2
}.
\label{eq:supp_A0_cartesian}
\end{equation}
For an isotropic sheet,
$\overline{\overline A}=A_g\overline{\overline I}$ and hence
\begin{equation}
A_0(\mathbf k_t)=A_g
\label{eq:supp_A0_iso}
\end{equation}
for every propagation direction.

Neglecting the explicit Schur-complement term in
Eq.~\eqref{eq:supp_schur} gives the dominant transverse
condition
\begin{equation}
\mathcal T_{\mathrm{TE}}=0.
\label{eq:supp_TE_reduced_local}
\end{equation}
This reduction is exact when
$A_{\times}=0$, and therefore whenever the propagation
direction coincides with a principal axis of
$\overline{\overline A}$. Away from a principal direction it is
the quasi-TE approximation to the exact hybrid eigenproblem,
with its accuracy characterised by
Eq.~\eqref{eq:supp_zeta}.

The local dyadic constitutive law itself is independent of
$\mathbf k_t$. To retain the same $k_t$-dependent correction
used for the scalar patch-array model, the Luukkonen correction
factor is applied only to the transverse scalar projection
$A_0$:
\begin{equation}
2Y_0^{\mathrm{TE}}
+
j\frac{2k_0}{\eta_0}
A_0(\mathbf k_t)
F_L(k_t)
=
0,
\label{eq:supp_qTE}
\end{equation}
where
\begin{equation}
F_L(k_t)
=
1-\frac{k_t^2}{2k_0^2}.
\label{eq:supp_FL}
\end{equation}
No tensorial extension of
$F_L$ to the off-diagonal hybridisation term is introduced.
Equation~\eqref{eq:supp_qTE} is therefore a reduced closure:
the complete anisotropy of the quasi-static sheet is retained
through the directional quadratic projection
$A_0(\mathbf k_t)$, whereas the explicit TE--TM
Schur-complement correction remains omitted.

\subsection{Closed-form evanescent decay constant}
\label{subsec:supp_gamma_derivation}

For a bound surface wave,
\begin{equation}
k_t^2
=
k_0^2+\gamma^2,
\qquad
\gamma>0.
\label{eq:supp_bound_relation}
\end{equation}
Using the scalar TE transverse-resonance relation corresponding
to Eq.~\eqref{eq:supp_qTE}, the common admittance factors cancel
to give
\begin{equation}
\gamma
=
k_0^2
A_0(\mathbf k_t)
F_L(k_t).
\label{eq:supp_gamma_intermediate}
\end{equation}
The $k_t$-dependent correction factor can be expressed in terms
of $\gamma$ through Eq.~\eqref{eq:supp_bound_relation}:
\begin{align}
F_L(k_t)
&=
1-
\frac{k_0^2+\gamma^2}{2k_0^2}
\nonumber\\
&=
\frac{1}{2}
\left(
1-\frac{\gamma^2}{k_0^2}
\right).
\label{eq:supp_FL_gamma}
\end{align}
Substituting Eq.~\eqref{eq:supp_FL_gamma} into
Eq.~\eqref{eq:supp_gamma_intermediate} gives
\begin{equation}
\gamma
=
\frac{
A_0
}{
2
}
\left(
k_0^2-\gamma^2
\right),
\label{eq:supp_gamma_step}
\end{equation}
where the argument
$\mathbf k_t$ of $A_0$ has been suppressed for compactness.
Rearrangement yields the quadratic equation
\begin{equation}
A_0\gamma^2
+
2\gamma
-
A_0k_0^2
=
0.
\label{eq:supp_gamma_quadratic}
\end{equation}
Its two algebraic roots are
\begin{equation}
\gamma_{\pm}
=
\frac{
-1
\pm
\sqrt{
1+(k_0A_0)^2
}
}{
A_0
}.
\label{eq:supp_gamma_roots}
\end{equation}
For the positive-definite capacitive sheets considered here,
$A_0>0$. The root with the minus sign is negative and is
therefore incompatible with exponential decay away from the
sheet. The physical root is consequently
\begin{equation}
\Gamma_{\mathrm{TL}}(k_0,\mathbf k_t)
=
\frac{
\sqrt{
1+
\left[
k_0A_0(\mathbf k_t)
\right]^2
}
-1
}{
A_0(\mathbf k_t)
}.
\label{eq:supp_GammaTL_1}
\end{equation}
Because Eq.~\eqref{eq:supp_GammaTL_1} was obtained after explicit
substitution of $F_L(k_t)$, the $k_t$-dependent correction is
already embedded in the closed-form root and must not be applied
a second time.

The corresponding quasi-TE residual is
\begin{equation}
R_0(k_0,\mathbf k_t)
=
\sqrt{k_t^2-k_0^2}
-
\Gamma_{\mathrm{TL}}(k_0,\mathbf k_t),
\label{eq:supp_R0_1}
\end{equation}
and the reduced anisotropic dispersion condition is
\begin{equation}
R_0(k_0,\mathbf k_t)=0.
\label{eq:supp_R0_zero}
\end{equation}
In the isotropic limit,
$A_0=A_g$, Eq.~\eqref{eq:supp_GammaTL_1} reduces identically to
the positive root of the scalar patch-array condition derived in
the main text.

\subsection{Directional dependence and principal-axis limit}
\label{subsec:supp_directional}

The directional character of the reduced model follows directly
from Eq.~\eqref{eq:supp_Aperp_principal}. In the principal frame,
\begin{equation}
A_0(\vartheta)
=
A_1\sin^2\vartheta
+
A_2\cos^2\vartheta.
\label{eq:supp_A0_angle}
\end{equation}
Thus the magnitude of the effective capacitive response sampled
by the TE field changes continuously as the propagation direction
rotates with respect to the principal axes.

For the positive-definite capacitance dyadics considered in this
work,
\begin{equation}
A_1>0,
\qquad
A_2>0,
\label{eq:supp_positive_eigenvalues}
\end{equation}
and therefore
\begin{equation}
A_0(\vartheta)>0
\qquad
\text{for all }\vartheta.
\label{eq:supp_positive_A0}
\end{equation}
The resulting quasi-TE isofrequency contours consequently remain
closed. In the weak-anisotropy regime they approach elliptical
contours whose principal directions coincide with the eigenvectors
of $\overline{\overline A}$. This is the positive-definite
counterpart of the sign-indefinite anisotropic-sheet regime, in
which one principal response changes sign and open hyperbolic
isofrequency contours may occur
~\cite{GomezDiaz2015,High2015,GomezDiazOME2015}.

Equations~\eqref{eq:supp_hybrid_det} and
\eqref{eq:supp_schur} provide the exact local anisotropic
eigencondition for the quasi-static sheet constitutive law.
Equation~\eqref{eq:supp_qTE}, together with
Eqs.~\eqref{eq:supp_GammaTL_1} and \eqref{eq:supp_R0_1}, specifies
the reduced quasi-TE closure with the prescribed
$k_t$-dependent transverse correction. Local TE--TM decoupling
is exact along a principal axis and in the isotropic limit.
The hybrid calculation described below retains the same
transverse correction and restores the cross-polarisation
response within every retained Floquet order.

\subsection{Hybrid TE--TM closure and full-zone isofrequency calculation}
\label{subsec:supp_hybrid_isofrequency}

For the isofrequency comparison in the main text, we use a
reduced hybrid extension of the quasi-TE model. The factor
$F_L$ continues to act only on the TE sheet entry, while
the TM and cross entries retain their quasi-static values.
This specifies the polarisation closure used in the
calculation; it is an additional modelling assumption beyond
the strictly local constitutive law of
Eq.~\eqref{eq:supp_Ys}.

We retain the fundamental order and the six orders of the
first reciprocal shell. For $p=0,\ldots,6$, let
\begin{equation}
\mathbf q_p=\mathbf k_t+\mathbf G_p,
\qquad
\mathbf G_0=\mathbf 0,
\qquad
\gamma_p=\sqrt{|\mathbf q_p|^2-k_0^2}>0.
\label{eq:supp_hyb_order_wavevectors}
\end{equation}
Here $\mathbf G_1,\ldots,\mathbf G_6$ are the six first-shell
reciprocal vectors. The projections
$A_{\parallel,p}$, $A_{\perp,p}$, and $A_{\times,p}$
are obtained from Eqs.~\eqref{eq:supp_Aparallel}--%
\eqref{eq:supp_Across} by replacing $\mathbf k_t$ with
$\mathbf q_p$. Thus every retained harmonic samples the
same dyadic in its own propagation-fixed basis.

Using the bound-wave admittance
$Y_{0,p}^{\mathrm{TE}}=-j\gamma_p/(k_0\eta_0)$,
the corrected TE diagonal is
\begin{equation}
\mathcal T_{\mathrm{TE},p}^{(L)}
=
j\frac{2k_0}{\eta_0}
\left[
A_{\perp,p}F_L(|\mathbf q_p|)
-\frac{\gamma_p}{k_0^2}
\right],
\label{eq:supp_hyb_TTE_corrected}
\end{equation}
whereas $\mathcal T_{\mathrm{TM},p}$ and
$\mathcal T_{\times,p}$ follow
Eqs.~\eqref{eq:supp_TTM} and \eqref{eq:supp_Tcross}
with the order-dependent projections.
Substituting these entries into the Schur complement
of Eq.~\eqref{eq:supp_schur}, and using
$|\mathbf q_p|^2=k_0^2+\gamma_p^2$, gives the
uncoupled hybrid residual
\begin{equation}
\mathcal F_p^{\mathrm{hyb}}
=
\gamma_p
-\frac{A_{\perp,p}}{2}(k_0^2-\gamma_p^2)
+\frac{k_0^2 A_{\times,p}^{\,2}\gamma_p}
{1+A_{\parallel,p}\gamma_p}.
\label{eq:supp_hyb_raw_residual}
\end{equation}
The local hybrid condition for order $p$ is
$\mathcal F_p^{\mathrm{hyb}}=0$. The last term restores
the longitudinal--transverse coupling; omitting it gives
the corrected quasi-TE condition.

To use the same decay-detuning normalisation as the
first-shell model, define
\begin{align}
\Gamma_p
&=
\Gamma_{\mathrm{TL}}(k_0,\mathbf q_p)
=
\frac{k_0^2 A_{\perp,p}}
{1+\sqrt{1+(k_0A_{\perp,p})^2}},
\label{eq:supp_hyb_Gamma_orders}
\\
\mathcal N_p
&=
1+\frac{A_{\perp,p}}{2}(\gamma_p+\Gamma_p)>0.
\label{eq:supp_hyb_normalisation}
\end{align}
The quantity $\Gamma_p$ remains the quasi-TE loaded decay
of Eq.~\eqref{eq:supp_GammaTL_1} in both calculations.
Its defining quadratic gives the identity
\begin{equation}
\gamma_p-\frac{A_{\perp,p}}{2}(k_0^2-\gamma_p^2)
=
\mathcal N_p(\gamma_p-\Gamma_p).
\label{eq:supp_hyb_factorisation}
\end{equation}
Consequently, the normalised detunings used in the two
models are
\begin{align}
D_p^{\mathrm{qTE}}
&=\gamma_p-\Gamma_p,
\label{eq:supp_hyb_DqTE}
\\
D_p^{\mathrm{hyb}}
&=
\frac{\mathcal F_p^{\mathrm{hyb}}}{\mathcal N_p}
=
\gamma_p-\Gamma_p
+
\frac{k_0^2 A_{\times,p}^{\,2}\gamma_p}
{\mathcal N_p(1+A_{\parallel,p}\gamma_p)}.
\label{eq:supp_hyb_Dhybrid}
\end{align}
Both detunings have units of inverse length. The division
by $\mathcal N_p$ preserves the uncoupled roots and puts
the hybrid residual in the normalisation of the existing
Floquet coupling ansatz.

We retain that same geometric coupling in both models:
\begin{equation}
v_p
=
\widetilde m_p(\mathbf k_t)
\sqrt{\Gamma_0\Gamma_p}\,
\eta_{0p}(\mathbf k_t),
\qquad p=1,\ldots,6,
\label{eq:supp_hyb_coupling}
\end{equation}
With
$\mathbf v=(v_1,\ldots,v_6)^{\mathsf T}$, the
TM-eliminated system is
\begin{equation}
\mathsf H^{(m)}(k_0,\mathbf k_t)
=
\begin{pmatrix}
D_0^{(m)} & -\mathbf v^\dagger\\
-\mathbf v &
\operatorname{diag}
\bigl(D_1^{(m)},\ldots,D_6^{(m)}\bigr)
\end{pmatrix},
\qquad m\in\{\mathrm{qTE},\mathrm{hyb}\}.
\label{eq:supp_hyb_seven_order_matrix}
\end{equation}
Inter-order coupling acts on the TE amplitudes, while the
TM response enters through the local Schur complement of
each order. The sideband block remains diagonal, as in
the reduced first-shell model. All six sidebands are
retained separately at a general $\mathbf k_t$.

Whenever the sideband detunings are nonzero, their
elimination gives
\begin{equation}
D_0^{(m)}
-\sum_{p=1}^{6}\frac{|v_p|^2}{D_p^{(m)}}=0.
\label{eq:supp_hyb_floquet_schur}
\end{equation}
For numerical evaluation we use the Hermitian matrix
in Eq.~\eqref{eq:supp_hyb_seven_order_matrix} directly,
including near a vanishing sideband detuning.
The frequency maps displayed in the main text are obtained
from the lowest eigenvalue condition
\begin{equation}
\lambda_{\min}
\!\left[\mathsf H^{(m)}(k_0,\mathbf k_t)\right]=0,
\qquad
0<k_0<|\mathbf k_t|,
\qquad
f^{(m)}(\mathbf k_t)=\frac{c_0k_0}{2\pi}.
\label{eq:supp_hyb_frequency_map}
\end{equation}
We solve this condition throughout the first Brillouin
zone and plot the level sets
$f^{(m)}(k_x,k_y)=f_\ast$. At $\Gamma$, the frequency map
uses the continuous lowest-branch limit $f\rightarrow0$.

The fundamental polarisation ratio follows from the
eliminated TM row:
\begin{equation}
\frac{E_{\parallel,0}}{E_{\perp,0}}
=
-\frac{A_{\times,0}\gamma_0}
{1+A_{\parallel,0}\gamma_0},
\qquad
\zeta_{\mathrm{hyb},0}
=
\frac{|A_{\times,0}|\gamma_0}
{1+A_{\parallel,0}\gamma_0}.
\label{eq:supp_hyb_fundamental_ratio}
\end{equation}
For an isotropic dyadic, $A_{\times,p}=0$ for every order,
so the quasi-TE and hybrid matrices coincide exactly.
This explains their agreement for the $C_{6v}$ example
within this reduced model. For a $C_1$ dyadic, the
cross projections are generally nonzero and modify both
the fundamental and sideband detunings. Even when
$\mathbf k_t$ lies along a principal axis and
$A_{\times,0}=0$, the shifted vectors $\mathbf q_p$
can have $A_{\times,p}\neq0$; the full first-shell
comparison can therefore retain a hybrid correction
along that direction.

\newpage
\section{Reciprocal-shell hierarchy and Floquet truncation}
\label{sec:supp_reciprocal_shells}

The nonlocal correction introduced in Sec.~\ref{subsec:sideband_self_energy} of the main text is constructed by
retaining the first nonzero shell of Floquet harmonics.  Here we first
identify the Floquet orders belonging to the first three reciprocal
shells and then compare their evanescent character at the
high-symmetry points $\Gamma$, $M$, and $K$.  This provides a simple
reciprocal-space justification for retaining the complete six-member
first shell while neglecting the more strongly evanescent higher
shells.

The triangular Bravais lattice is associated with the regular
hexagonal unit cell of circumradius
\begin{equation}
R_h=11.547~\mathrm{mm}.
\end{equation}
For the Wigner--Seitz hexagon of the triangular lattice, the
circumradius and lattice constant satisfy
\begin{equation}
R_h=\frac{a}{\sqrt{3}},
\qquad
a=\sqrt{3}\,R_h=20.000~\mathrm{mm}.
\end{equation}
The lattice constant $a$ determines the reciprocal-lattice scale,
whereas $R_h$ is used below only as a characteristic geometrical
length against which the evanescent decay lengths may be compared.

\subsection{First three reciprocal shells}
\label{sec:supp_shells}

We choose the primitive direct-lattice vectors as
\begin{equation}
\mathbf a_1
=
a
\begin{pmatrix}
1\\
0
\end{pmatrix},
\qquad
\mathbf a_2
=
a
\begin{pmatrix}
1/2\\
\sqrt{3}/2
\end{pmatrix}.
\label{eq:supp_direct_vectors}
\end{equation}
The primitive reciprocal vectors are defined by
\begin{equation}
\mathbf a_i\cdot\mathbf b_j
=
2\pi\delta_{ij},
\label{eq:supp_reciprocal_definition}
\end{equation}
which gives
\begin{equation}
\mathbf b_1
=
\frac{2\pi}{a}
\begin{pmatrix}
1\\
-1/\sqrt{3}
\end{pmatrix},
\qquad
\mathbf b_2
=
\frac{2\pi}{a}
\begin{pmatrix}
0\\
2/\sqrt{3}
\end{pmatrix}.
\label{eq:supp_b_vectors}
\end{equation}
Their common magnitude is
\begin{equation}
G_0
=
|\mathbf b_1|
=
|\mathbf b_2|
=
\frac{4\pi}{\sqrt{3}a}
=
0.36275987~\mathrm{mm}^{-1}.
\label{eq:supp_G0}
\end{equation}

An arbitrary reciprocal-lattice vector is
\begin{equation}
\mathbf G_{mn}
=
m\mathbf b_1+n\mathbf b_2,
\qquad
m,n\in\mathbb Z.
\label{eq:supp_Gmn}
\end{equation}
Using Eq.~\eqref{eq:supp_b_vectors},
\begin{equation}
|\mathbf G_{mn}|^2
=
G_0^2
\left(
m^2-mn+n^2
\right).
\label{eq:supp_Gmn_mag}
\end{equation}
Defining
\begin{equation}
s_{mn}
=
m^2-mn+n^2,
\label{eq:supp_smn}
\end{equation}
the reciprocal-shell radius is therefore
\begin{equation}
|\mathbf G_{mn}|
=
G_0\sqrt{s_{mn}}.
\label{eq:supp_shell_radius}
\end{equation}

The first nonzero shell is obtained for $s_{mn}=1$.  Its six Floquet
orders are
\begin{equation}
(m,n)
=
(1,0),\,
(0,1),\,
(1,1),\,
(-1,0),\,
(0,-1),\,
(-1,-1),
\label{eq:supp_first_shell_orders}
\end{equation}
corresponding to
\begin{align}
\mathbf G_1 &= \mathbf b_1,
&
\mathbf G_2 &= \mathbf b_2,
&
\mathbf G_3 &= \mathbf b_1+\mathbf b_2,
\nonumber\\
\mathbf G_4 &= -\mathbf b_1,
&
\mathbf G_5 &= -\mathbf b_2,
&
\mathbf G_6 &= -(\mathbf b_1+\mathbf b_2).
\label{eq:supp_first_shell_vectors}
\end{align}
All six have the same reciprocal-space magnitude,
\begin{equation}
|\mathbf G_p|=G_0,
\qquad
p=1,\ldots,6.
\label{eq:supp_first_shell_radius}
\end{equation}

The second shell corresponds to $s_{mn}=3$ and contains
\begin{equation}
(m,n)
=
(1,-1),\,
(-1,1),\,
(2,1),\,
(1,2),\,
(-2,-1),\,
(-1,-2),
\label{eq:supp_second_shell_orders}
\end{equation}
with
\begin{equation}
|\mathbf G|_{\mathrm{II}}
=
\sqrt{3}\,G_0.
\label{eq:supp_second_shell_radius}
\end{equation}

The third shell corresponds to $s_{mn}=4$ and contains
\begin{equation}
(m,n)
=
(2,0),\,
(0,2),\,
(2,2),\,
(-2,0),\,
(0,-2),\,
(-2,-2),
\label{eq:supp_third_shell_orders}
\end{equation}
with
\begin{equation}
|\mathbf G|_{\mathrm{III}}
=
2G_0.
\label{eq:supp_third_shell_radius}
\end{equation}

The first three shell radii are consequently
\begin{equation}
|\mathbf G|_{\mathrm I}=G_0,
\qquad
|\mathbf G|_{\mathrm{II}}=\sqrt{3}G_0,
\qquad
|\mathbf G|_{\mathrm{III}}=2G_0.
\label{eq:supp_three_shells}
\end{equation}

\begin{figure}[t]
    \centering
    \includegraphics[width=0.82\linewidth]
    {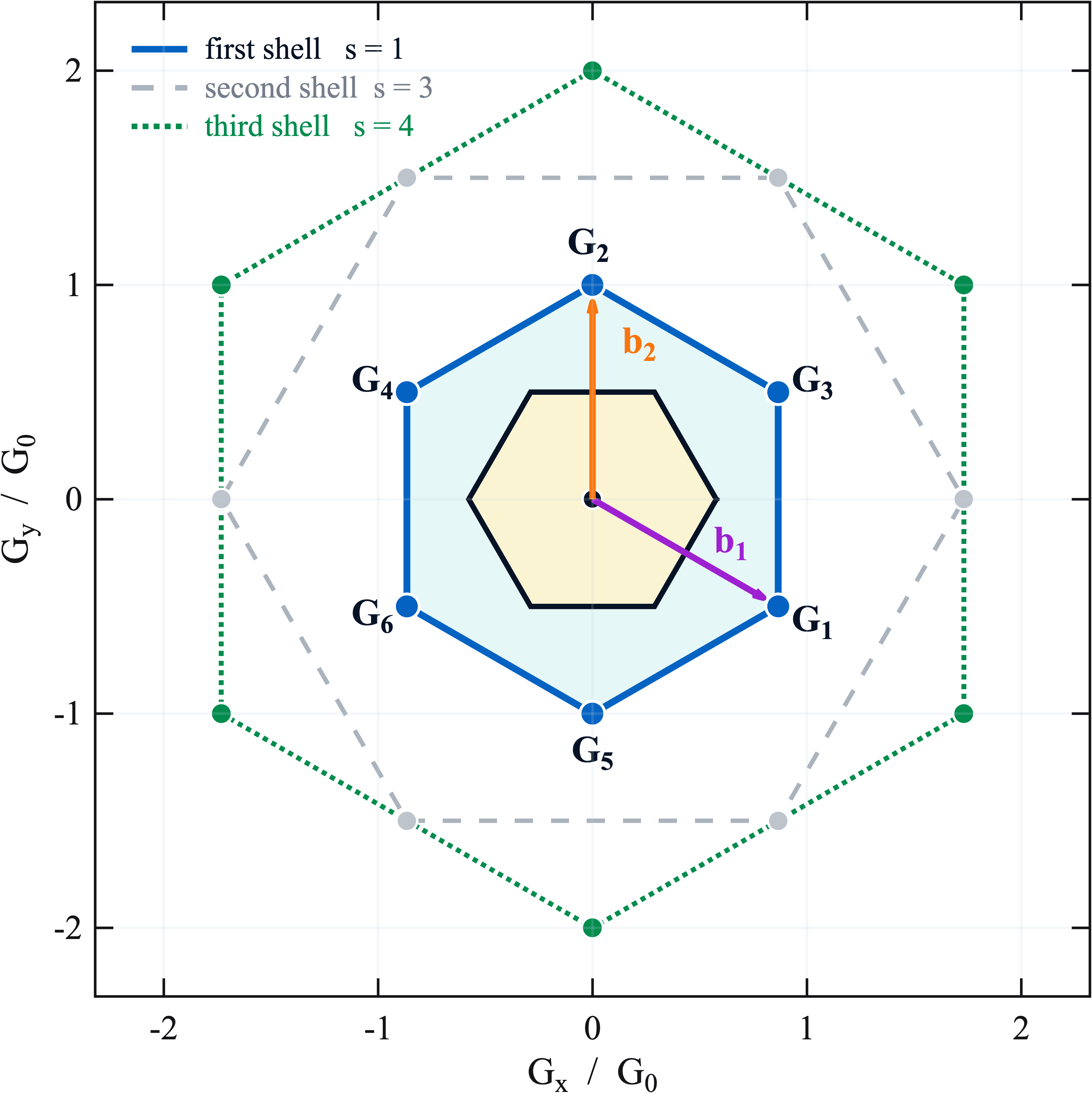}
    \caption{
    First three reciprocal shells of the triangular Bravais lattice,
    normalized by $G_0=|\mathbf b_1|=|\mathbf b_2|$.
    The solid blue, dashed gray, and dotted green contours correspond
    respectively to the first, second, and third shells.
    The six vectors $\mathbf G_1,\ldots,\mathbf G_6$ form the complete
    first nonzero reciprocal shell retained in the Floquet model.
    }
    \label{fig:supp_reciprocal_shells}
\end{figure}

The corresponding reciprocal-space construction is shown in
Fig.~\ref{fig:supp_reciprocal_shells}.

\subsection{Selection of the six first-shell Floquet harmonics}
\label{sec:supp_first_shell_selection}

A Bloch field of the periodic sheet can be expanded in Floquet
harmonics as
\begin{equation}
\mathbf E_t(\boldsymbol\rho,z)
=
\sum_{m,n\in\mathbb Z}
\mathbf E_{mn}(z)
\exp
\left[
j
\left(
\mathbf k_t+\mathbf G_{mn}
\right)
\cdot\boldsymbol\rho
\right].
\label{eq:supp_floquet_expansion}
\end{equation}
The fundamental Floquet harmonic corresponds to
$\mathbf G_{00}=0$, while the nonzero reciprocal vectors generate the
higher Floquet harmonics of the same Bloch field.

For a Bloch wavevector $\mathbf k_t$, the transverse wavenumber of the
harmonic associated with $\mathbf G_{mn}$ is
\begin{equation}
\mathbf q_{mn}
=
\mathbf k_t+\mathbf G_{mn},
\qquad
q_{mn}
=
|\mathbf q_{mn}|.
\label{eq:supp_qmn}
\end{equation}
Its free-space normal wavenumber is
\begin{equation}
k_{z,mn}
=
\sqrt{k_0^2-q_{mn}^2},
\qquad
k_0=\frac{2\pi f}{c_0}.
\label{eq:supp_kz}
\end{equation}
For $q_{mn}>k_0$, the harmonic is evanescent.  Defining
\begin{equation}
\gamma_{mn}
=
\sqrt{q_{mn}^2-k_0^2},
\label{eq:supp_gamma}
\end{equation}
the normal wavenumber can be written as
\begin{equation}
k_{z,mn}
=
\pm j\gamma_{mn},
\end{equation}
where the appropriate sign is chosen to enforce decay away from the
sheet.  The field envelope therefore varies as
\begin{equation}
e^{-\gamma_{mn}|z|},
\end{equation}
with $1/e$ decay length
\begin{equation}
\ell_{mn}
=
\frac{1}{\gamma_{mn}}.
\label{eq:supp_decay}
\end{equation}

A large $\gamma_{mn}$, or equivalently a small $\ell_{mn}$,
corresponds to a deeply evanescent and strongly confined Floquet
harmonic.  Conversely, as
\begin{equation}
q_{mn}\rightarrow k_0^{+},
\end{equation}
one has
\begin{equation}
\gamma_{mn}\rightarrow0^{+},
\qquad
\ell_{mn}\rightarrow\infty,
\end{equation}
and the harmonic becomes progressively more weakly evanescent.

The equality
\begin{equation}
q_{mn}=k_0
\end{equation}
defines the free-space propagation threshold of that Floquet harmonic,
\begin{equation}
f_{\mathrm{th},mn}
=
\frac{c_0q_{mn}}{2\pi}.
\label{eq:supp_fth}
\end{equation}
This frequency is not a resonance of the metasurface; it only marks
the transition of the corresponding Floquet harmonic from evanescent
to propagating.

To compare the first three reciprocal shells, we evaluate the
least-confined Floquet harmonic from each shell at the three
high-symmetry points $\Gamma$, $M$, and $K$.

At $\Gamma$,
\begin{equation}
\mathbf k_t=0,
\end{equation}
so that $q_{mn}=|\mathbf G_{mn}|$.  Therefore
\begin{equation}
q_{\mathrm I}(\Gamma)
=
G_0,
\qquad
q_{\mathrm{II}}(\Gamma)
=
\sqrt{3}G_0,
\qquad
q_{\mathrm{III}}(\Gamma)
=
2G_0.
\label{eq:supp_Gamma_q}
\end{equation}
The corresponding propagation thresholds are
\begin{equation}
f_{\mathrm{th,I}}(\Gamma)
=
17.31~\mathrm{GHz},
\qquad
f_{\mathrm{th,II}}(\Gamma)
=
29.98~\mathrm{GHz},
\qquad
f_{\mathrm{th,III}}(\Gamma)
=
34.62~\mathrm{GHz}.
\label{eq:supp_Gamma_threshold}
\end{equation}
In particular, all six members of the first shell have the same
transverse wavenumber $q_p=G_0$ at $\Gamma$ and therefore the same
free-space evanescent decay constant.

For a representative $M$ point,
\begin{equation}
\mathbf M
=
\frac{\mathbf b_1}{2},
\label{eq:supp_M_b}
\end{equation}
the first-shell vector $\mathbf G_4=-\mathbf b_1$ gives
\begin{equation}
|\mathbf M+\mathbf G_4|
=
\frac{G_0}{2}.
\label{eq:supp_M_first}
\end{equation}
The minimum values obtained from the second and third shells are
\begin{equation}
q_{\mathrm{II,min}}(M)
=
\frac{\sqrt{7}}{2}G_0,
\qquad
q_{\mathrm{III,min}}(M)
=
\frac{3}{2}G_0.
\label{eq:supp_M_higher}
\end{equation}
Hence
\begin{equation}
f_{\mathrm{th,I}}(M)
=
8.65~\mathrm{GHz},
\qquad
f_{\mathrm{th,II}}(M)
=
22.9~\mathrm{GHz},
\qquad
f_{\mathrm{th,III}}(M)
=
25.96~\mathrm{GHz}.
\label{eq:supp_M_threshold}
\end{equation}

For a representative $K$ point,
\begin{equation}
\mathbf K
=
\frac{2\mathbf b_1+\mathbf b_2}{3},
\label{eq:supp_K}
\end{equation}
the two first-shell vectors
\begin{equation}
\mathbf G_4=-\mathbf b_1,
\qquad
\mathbf G_6=-(\mathbf b_1+\mathbf b_2)
\end{equation}
give
\begin{equation}
|\mathbf K+\mathbf G_4|
=
|\mathbf K+\mathbf G_6|
=
\frac{G_0}{\sqrt{3}}.
\label{eq:supp_K_first}
\end{equation}
For the second and third shells,
\begin{equation}
q_{\mathrm{II,min}}(K)
=
\frac{2G_0}{\sqrt{3}},
\qquad
q_{\mathrm{III,min}}(K)
=
G_0\sqrt{\frac{7}{3}}.
\label{eq:supp_K_higher}
\end{equation}
The corresponding thresholds are
\begin{equation}
f_{\mathrm{th,I}}(K)
=
9.99~\mathrm{GHz},
\qquad
f_{\mathrm{th,II}}(K)
=
19.99~\mathrm{GHz},
\qquad
f_{\mathrm{th,III}}(K)
=
26.44~\mathrm{GHz}.
\label{eq:supp_K_threshold}
\end{equation}

The comparison is summarized in Table~\ref{tab:supp_thresholds}.

\begin{table}[t]
\caption{
Propagation threshold of the least-confined Floquet harmonic from each
of the first three reciprocal shells at the high-symmetry points,
calculated for $a=20~\mathrm{mm}$.
}
\label{tab:supp_thresholds}
\begin{ruledtabular}
\begin{tabular}{cccc}
 & $\Gamma$ & $M$ & $K$ \\
\hline
First shell
& $17.31$ GHz & $8.65$ GHz & $9.99$ GHz \\
Second shell
& $29.98$ GHz & $22.9$ GHz & $19.99$ GHz \\
Third shell
& $34.62$ GHz & $25.96$ GHz & $26.44$ GHz
\end{tabular}
\end{ruledtabular}
\end{table}

The first-shell harmonics are therefore the first higher Floquet
orders to approach weak evanescence near the Brillouin-zone boundary.
The second and third shells remain substantially farther from their
propagation thresholds and consequently remain more strongly
evanescent over the same frequency range.

The difference can also be seen directly from the decay lengths.
For example, at $f=9~\mathrm{GHz}$ the least-confined harmonic of
each shell has
\begin{equation}
\begin{array}{c|ccc}
 & \Gamma & M & K \\ \hline
\text{first shell}
&
3.23~\mathrm{mm}
&
\mathrm{propagating}
&
10.99~\mathrm{mm}
\\
\text{second shell}
&
1.67~\mathrm{mm}
&
2.27~\mathrm{mm}
&
2.67~\mathrm{mm}
\\
\text{third shell}
&
1.43~\mathrm{mm}
&
1.96~\mathrm{mm}
&
1.92~\mathrm{mm}
\end{array}.
\label{eq:supp_decay_HSP}
\end{equation}
Compared with the characteristic hexagonal radius
$R_h=11.547~\mathrm{mm}$, the first-shell harmonics near $M$ and
$K$ possess considerably longer decay lengths than those of the
second and third shells.  They therefore provide the leading
higher-harmonic correction to the fundamental $\mathbf G=0$ channel.

The evanescent hierarchy is used here as a spectral truncation
criterion; it does not imply that the higher-shell harmonics vanish
identically.  Their actual contribution also depends on the
geometry-dependent coupling to the fundamental harmonic.
Nevertheless, the substantially larger transverse wavenumbers and
shorter decay lengths of the second and higher shells identify them as
higher-order corrections over the frequency range considered here.

Finally, the complete first shell is retained rather than only the
harmonic that is least evanescent for one chosen propagation direction.
At $\Gamma$, all six first-shell vectors have the same
$q_p=G_0$.  At the representative $M$ point of
Eq.~\eqref{eq:supp_M_b}, $\mathbf G_4$ is the least-confined
first-shell harmonic.  Moving to the other geometrically related
$M$ points of the hexagonal Brillouin zone selects the other members
of the same first shell.  Likewise, at the representative $K$ point
of Eq.~\eqref{eq:supp_K}, the pair $\mathbf G_4$ and
$\mathbf G_6$ has the smallest shifted transverse wavenumber, while
the corresponding pairs are permuted among the first-shell vectors
at the other $K$ vertices.

Accordingly, the truncation is performed by reciprocal-shell radius,
rather than by selecting only those harmonics that are least
evanescent along a particular propagation direction.  The six vectors
\begin{equation}
\left\{
\pm\mathbf b_1,\,
\pm\mathbf b_2,\,
\pm(\mathbf b_1+\mathbf b_2)
\right\}
\end{equation}
form the complete nearest nonzero reciprocal shell.  Retaining this
complete six-member set therefore provides a magnitude-complete
first-shell truncation without artificially privileging a particular
propagation direction, while the more deeply evanescent second and
higher shells are neglected as higher-order spectral corrections.

\newpage

\section{Geometry-derived first-shell Floquet correction}
\label{sec:supp_first_shell_correction}

This section gives the detailed construction of the nonlocal correction used
in the main text.  The local sheet response is supplied by the equivalent
surface-capacitance dyadic
$\overline{\overline C}_{\mathrm{eqv}}$, whereas the leading spatially
dispersive correction additionally uses the reconstructed metal boundary
$\partial\Sigma$.  The complete first reciprocal shell is retained as six
separate Floquet orders.  No positive-half-shell multiplicity and no fitted
per-order coupling coefficient are introduced.

The derivation proceeds in four stages.  First, the same local dyadic is
projected onto the TE direction of every shifted Floquet wavevector.  Second,
the complex coupling between the fundamental order and each sideband is
computed by an exact Fourier integral over the polygonal boundary.  Third, a
directional-turning functional assigns a geometry-derived amplitude to each
of the six orders.  Finally, the six amplitudes are eliminated by a Schur
complement, with a little-group projection applied when the propagation path
is a mirror axis of the complete unit-cell geometry.

\subsection{Complete first shell and TE-projected local quantities}
\label{subsec:supp_first_shell}

Let the triangular Bravais lattice be generated by
\begin{equation}
\mathbf a_1=a
\begin{pmatrix}
1\\[1mm]0
\end{pmatrix},
\qquad
\mathbf a_2=a
\begin{pmatrix}
1/2\\[1mm]\sqrt{3}/2
\end{pmatrix}.
\label{eq:supp_direct_basis}
\end{equation}
Writing
\begin{equation}
\mathbf A_{\mathrm{lat}}
=
\begin{pmatrix}
\mathbf a_1&\mathbf a_2
\end{pmatrix},
\qquad
\mathbf B_{\mathrm{lat}}
=2\pi\mathbf A_{\mathrm{lat}}^{-\mathsf T}
=
\begin{pmatrix}
\mathbf b_1&\mathbf b_2
\end{pmatrix},
\label{eq:supp_reciprocal_matrix}
\end{equation}
gives
\begin{equation}
\mathbf b_1
=\frac{2\pi}{a}
\begin{pmatrix}
1\\[1mm]-1/\sqrt{3}
\end{pmatrix},
\qquad
\mathbf b_2
=\frac{2\pi}{a}
\begin{pmatrix}
0\\[1mm]2/\sqrt{3}
\end{pmatrix},
\label{eq:supp_reciprocal_basis}
\end{equation}
so that $\mathbf a_i\!\cdot\!\mathbf b_j=2\pi\delta_{ij}$.  Their
magnitudes and mutual scalar product are
\begin{equation}
|\mathbf b_1|=|\mathbf b_2|
=|\mathbf b_1+\mathbf b_2|
=G_0,
\qquad
G_0=\frac{4\pi}{\sqrt{3}a},
\qquad
\mathbf b_1\!\cdot\!\mathbf b_2=-\frac{G_0^2}{2}.
\label{eq:supp_G0_coupling}
\end{equation}
Consequently, the six shortest nonzero reciprocal vectors form the first
shell~\cite{Brillouin1953,Munk2005}.  We use the fixed ordering
\begin{align}
\mathbf G_1&=+\mathbf b_1,
&\mathbf G_2&=+\mathbf b_2,
&\mathbf G_3&=+(\mathbf b_1+\mathbf b_2),
\nonumber\\
\mathbf G_4&=-\mathbf b_1,
&\mathbf G_5&=-\mathbf b_2,
&\mathbf G_6&=-(\mathbf b_1+\mathbf b_2).
\label{eq:supp_six_orders}
\end{align}
The selected $K$ point and the straight $\Gamma$--$K$ path are
\begin{equation}
\mathbf K=\frac{2\mathbf b_1+\mathbf b_2}{3}
=\frac{4\pi}{3a}\widehat{\mathbf x},
\qquad
\mathbf k_t(s)=s\mathbf K,
\qquad 0\leq s\leq1.
\label{eq:supp_GammaK_path}
\end{equation}
For notational uniformity, introduce the fundamental order
$\mathbf G_0=\mathbf0$.  The transverse wavevector of order
$a\in\{0,1,\ldots,6\}$ is then
\begin{equation}
\mathbf q_a=\mathbf k_t+\mathbf G_a,
\qquad
\mathbf q_0=\mathbf k_t.
\label{eq:supp_qa}
\end{equation}

\begin{figure*}[t]
  \centering
  \includegraphics[width=\textwidth]
  {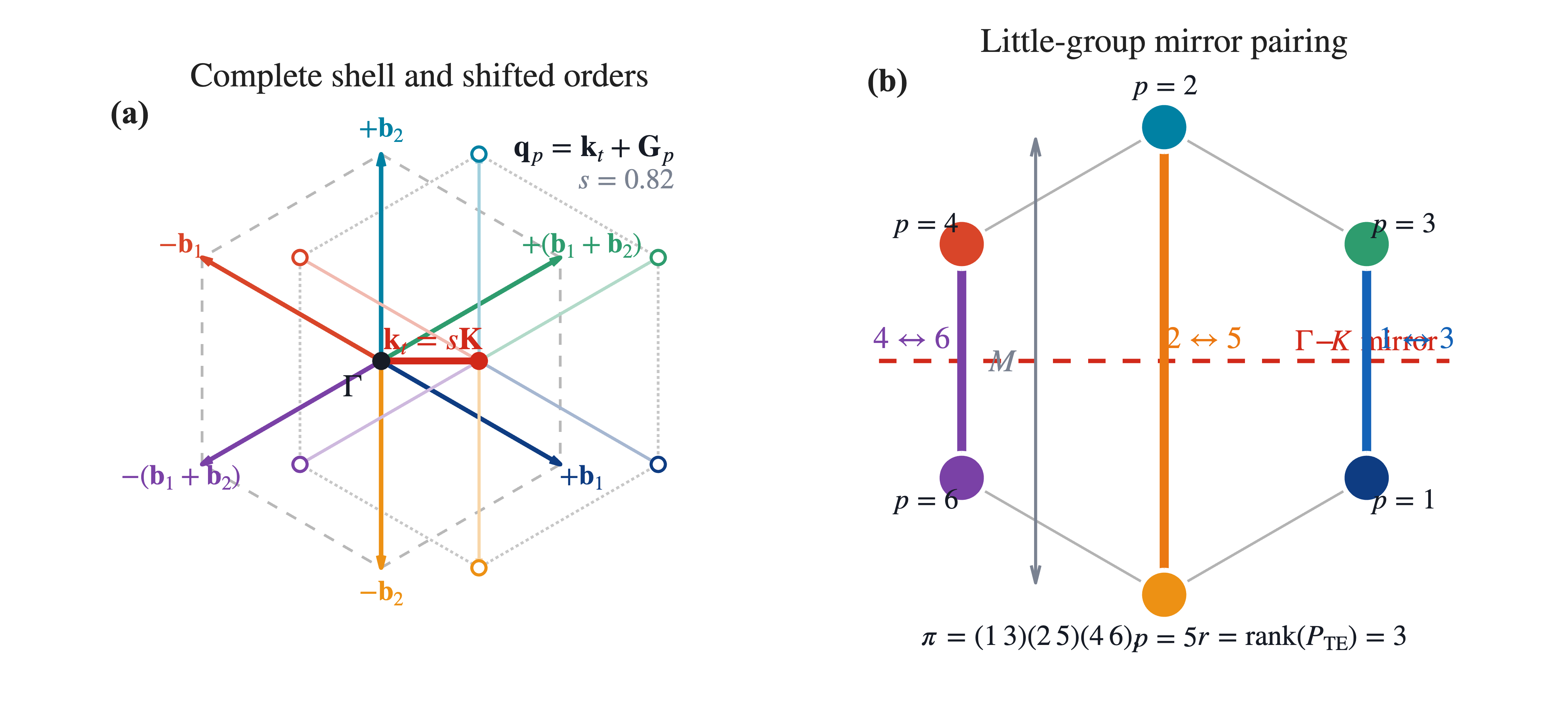}
  \caption{
  Complete first reciprocal shell and its mirror action on the
  $\Gamma$--$K$ path.
  (a) The six vectors
  $\mathbf G_p\in\{\pm\mathbf b_1,\pm\mathbf b_2,
  \pm(\mathbf b_1+\mathbf b_2)\}$ and the shifted Floquet
  wavevectors $\mathbf q_p=\mathbf k_t+\mathbf G_p$ at the
  representative point $\mathbf k_t=0.82\,\mathbf K$.
  (b) Reflection about the $\Gamma$--$K$ propagation axis pairs
  the retained orders according to $\pi=(1\,3)(2\,5)(4\,6)$.
  As derived in Sec.~\ref{subsec:supp_mirror_projection}, the
  associated TE-compatible projector has rank three.
  }
  \label{fig:supp_first_shell_mirror}
\end{figure*}

The complete first shell and the corresponding shifted
Floquet wavevectors are summarized in
Fig.~\ref{fig:supp_first_shell_mirror}.  Panel (b) also
previews the mirror pairing used later in
Sec.~\ref{subsec:supp_mirror_projection}.

For an evanescent harmonic, the free-space normal decay is evaluated on the
branch with positive real part,
\begin{equation}
\gamma_{h_a}
=\sqrt{|\mathbf q_a|^2-k_0^2},
\qquad
\operatorname{Re}\gamma_{h_a}>0.
\label{eq:supp_gammaha}
\end{equation}
Thus $\gamma_{h_0}=\sqrt{k_t^2-k_0^2}$ on the bound fundamental
branch.

Every order samples the same local sheet dyadic through its own TE
polarization.  Specifically,
\begin{equation}
\widehat{\mathbf e}_a
=\widehat{\mathbf e}_{\mathrm{TE}}(\mathbf q_a)
=\frac{\widehat{\mathbf z}\times\mathbf q_a}{|\mathbf q_a|}
=\frac{1}{|\mathbf q_a|}
\begin{pmatrix}
-q_{a,y}\\[1mm]q_{a,x}
\end{pmatrix},
\label{eq:supp_ea}
\end{equation}
and
\begin{equation}
\overline{\overline A}
=\frac{\overline{\overline C}_{\mathrm{eqv}}}{2\varepsilon_0},
\qquad
A_a
=\widehat{\mathbf e}_a\!\cdot\!
\overline{\overline A}\!\cdot\!
\widehat{\mathbf e}_a.
\label{eq:supp_Aa}
\end{equation}
Because $\overline{\overline A}$ is real symmetric and positive definite,
$A_a>0$ for every nonzero $\mathbf q_a$.  At $\Gamma$, the direction of
$\widehat{\mathbf e}_0$ is defined by the one-sided path limit
\begin{equation}
\widehat{\mathbf e}_0(\Gamma)
=\lim_{s\downarrow0}
\frac{\widehat{\mathbf z}\times\mathbf k_t(s)}{|\mathbf k_t(s)|}
=\widehat{\mathbf z}\times\widehat{\mathbf K}.
\label{eq:supp_Gamma_limit}
\end{equation}
This removes the apparent $0/0$ ambiguity without assigning an arbitrary
polarization at $\Gamma$.

Let $\Gamma_{\mathrm{TL}}(k_0,A)$ denote the sheet-loaded decay obtained
from the local transverse-resonance condition in the main text.  With the
Luukkonen obliquity factor used there, it is the positive solution of
\begin{equation}
A\Gamma^2+2\Gamma-Ak_0^2=0.
\label{eq:supp_Gamma_quadratic}
\end{equation}
For $A>0$, the two algebraic roots are
\begin{equation}
\Gamma_{\pm}
=\frac{-1\pm\sqrt{1+(k_0A)^2}}{A}.
\label{eq:supp_Gamma_roots}
\end{equation}
The decaying bound branch requires $\Gamma>0$, and therefore
\begin{equation}
\Gamma_{\mathrm{TL}}(k_0,A)
=\frac{\sqrt{1+(k_0A)^2}-1}{A}
=\frac{k_0^2A}{1+\sqrt{1+(k_0A)^2}}.
\label{eq:supp_GammaTL}
\end{equation}
The second form follows by rationalizing the numerator and is numerically
stable when $k_0A\ll1$.  In particular,
\begin{equation}
\Gamma_{\mathrm{TL}}(k_0,A)
=\frac{k_0^2A}{2}
-\frac{k_0^4A^3}{8}
+O\!\left(k_0^6A^5\right).
\label{eq:supp_Gamma_small}
\end{equation}
We henceforth use
\begin{equation}
\Gamma_0
\equiv\Gamma_{\mathrm{TL}}(k_0,A_0),
\qquad
\Gamma_p
\equiv\Gamma_{\mathrm{TL}}(k_0,A_p),
\qquad p=1,\ldots,6.
\label{eq:supp_Gamma0p}
\end{equation}
The local fundamental residual is
\begin{equation}
R_0(k_0,\mathbf k_t)
=\gamma_{h_0}-\Gamma_0
=\sqrt{k_t^2-k_0^2}-\Gamma_0.
\label{eq:supp_R0}
\end{equation}

\subsection{Exact complex boundary Fourier overlap}
\label{subsec:supp_boundary_overlap}

After unioning the metallic tessellation cells and clipping the result to the
unit cell, the metal--gap interface is reconstructed as a collection of
closed polygonal loops.  Duplicate consecutive vertices and vertices lying
on a straight continuation are removed.  The resulting set of open straight
edges is denoted by $\mathcal E(\partial\Sigma)$.  For each edge
$e\in\mathcal E(\partial\Sigma)$, let
\begin{equation}
\mathbf r_e^{-},\ \mathbf r_e^{+}
\quad\text{be its endpoints},
\qquad
\mathbf d_e=\mathbf r_e^{+}-\mathbf r_e^{-},
\qquad
L_e=|\mathbf d_e|,
\label{eq:supp_edge_geometry}
\end{equation}
and define
\begin{equation}
\widehat{\mathbf t}_e=\frac{\mathbf d_e}{L_e},
\qquad
\widehat{\mathbf n}_e
=
\begin{pmatrix}
t_{e,y}\\[1mm]-t_{e,x}
\end{pmatrix}.
\label{eq:supp_edge_frame}
\end{equation}
The normal is constant on the interior of a polygon edge.  Reversing the
orientation of a complete loop changes both factors of
$\widehat{\mathbf n}_e$ below and consequently leaves the overlap unchanged.

The electromagnetic field behaviour near conducting edges
and the associated concentration of induced charge motivate
a boundary-based description of the capacitive
response~\cite{Jackson1998,Meixner1972}.
Within the present reduced model, the normal projections
specify geometric weights for the assumed edge-charge
profiles.

Let $\ell\in[0,L]$ parameterize the reconstructed boundary,
where $L$ is its total length and
$\mathbf r(\ell)\in\partial\Sigma$.
For several boundary loops, their arclength
parameterizations are concatenated in a fixed order.
Consistently with the main text, we define the
dimensionless model profiles
\begin{equation}
    \widetilde{\rho}_a(\ell;\mathbf k_t)
    =
    \bigl[
        \widehat{\mathbf e}_a\cdot
        \widehat{\mathbf n}(\ell)
    \bigr]
    e^{-i(\mathbf k_t+\mathbf G_a)\cdot\mathbf r(\ell)},
    \qquad a=0,\ldots,6,
    \label{eq:supp_model_edge_charge_profiles}
\end{equation}
where $\mathbf G_0=\mathbf0$ and
$\mathbf G_p$, $p=1,\ldots,6$, are the retained
first-shell reciprocal vectors.
Here $\widehat{\mathbf n}$ is the in-plane boundary
normal,
$\widehat{\mathbf e}_0
\equiv\widehat{\mathbf e}_{\mathrm{TE}}$,
and $\widehat{\mathbf e}_p$ is the TE polarization
relative to $\mathbf k_t+\mathbf G_p$.
The polarization vectors and $\mathbf k_t$ are real.
The modal amplitudes are factored out; the profile
shapes constitute the adopted edge-projection
approximation.

The full-boundary cross moment between the fundamental
order and sideband $p$ is
\begin{align}
    B_{0p}
    &=
    \int_0^L
    \widetilde{\rho}_0^{\,*}(\ell;\mathbf k_t)
    \widetilde{\rho}_p(\ell;\mathbf k_t)
    \,\mathrm d\ell
    \nonumber\\
    &=
    \oint_{\partial\Sigma}
    (\widehat{\mathbf e}_0\cdot\widehat{\mathbf n})
    (\widehat{\mathbf e}_p\cdot\widehat{\mathbf n})
    e^{-i\mathbf G_p\cdot\mathbf r}
    \,\mathrm d\ell,
    \label{eq:supp_B0p_continuous}
\end{align}
and the squared boundary-profile norms are
\begin{align}
    N_a
    &=
    \int_0^L
    \left|
        \widetilde{\rho}_a(\ell;\mathbf k_t)
    \right|^2
    \,\mathrm d\ell
    \nonumber\\
    &=
    \oint_{\partial\Sigma}
    (\widehat{\mathbf e}_a\cdot\widehat{\mathbf n})^2
    \,\mathrm d\ell,
    \qquad a\in\{0,p\}.
    \label{eq:supp_Na_continuous}
\end{align}
The common Bloch phase cancels in the cross moment,
leaving the reciprocal phase
$e^{-i\mathbf G_p\cdot\mathbf r}$.
The phase factors do not contribute to the squared norms.

For nonzero profile norms, the normalized overlap is
\begin{equation}
    \eta_{0p}(\mathbf k_t)
    =
    \frac{B_{0p}}{\sqrt{N_0N_p}},
    \label{eq:supp_eta_definition}
\end{equation}
which measures the complex correlation between the
two assumed charge-profile shapes.

The full-boundary quantities $B_{0p}$ and $N_a$
correspond to $B_{0p}(L)$ and $N_a(L)$ in the
cumulative arclength notation of the main text.

\subsubsection{Floquet orthogonality and boundary-weighted overlap}
\label{subsubsec:supp_floquet_orthogonality}

We distinguish the orthogonality of Floquet harmonics over
a complete primitive cell from the boundary-weighted overlap
used in the present coupling model.

For a real tangential Bloch wavevector $\mathbf k_t$,
the fundamental and sideband harmonic functions at the
sheet plane are
\begin{equation}
    \boldsymbol{\psi}_0(\mathbf r)
    =
    \widehat{\mathbf e}_0
    e^{-i\mathbf k_t\cdot\mathbf r},
    \qquad
    \boldsymbol{\psi}_p(\mathbf r)
    =
    \widehat{\mathbf e}_p
    e^{-i(\mathbf k_t+\mathbf G_p)\cdot\mathbf r}.
    \label{eq:supp_orth_harmonics}
\end{equation}
Here the TE polarization vectors are real.
Their unweighted area inner product over a complete
primitive cell $U$ is
\begin{align}
    \langle\boldsymbol{\psi}_0,
           \boldsymbol{\psi}_p\rangle_U
    &=
    \int_U
    \boldsymbol{\psi}_0^*(\mathbf r)
    \cdot\boldsymbol{\psi}_p(\mathbf r)
    \,\mathrm dA
    \nonumber\\
    &=
    (\widehat{\mathbf e}_0\cdot\widehat{\mathbf e}_p)
    \int_U e^{-i\mathbf G_p\cdot\mathbf r}\,\mathrm dA.
    \label{eq:supp_orth_area_product}
\end{align}
The common Bloch phase cancels in this product.

Let $\mathbf a_1,\mathbf a_2$ be the primitive lattice
vectors and $\mathbf b_1,\mathbf b_2$ their reciprocal
vectors, with
$\mathbf a_i\cdot\mathbf b_j=2\pi\delta_{ij}$.
We write
\begin{equation}
    \mathbf G_p=m_1\mathbf b_1+m_2\mathbf b_2,
    \qquad
    \mathbf r=t_1\mathbf a_1+t_2\mathbf a_2,
    \qquad
    0\leq t_1,t_2<1,
\end{equation}
where $m_1,m_2$ are integers. Consequently,
\begin{align}
    \int_U e^{-i\mathbf G_p\cdot\mathbf r}\,\mathrm dA
    &=
    |U|
    \int_0^1 e^{-i2\pi m_1t_1}\,\mathrm dt_1
    \int_0^1 e^{-i2\pi m_2t_2}\,\mathrm dt_2
    \nonumber\\
    &=
    |U|\,\delta_{m_1,0}\delta_{m_2,0}.
    \label{eq:supp_orth_cell_integral}
\end{align}
For every nonzero integer $m$,
\begin{equation}
    \int_0^1 e^{-i2\pi mt}\,\mathrm dt
    =
    \frac{e^{-i2\pi m}-1}{-i2\pi m}
    =
    0.
\end{equation}
Thus, for $\mathbf G_p\neq\mathbf0$,
\begin{equation}
    \langle\boldsymbol{\psi}_0,
           \boldsymbol{\psi}_p\rangle_U=0.
    \label{eq:supp_orth_area_result}
\end{equation}
Although this derivation uses a primitive parallelogram,
the integrand is lattice-periodic, so the same result holds
over the hexagonal Wigner--Seitz cell.

The boundary cross moment uses a different inner product.
Let $\ell\in[0,L]$ parameterize the reconstructed boundary,
with $\mathbf r(\ell)\in\partial\Sigma$, and define
the normal traces
\begin{equation}
    g_a(\ell)
    =
    \widehat{\mathbf n}(\ell)\cdot
    \boldsymbol{\psi}_a\bigl(\mathbf r(\ell)\bigr),
    \qquad a\in\{0,p\}.
    \label{eq:supp_orth_normal_traces}
\end{equation}

These normal traces coincide with the dimensionless
model edge-charge profiles defined above:
\begin{equation}
    g_a(\ell)
    =
    \widetilde{\rho}_a(\ell;\mathbf k_t),
    \qquad a\in\{0,p\}.
    \label{eq:supp_trace_charge_identity}
\end{equation}
Thus the boundary inner product used below is precisely
the model charge-profile overlap introduced in the
main text.

For several boundary loops, the following integrals denote
the sum over their arclength parameterizations.
Substituting Eq.~\eqref{eq:supp_orth_harmonics} gives
\begin{align}
    B_{0p}
    &=
    \int_0^L g_0^*(\ell')g_p(\ell')\,\mathrm d\ell'
    \nonumber\\
    &=
    \int_0^L
    \bigl(\widehat{\mathbf e}_0\cdot
          \widehat{\mathbf n}(\ell')\bigr)
    \bigl(\widehat{\mathbf e}_p\cdot
          \widehat{\mathbf n}(\ell')\bigr)
    e^{-i\mathbf G_p\cdot\mathbf r(\ell')}
    \,\mathrm d\ell'.
    \label{eq:supp_orth_boundary_product}
\end{align}
Likewise,
\begin{equation}
    N_a=\int_0^L |g_a(\ell')|^2\,\mathrm d\ell',
    \qquad
    \eta_{0p}
    =
    \frac{\langle g_0,g_p\rangle_{\partial\Sigma}}
    {\|g_0\|_{\partial\Sigma}\|g_p\|_{\partial\Sigma}}.
    \label{eq:supp_orth_boundary_normalization}
\end{equation}
The boundary integral follows the metal--gap interface and
contains its local normal projections. The complete-cell
cancellation in Eq.~\eqref{eq:supp_orth_cell_integral}
therefore does not apply to this integral.

To display the geometric weighting explicitly, introduce
the boundary Fourier tensor
\begin{equation}
    \overline{\overline F}_{\partial\Sigma}(\mathbf G)
    =
    \oint_{\partial\Sigma}
    \widehat{\mathbf n}\widehat{\mathbf n}^{\mathsf T}
    e^{-i\mathbf G\cdot\mathbf r}\,\mathrm d\ell.
    \label{eq:supp_orth_boundary_tensor}
\end{equation}
Then
\begin{equation}
    B_{0p}
    =
    \widehat{\mathbf e}_0^{\mathsf T}
    \overline{\overline F}_{\partial\Sigma}(\mathbf G_p)
    \widehat{\mathbf e}_p.
    \label{eq:supp_orth_tensor_cross}
\end{equation}
This tensor contains the spatial distribution and orientation
of the metal edges. Its coefficients at nonzero reciprocal
vectors need not vanish, although particular geometries,
polarizations, or symmetry selection rules can make an
individual contraction zero.

Orthogonality of the Floquet basis consequently does not
require the patterned interaction operator to be diagonal
in that basis. The fundamental and sideband harmonics can
contribute to the same Bloch eigenmode through nonzero
inter-order matrix elements.

Within the present reduced closure, $\eta_{0p}$ is a
boundary-weighted geometric measure used to approximate
fundamental-to-sideband coupling. Its possible nonzero value
is consistent with complete-cell Floquet orthogonality.
The integral is evaluated exactly for the prescribed
polygonal boundary, while its use as an electromagnetic
coupling measure remains part of the adopted
edge-projection approximation.

\subsubsection{Analytical evaluation on polygon edges}

For completeness, Eq.~\eqref{eq:supp_B0p_continuous} can be evaluated
analytically on every straight edge.  Parameterize edge $e$ by
\begin{equation}
\mathbf r_e(u)=\mathbf r_e^{-}+u\mathbf d_e,
\qquad
0\leq u\leq1,
\qquad
\mathrm d\ell=L_e\,\mathrm du.
\label{eq:supp_edge_parameter}
\end{equation}
Writing $x_{ep}=\mathbf G_p\!\cdot\!\mathbf d_e$, its exponential
integral is
\begin{align}
I_e(\mathbf G_p)
&=\int_e e^{-i\mathbf G_p\cdot\mathbf r}\,\mathrm d\ell
\nonumber\\
&=L_e e^{-i\mathbf G_p\cdot\mathbf r_e^{-}}
\int_0^1e^{-iu x_{ep}}\,\mathrm du
\nonumber\\
&=L_e e^{-i\mathbf G_p\cdot\mathbf r_e^{-}}
\frac{1-e^{-ix_{ep}}}{ix_{ep}}.
\label{eq:supp_edge_integral_endpoints}
\end{align}
Introducing the edge midpoint
$\mathbf r_e^{c}=(\mathbf r_e^{-}+\mathbf r_e^{+})/2$ and using
\begin{equation}
1-e^{-ix}=2ie^{-ix/2}\sin(x/2),
\label{eq:supp_exponential_identity}
\end{equation}
gives the equivalent midpoint form
\begin{equation}
I_e(\mathbf G_p)
=L_e e^{-i\mathbf G_p\cdot\mathbf r_e^{c}}
\operatorname{sinc}\!\left(\frac{x_{ep}}{2}\right),
\qquad
\operatorname{sinc}(y)=\frac{\sin y}{y}.
\label{eq:supp_edge_integral_sinc}
\end{equation}
The removable singularity is fixed by
$\operatorname{sinc}(0)=1$, so that
$I_e(\mathbf G_p)=L_e e^{-i\mathbf G_p\cdot\mathbf r_e^c}$ whenever
$\mathbf G_p\!\cdot\!\mathbf d_e=0$.  This form avoids numerical
quadrature and the cancellation that would otherwise occur when
$|x_{ep}|\ll1$.

Since the polarization and normal vectors are constant on each edge,
Eqs.~\eqref{eq:supp_B0p_continuous} and
\eqref{eq:supp_Na_continuous} reduce to the finite sums
\begin{align}
B_{0p}
&=\sum_{e\in\mathcal E(\partial\Sigma)}
(\widehat{\mathbf e}_0\!\cdot\!\widehat{\mathbf n}_e)
(\widehat{\mathbf e}_p\!\cdot\!\widehat{\mathbf n}_e)
I_e(\mathbf G_p),
\label{eq:supp_B0p_edges}
\\
N_a
&=\sum_{e\in\mathcal E(\partial\Sigma)}
L_e(\widehat{\mathbf e}_a\!\cdot\!\widehat{\mathbf n}_e)^2.
\label{eq:supp_Na_edges}
\end{align}
The normalized contribution of a single edge is therefore
\begin{equation}
\mathrm d\eta_{0p}^{(e)}
=
\frac{
(\widehat{\mathbf e}_0\!\cdot\!\widehat{\mathbf n}_e)
(\widehat{\mathbf e}_p\!\cdot\!\widehat{\mathbf n}_e)
I_e(\mathbf G_p)
}{\sqrt{N_0N_p}},
\qquad
\eta_{0p}=\sum_e\mathrm d\eta_{0p}^{(e)}.
\label{eq:supp_edge_eta_contribution}
\end{equation}
This is the quantity whose phase colors the boundary in the six overlap maps
of the main text.  If the edges are traversed in the order
$e_1,e_2,\ldots,e_{N_e}$, the displayed cumulative phasor is
\begin{equation}
z_j^{(p)}
=\sum_{n=1}^{j}\mathrm d\eta_{0p}^{(e_n)},
\qquad
z_0^{(p)}=0,
\qquad
z_{N_e}^{(p)}=\eta_{0p}.
\label{eq:supp_cumulative_phasor}
\end{equation}

The bound $|\eta_{0p}|\leq1$ follows directly from
Cauchy--Schwarz.  Indeed, define the boundary functions
\begin{equation}
f_0(\mathbf r)
=\widehat{\mathbf e}_0\!\cdot\!\widehat{\mathbf n}(\mathbf r),
\qquad
f_p(\mathbf r)
=(\widehat{\mathbf e}_p\!\cdot\!\widehat{\mathbf n}(\mathbf r))
e^{-i\mathbf G_p\cdot\mathbf r}.
\label{eq:supp_boundary_functions}
\end{equation}
Because $|e^{-i\mathbf G_p\cdot\mathbf r}|=1$,
\begin{equation}
\|f_0\|_{L^2(\partial\Sigma)}^2=N_0,
\qquad
\|f_p\|_{L^2(\partial\Sigma)}^2=N_p,
\qquad
B_{0p}=\langle f_0,f_p\rangle.
\label{eq:supp_boundary_norms}
\end{equation}
Hence
\begin{equation}
|B_{0p}|^2
\leq\|f_0\|_{L^2}^2\|f_p\|_{L^2}^2
=N_0N_p,
\qquad
|\eta_{0p}|\leq1.
\label{eq:supp_eta_bound}
\end{equation}

The phase in Eq.~\eqref{eq:supp_eta_definition} also records the choice of
spatial origin.  If the complete patch is translated by $\mathbf d$, then
the change of variables $\mathbf r'=\mathbf r+\mathbf d$ gives
\begin{equation}
B_{0p}[\Sigma+\mathbf d]
=e^{-i\mathbf G_p\cdot\mathbf d}B_{0p}[\Sigma],
\qquad
N_a[\Sigma+\mathbf d]=N_a[\Sigma],
\label{eq:supp_translation_BN}
\end{equation}
and consequently
\begin{equation}
\eta_{0p}[\Sigma+\mathbf d]
=e^{-i\mathbf G_p\cdot\mathbf d}\eta_{0p}[\Sigma].
\label{eq:supp_translation_eta}
\end{equation}
Its magnitude is translation invariant, but its phase is not.  The patch is
therefore centered at the symmetry origin before the mirror-related sidebands are combined. The complex edge contributions are summed before taking the modulus, retaining constructive and destructive interference along the boundary in each overlap. The resulting complex overlaps are used to construct the symmetry-adapted coupling in Sec.~\ref{subsec:supp_mirror_projection}

\subsection{Directional turning and the six geometry-derived multipliers}
\label{subsec:supp_directional_turning}

Let $\mathbf r_v$ be a vertex of a reconstructed polygonal boundary loop,
with adjacent vertices $\mathbf r_{v-1}$ and $\mathbf r_{v+1}$.  The
one-sided unit tangents are
\begin{equation}
\widehat{\mathbf t}_v^{-}
=\frac{\mathbf r_v-\mathbf r_{v-1}}
{|\mathbf r_v-\mathbf r_{v-1}|},
\qquad
\widehat{\mathbf t}_v^{+}
=\frac{\mathbf r_{v+1}-\mathbf r_v}
{|\mathbf r_{v+1}-\mathbf r_v|},
\label{eq:supp_vertex_tangents}
\end{equation}
and the corresponding one-sided normals can be taken as
\begin{equation}
\widehat{\mathbf n}_v^{\pm}
=
\begin{pmatrix}
t_{v,y}^{\pm}\\[1mm]-t_{v,x}^{\pm}
\end{pmatrix}.
\label{eq:supp_vertex_normals}
\end{equation}
The exterior signed turning from the incoming to the outgoing edge is
\begin{equation}
\Delta\theta_v
=\operatorname{atan2}
\!\left[
\widehat{\mathbf z}\!\cdot\!
(\widehat{\mathbf t}_v^{-}\!\times\!\widehat{\mathbf t}_v^{+}),
\widehat{\mathbf t}_v^{-}\!\cdot\!\widehat{\mathbf t}_v^{+}
\right],
\qquad
-\pi<\Delta\theta_v\leq\pi.
\label{eq:supp_turning_angle}
\end{equation}
For a counterclockwise-oriented simple closed polygonal loop, the
signed turning theorem gives $\sum_v\Delta\theta_v=2\pi$.
Moreover, $\sum_v|\Delta\theta_v|$ is the polygonal total curvature
of the loop. Milnor's total-curvature theorem states that this
quantity is at least $2\pi$, with equality if and only if the loop is
convex~\cite{Milnor1950}. Hence, for a simple convex loop,
\begin{equation}
\sum_v\Delta\theta_v=2\pi,
\qquad
\sum_v|\Delta\theta_v|=2\pi.
\label{eq:supp_convex_turning}
\end{equation}
For a nonconvex or multicomponent boundary, positive and negative
signed turns can cancel in the first sum but not in the second. The
normalised total absolute turning is therefore defined by
\begin{equation}
T_{\mathrm{abs}}
=\frac{1}{2\pi}\sum_v|\Delta\theta_v|.
\label{eq:supp_Tabs}
\end{equation}
Consequently, $T_{\mathrm{abs}}\geq1$ for a single simple closed
boundary loop, with equality if and only if the loop is convex.
Additional boundary components contribute their own positive total
curvature. Unlike the signed rotation index, $T_{\mathrm{abs}}$ is a
shape-sensitive geometric quantity rather than a topological
invariant.

The turning at a vertex contributes to sideband $p$ only to the extent that
both the fundamental and sideband TE directions load its adjacent edges.
For $a\in\{0,p\}$, define
\begin{equation}
S_a(v;\mathbf k_t)
=\frac12\left[
(\widehat{\mathbf e}_a\!\cdot\!\widehat{\mathbf n}_v^{-})^2
+
(\widehat{\mathbf e}_a\!\cdot\!\widehat{\mathbf n}_v^{+})^2
\right].
\label{eq:supp_Sa}
\end{equation}
Since all vectors entering a scalar product are unit vectors,
\begin{equation}
0\leq S_a(v;\mathbf k_t)\leq1.
\label{eq:supp_Sa_bound}
\end{equation}
The symmetric geometric-mean participation of the two TE directions is
\begin{equation}
w_p(v;\mathbf k_t)
=2\sqrt{S_0(v;\mathbf k_t)S_p(v;\mathbf k_t)}.
\label{eq:supp_vertex_weight}
\end{equation}
It vanishes if either TE direction is tangent to both adjacent edges, it is
unchanged by reversing a boundary normal, and it treats the two coupled
orders symmetrically.  The vertex contribution displayed by the color and
size of the markers in the main-text turning maps is
\begin{equation}
c_v^{(p)}(\mathbf k_t)
=\frac{|\Delta\theta_v|}{2\pi}
w_p(v;\mathbf k_t)
=\frac{|\Delta\theta_v|}{2\pi}
2\sqrt{S_0(v;\mathbf k_t)S_p(v;\mathbf k_t)}.
\label{eq:supp_cv}
\end{equation}
Summing all vertex contributions gives the raw directional turning
\begin{equation}
T_p^{\mathrm{raw}}(\mathbf k_t)
=\sum_v c_v^{(p)}(\mathbf k_t)
=\frac{1}{2\pi}\sum_v
|\Delta\theta_v|
2\sqrt{S_0(v;\mathbf k_t)S_p(v;\mathbf k_t)}.
\label{eq:supp_Traw}
\end{equation}

The six raw values distribute the boundary turning anisotropically, but their
mean can vary with the instantaneous set of TE directions.  We therefore
define
\begin{equation}
\langle T^{\mathrm{raw}}\rangle_6
=\frac16\sum_{r=1}^{6}T_r^{\mathrm{raw}}
\label{eq:supp_Traw_mean}
\end{equation}
and, for $\langle T^{\mathrm{raw}}(\mathbf k_t)\rangle_6>0$,
use the mean-preserving normalization
\begin{equation}
T_p(\mathbf k_t)
=T_{\mathrm{abs}}
\frac{T_p^{\mathrm{raw}}(\mathbf k_t)}
{\langle T^{\mathrm{raw}}(\mathbf k_t)\rangle_6}.
\label{eq:supp_Tp}
\end{equation}
The normalization fixes the shell-average scale exactly, because
\begin{align}
\frac16\sum_{p=1}^{6}T_p
&=\frac16\sum_{p=1}^{6}
T_{\mathrm{abs}}
\frac{T_p^{\mathrm{raw}}}
{\langle T^{\mathrm{raw}}\rangle_6}
\nonumber\\
&=\frac{T_{\mathrm{abs}}}
{\langle T^{\mathrm{raw}}\rangle_6}
\left(\frac16\sum_{p=1}^{6}T_p^{\mathrm{raw}}\right)
\nonumber\\
&=T_{\mathrm{abs}}.
\label{eq:supp_Tp_mean_proof}
\end{align}
Hence $T_{\mathrm{abs}}$ fixes the total geometric-complexity scale, while
$T_p^{\mathrm{raw}}/\langle T^{\mathrm{raw}}\rangle_6$ assigns that
scale among the six reciprocal directions.

The pointwise amplitude multiplier is
\begin{equation}
\widetilde m_p(\mathbf k_t)
=\sqrt{2T_p(\mathbf k_t)}.
\label{eq:supp_m_tilde}
\end{equation}
The square root converts the normalized participation
weight into a coupling-amplitude multiplier.
The factor of two is the fixed normalization adopted
for the present six-order closure.
Together with Eq.~\eqref{eq:supp_Tp_mean_proof}, it gives
\begin{equation}
\frac16\sum_{p=1}^{6}
\widetilde m_p^{\,2}(\mathbf k_t)
=2T_{\mathrm{abs}}.
\label{eq:supp_m_shell_mean}
\end{equation}
This normalization is applied once through
Eq.~\eqref{eq:supp_m_tilde}; the six-order sum and
subsequent mirror projection use these amplitudes directly.
No adjustable scale multiplies the turning multiplier.

For the dispersion calculation along
$\mathbf k_t(s)=s\mathbf K$, the directional multiplier
$\widetilde m_p[\mathbf k_t(s)]$ is evaluated at every
sampled Bloch wavevector and used directly in the
fundamental-to-sideband coupling.
The turning multiplier and boundary Fourier overlap
$\eta_{0p}[\mathbf k_t(s)]$ are evaluated at the same
wavevector throughout the complete
$\Gamma$--$\mathrm K$ path.

The reconstructed geometry remains fixed.
The wavevector dependence of the turning multipliers
arises from the TE polarization projections entering
the vertex participation factors.
The shell-mean normalization therefore holds
pointwise along the path.
At $\Gamma$, the fundamental TE polarization and the
resulting multipliers are defined by the one-sided
limit $s\to0^+$ along the selected propagation direction.

\begin{figure*}[t]
  \centering
  \includegraphics[width=\textwidth]
  {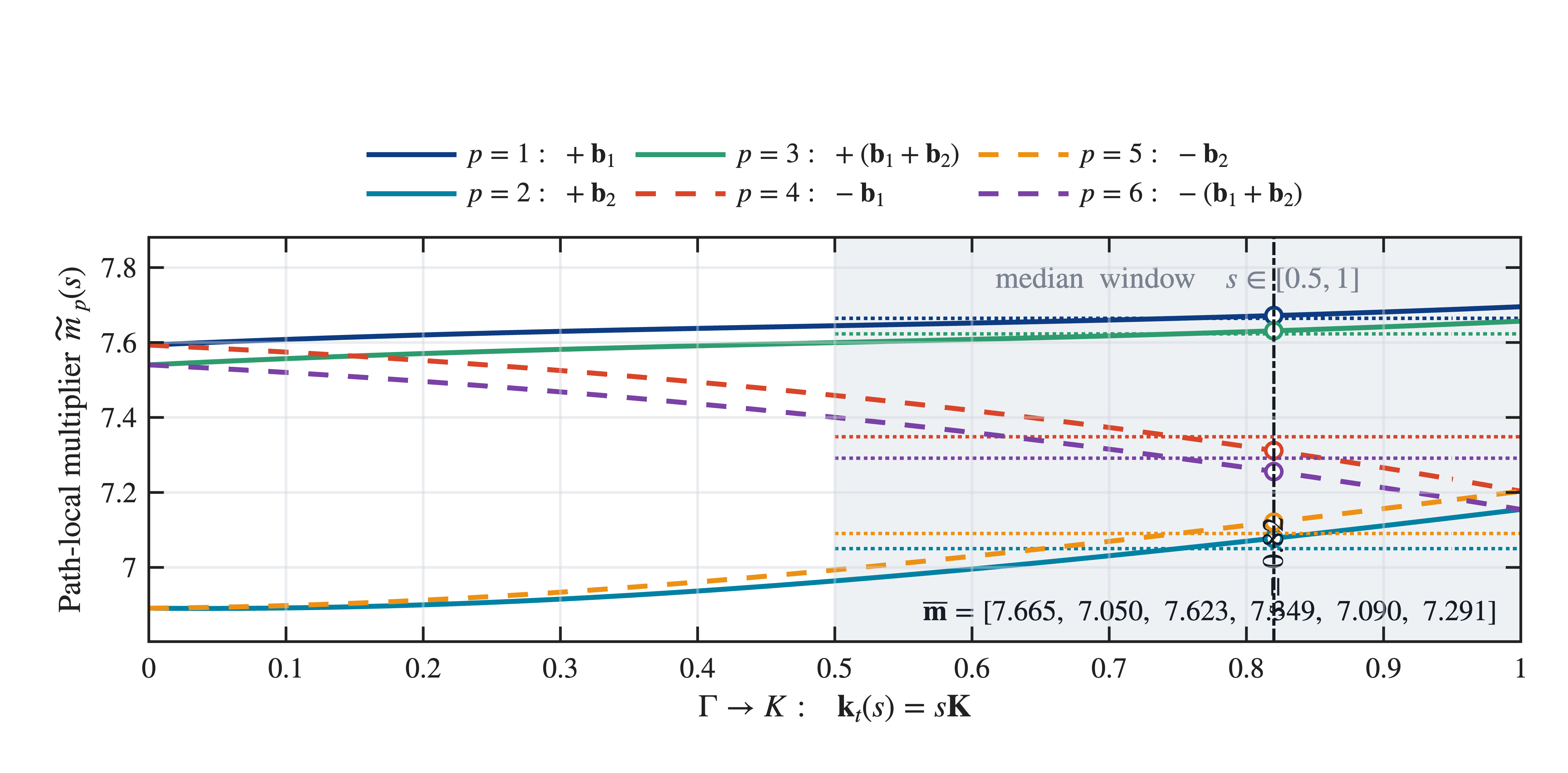}
  \caption{
Directional turning multipliers for the representative
unit cell used in the main-text edgewise maps, along
$\mathbf k_t(s)=s\mathbf K$.
Solid and dashed curves show the pointwise quantities
$\widetilde m_p[\mathbf k_t(s)]$ for the positive and
negative first-shell orders, respectively.
For an optional wavevector-independent approximation,
one constant per order may be defined by
$\overline m_p=
\operatorname*{median}_{s\in[1/2,1]}
\widetilde m_p[\mathbf k_t(s)]$.
The shaded region marks this prescribed sampling
interval; horizontal dotted segments and the displayed
vector give the resulting constants.
Open markers and the vertical dash-dotted line identify
$s=0.82$, used for the representative main-text
visualizations.
The dispersion results reported here use the full
pointwise multipliers.
}
  \label{fig:supp_multiplier_freezing}
\end{figure*}

For the representative unit cell used in the
main-text edgewise maps, the six pointwise multipliers
are shown in Fig.~\ref{fig:supp_multiplier_freezing}.
The point $s=0.82$ identifies the representative
wavevector used for those visualizations; the
dispersion calculation uses the corresponding
pointwise quantities at every sampled wavevector.

Finally, consider a positive uniform dilation of the reconstructed
boundary about an arbitrary fixed point $\mathbf r_c$,
\begin{equation}
\mathbf r_v^{(\lambda)}
=
\mathbf r_c+\lambda(\mathbf r_v-\mathbf r_c),
\qquad \lambda>0,
\end{equation}
while keeping the reciprocal lattice and the dimensionless Bloch-path
coordinate $s$ fixed. The adjacent edge vectors consequently transform as
\begin{equation}
\mathbf d_v^{\pm,(\lambda)}
=
\lambda\mathbf d_v^\pm,
\qquad
\widehat{\mathbf t}_v^{\pm,(\lambda)}
=
\frac{\lambda\mathbf d_v^\pm}
{|\lambda\mathbf d_v^\pm|}
=
\widehat{\mathbf t}_v^\pm .
\label{eq:supp_scale_tangent}
\end{equation}
The unit normals and turning angles are therefore unchanged. Moreover,
the TE directions $\widehat{\mathbf e}_a$ are fixed by the reciprocal
lattice and the selected Bloch point and do not depend on the lengths
of the polygon edges. Hence
\begin{equation}
\widehat{\mathbf n}_v^{\pm,(\lambda)}
=
\widehat{\mathbf n}_v^\pm,
\qquad
\Delta\theta_v^{(\lambda)}
=
\Delta\theta_v,
\qquad
S_a^{(\lambda)}(v;\mathbf k_t)
=
S_a(v;\mathbf k_t).
\label{eq:supp_scale_local}
\end{equation}
It follows successively that
\begin{equation}
T_{\mathrm{abs}}^{(\lambda)}
=
T_{\mathrm{abs}},
\qquad
T_p^{\mathrm{raw},(\lambda)}
=
T_p^{\mathrm{raw}},
\qquad
T_p^{(\lambda)}
=
T_p,
\qquad
\widetilde m_p^{(\lambda)}
=
\widetilde m_p .
\label{eq:supp_scale_turning}
\end{equation}
Equation~\eqref{eq:supp_scale_turning} holds at every
Bloch wavevector along the selected path and therefore
applies directly to the pointwise multipliers used
in the dispersion calculation.
The optional median constants illustrated in
Fig.~\ref{fig:supp_multiplier_freezing} inherit the
same invariance for a fixed sampling interval.

Thus, the directional-turning multiplier contains no metric
edge-length scale. A dilation of the patch at fixed lattice still
changes the boundary Fourier overlap through
$\mathbf G_p\!\cdot\!\mathbf r$, and it also changes
$\overline{\overline C}_{\mathrm{eqv}}$ and the associated decay
constants. The scale invariance established here therefore applies
only to the turning multiplier, not to the complete electromagnetic
dispersion model.

\subsection{Six-order Schur reduction}
\label{subsec:supp_schur_reduction}

Let $a_0$ denote the fundamental TE amplitude and
$a_p$, $p=1,\ldots,6$, the six first-shell amplitudes.  For each retained
order define the detuning
\begin{equation}
D_p
=\gamma_{h_p}-\Gamma_p
\label{eq:supp_Dp}
\end{equation}
and the complex fundamental-to-sideband coupling
\begin{equation}
v_p(k_0,\mathbf k_t)
=
\widetilde m_p(\mathbf k_t)
\sqrt{\Gamma_0\Gamma_p}\,
\eta_{0p}(\mathbf k_t).
\label{eq:supp_vp}
\end{equation}
Here $\Gamma_0\equiv\Gamma_{\mathrm{TL}}$ is the
fundamental sheet-loaded decay used in the main text.
At each Bloch wavevector,
$\widetilde m_p(\mathbf k_t)$ follows from
Eq.~\eqref{eq:supp_m_tilde}, and
$\eta_{0p}(\mathbf k_t)$ follows from the boundary
Fourier integral.
The decays $\Gamma_0$ and $\Gamma_p$ are obtained from
the corresponding TE projections of the same local dyadic.
All quantities are evaluated at the same
$\mathbf k_t$, with their frequency dependence retained
where applicable.

Collect the sideband amplitudes and couplings into
\begin{equation}
\mathbf a_h
=
\begin{pmatrix}
a_1&\cdots&a_6
\end{pmatrix}^{\mathsf T},
\qquad
\mathbf v
=
\begin{pmatrix}
v_1&\cdots&v_6
\end{pmatrix}^{\mathsf T},
\label{eq:supp_vectors}
\end{equation}
and introduce the diagonal retained sideband block
\begin{equation}
\mathbf D
=\operatorname{diag}(D_1,\ldots,D_6).
\label{eq:supp_Dmatrix}
\end{equation}
The scalar TE-projected fundamental-plus-six-order system is
\begin{equation}
\begin{pmatrix}
R_0&-\mathbf v^{\dagger}\\
-\mathbf v&\mathbf D
\end{pmatrix}
\begin{pmatrix}
a_0\\[1mm]\mathbf a_h
\end{pmatrix}
=\mathbf0.
\label{eq:supp_seven_system}
\end{equation}
Written row by row, Eq.~\eqref{eq:supp_seven_system} is
\begin{align}
R_0a_0-\sum_{p=1}^{6}v_p^*a_p&=0,
\label{eq:supp_fundamental_row}
\\
D_pa_p-v_pa_0&=0,
\qquad p=1,\ldots,6.
\label{eq:supp_sideband_rows}
\end{align}
If $D_p\neq0$, each sideband amplitude follows directly from
Eq.~\eqref{eq:supp_sideband_rows},
\begin{equation}
a_p=\frac{v_p}{D_p}a_0.
\label{eq:supp_ap_eliminated}
\end{equation}
Substitution into Eq.~\eqref{eq:supp_fundamental_row} gives
\begin{align}
0
&=R_0a_0
-\sum_{p=1}^{6}v_p^*
\left(\frac{v_p}{D_p}a_0\right)
\nonumber\\
&=
\left[
R_0-\sum_{p=1}^{6}\frac{|v_p|^2}{D_p}
\right]a_0.
\label{eq:supp_substitution_independent}
\end{align}
A nontrivial fundamental amplitude therefore requires
\begin{equation}
R_0-\Sigma_{\mathrm{sb}}=0,
\qquad
\Sigma_{\mathrm{sb}}
=\mathbf v^{\dagger}\mathbf D^{-1}\mathbf v
=\sum_{p=1}^{6}\frac{|v_p|^2}{D_p}.
\label{eq:supp_sigma_ind_basic}
\end{equation}
Using Eqs.~\eqref{eq:supp_Dp} and \eqref{eq:supp_vp},
the unprojected six-order self-energy becomes
\begin{equation}
\Sigma_{\mathrm{sb}}(k_0,\mathbf k_t)
=\Gamma_0
\sum_{p=1}^{6}
\frac{
\widetilde m_p^{\,2}(\mathbf k_t)\,
\Gamma_p\,|\eta_{0p}(\mathbf k_t)|^2
}{
\gamma_{h_p}-\Gamma_p
}.
\label{eq:supp_sigma_ind_expanded}
\end{equation}
All decay quantities in this expression are evaluated
at the same $(k_0,\mathbf k_t)$.
The corrected unprojected dispersion equation is
\begin{equation}
\sqrt{k_t^2-k_0^2}
-\Gamma_0(k_0,\mathbf k_t)
-\Sigma_{\mathrm{sb}}(k_0,\mathbf k_t)
=0.
\label{eq:supp_disp_independent}
\end{equation}

The dimensions are consistent:
$[\Gamma_0]=[\Gamma_p]=[\gamma_{h_p}]=\mathrm{m}^{-1}$,
$[v_p]=\mathrm{m}^{-1}$, and hence
$[|v_p|^2/D_p]=[\Sigma_{\mathrm{sb}}]=\mathrm{m}^{-1}$.
Setting $\mathbf v=\mathbf0$ recovers the local residual
$R_0=0$.

Equation~\eqref{eq:supp_Dmatrix} is the defining reduced-closure
approximation.  Before this approximation, a general first-shell operator
would have the form
\begin{equation}
\mathbf D_{hh}^{\mathrm{exact}}
=
\begin{pmatrix}
D_{11}&D_{12}&\cdots&D_{16}\\
D_{21}&D_{22}&\cdots&D_{26}\\
\vdots&\vdots&\ddots&\vdots\\
D_{61}&D_{62}&\cdots&D_{66}
\end{pmatrix},
\label{eq:supp_D_exact}
\end{equation}
and its Schur self-energy would be
\begin{equation}
\Sigma_{\mathrm{sb}}^{\mathrm{exact}}
=\mathbf v^{\dagger}
(\mathbf D_{hh}^{\mathrm{exact}})^{-1}\mathbf v.
\label{eq:supp_sigma_exact}
\end{equation}
The present model replaces Eq.~\eqref{eq:supp_D_exact} by
\begin{equation}
\mathbf D_{hh}^{\mathrm{exact}}
\longrightarrow
\mathbf D=\operatorname{diag}
(\gamma_{h_1}-\Gamma_1,\ldots,
\gamma_{h_6}-\Gamma_6).
\label{eq:supp_diagonal_closure}
\end{equation}
Thus, fundamental-to-sideband coupling is retained, whereas explicit
sideband-to-sideband matrix elements are not calculated.  The resulting
model is a geometry-derived reduced closure, not an exact reduction of the
full periodic Maxwell operator.

\subsection{Mirror projection on the $\Gamma$--$K$ path}
\label{subsec:supp_mirror_projection}

When reflection about the propagation direction is not a symmetry of the
complete patch-plus-lattice geometry, the six orders are treated as the
independent channels of Eq.~\eqref{eq:supp_sigma_ind_expanded}.  When the
$\Gamma$--$K$ path is a mirror axis of the complete unit-cell geometry, the six retained orders can be organized into three symmetry-adapted pairs associated with the mirror subgroup $C_s=\{E,M\}$.

Along the selected $\Gamma$--$K$ path, define
\begin{equation}
\widehat{\mathbf u}=\frac{\mathbf K}{|\mathbf K|},
\qquad
\mathbf M=2\widehat{\mathbf u}\widehat{\mathbf u}^{\mathsf T}
-\mathbf I_2.
\label{eq:supp_M}
\end{equation}
This gives the in-plane reflection about the propagation
axis, including its one-sided limit at $\Gamma$.  To determine its
action on a TE vector, introduce the $90^{\circ}$ rotation matrix
\begin{equation}
\mathbf J=
\begin{pmatrix}
0&-1\\[1mm]1&0
\end{pmatrix},
\qquad
\widehat{\mathbf e}_{\mathrm{TE}}(\mathbf q)
=\frac{\mathbf J\mathbf q}{|\mathbf q|}.
\label{eq:supp_J}
\end{equation}
For any two-dimensional orthogonal matrix $\mathbf R$,
\begin{equation}
\mathbf R\mathbf J
=\det(\mathbf R)\mathbf J\mathbf R.
\label{eq:supp_RJ_identity}
\end{equation}
Using $|\mathbf R\mathbf q|=|\mathbf q|$, it follows that
\begin{align}
\mathbf R\widehat{\mathbf e}_{\mathrm{TE}}(\mathbf q)
&=\frac{\mathbf R\mathbf J\mathbf q}{|\mathbf q|}
\nonumber\\
&=\det(\mathbf R)
\frac{\mathbf J\mathbf R\mathbf q}{|\mathbf R\mathbf q|}
\nonumber\\
&=\det(\mathbf R)
\widehat{\mathbf e}_{\mathrm{TE}}(\mathbf R\mathbf q).
\label{eq:supp_TE_orthogonal_action}
\end{align}
Since $\det(\mathbf M)=-1$,
\begin{equation}
\mathbf M\widehat{\mathbf e}_{\mathrm{TE}}(\mathbf q)
=-\widehat{\mathbf e}_{\mathrm{TE}}(\mathbf M\mathbf q).
\label{eq:supp_TE_mirror_action}
\end{equation}
The minus sign is essential: the TE polarization is mirror odd when its
wavevector lies in the mirror plane.

The first reciprocal shell is closed under $M$.  Hence there is a
permutation $\pi$ satisfying
\begin{equation}
\mathbf M\mathbf G_p=\mathbf G_{\pi(p)}.
\label{eq:supp_mirror_permutation_definition}
\end{equation}
For the coordinate choice of Eqs.~\eqref{eq:supp_reciprocal_basis} and
\eqref{eq:supp_GammaK_path}, $\mathbf M=\operatorname{diag}(1,-1)$.
Direct application gives
\begin{align}
\mathbf M\mathbf b_1&=\mathbf b_1+\mathbf b_2,
&
\mathbf M(\mathbf b_1+\mathbf b_2)&=\mathbf b_1,
\nonumber\\
\mathbf M\mathbf b_2&=-\mathbf b_2,
&
\mathbf M(-\mathbf b_2)&=\mathbf b_2,
\nonumber\\
\mathbf M(-\mathbf b_1)&=-(\mathbf b_1+\mathbf b_2),
&
\mathbf M[-(\mathbf b_1+\mathbf b_2)]&=-\mathbf b_1.
\label{eq:supp_mirror_pairs_vectors}
\end{align}
With the ordering in Eq.~\eqref{eq:supp_six_orders}, the permutation is
therefore
\begin{equation}
\pi=(1\ 3)(2\ 5)(4\ 6).
\label{eq:supp_pi_cycles}
\end{equation}

These three mirror-related pairs are illustrated geometrically
in Fig.~\ref{fig:supp_first_shell_mirror}(b).

Its matrix representation is
\begin{equation}
\boldsymbol\Pi_M
=
\begin{pmatrix}
0&0&1&0&0&0\\
0&0&0&0&1&0\\
1&0&0&0&0&0\\
0&0&0&0&0&1\\
0&1&0&0&0&0\\
0&0&0&1&0&0
\end{pmatrix},
\qquad
\boldsymbol\Pi_M^2=\mathbf I_6.
\label{eq:supp_Pi_matrix}
\end{equation}

Because the polarization sign in Eq.~\eqref{eq:supp_TE_mirror_action} is
part of the modal basis, the representation of the mirror on the six TE
coefficients is
\begin{equation}
\mathbf D_{\mathrm{TE}}(M)=-\boldsymbol\Pi_M.
\label{eq:supp_DTE_M}
\end{equation}
The fundamental TE mode is odd under the same reflection, so its
$C_s$ character is
\begin{equation}
\chi_{\mathrm{TE}}(E)=1,
\qquad
\chi_{\mathrm{TE}}(M)=-1.
\label{eq:supp_TE_character}
\end{equation}
The character projector onto the sideband sector transforming like the
fundamental is consequently
\begin{align}
\mathbf P_{\mathrm{TE}}
&=\frac{1}{|C_s|}
\sum_{g\in C_s}
\chi_{\mathrm{TE}}^*(g)\mathbf D_{\mathrm{TE}}(g)
\nonumber\\
&=\frac12
\left[
\mathbf I_6
+(-1)(-\boldsymbol\Pi_M)
\right]
\nonumber\\
&=\frac12(\mathbf I_6+\boldsymbol\Pi_M).
\label{eq:supp_PTE_derivation}
\end{align}
Thus the apparently symmetric coefficient combinations represent the
mirror-odd physical TE sector; the physical minus sign has already been
carried by the transformed TE basis vectors.

An orthonormal basis for $\operatorname{range}(\mathbf P_{\mathrm{TE}})$ is
\begin{equation}
\mathbf U
=\frac{1}{\sqrt2}
\begin{pmatrix}
1&0&0\\
0&1&0\\
1&0&0\\
0&0&1\\
0&1&0\\
0&0&1
\end{pmatrix},
\qquad
\mathbf U^{\dagger}\mathbf U=\mathbf I_3,
\qquad
\mathbf U\mathbf U^{\dagger}=\mathbf P_{\mathrm{TE}}.
\label{eq:supp_U_explicit}
\end{equation}
Hence the six first-shell orders reduce to three TE-compatible mirror
channels.  Write
\begin{equation}
\mathbf a_h=\mathbf U\mathbf b,
\qquad
\mathbf b\in\mathbb C^3.
\label{eq:supp_ah_Ub}
\end{equation}
Substitution into Eq.~\eqref{eq:supp_seven_system}, followed by left
multiplication of the sideband rows by $\mathbf U^{\dagger}$, gives the
reduced system
\begin{equation}
\begin{pmatrix}
R_0&-(\mathbf U^{\dagger}\mathbf v)^{\dagger}\\
-\mathbf U^{\dagger}\mathbf v&
\mathbf U^{\dagger}\mathbf D\mathbf U
\end{pmatrix}
\begin{pmatrix}
a_0\\[1mm]\mathbf b
\end{pmatrix}
=\mathbf0.
\label{eq:supp_reduced_system}
\end{equation}
Define
\begin{equation}
\mathbf v_r=\mathbf U^{\dagger}\mathbf v,
\qquad
\mathbf D_r=\mathbf U^{\dagger}\mathbf D\mathbf U.
\label{eq:supp_vr_Dr}
\end{equation}
The reduced sideband equation is
\begin{equation}
\mathbf D_r\mathbf b=\mathbf v_r a_0,
\qquad
\mathbf b=\mathbf D_r^{-1}\mathbf v_r a_0,
\label{eq:supp_b_eliminated}
\end{equation}
and the corresponding Schur complement is
\begin{equation}
\Sigma_{\mathrm{sb}}^{\mathrm{proj}}
=\mathbf v_r^{\dagger}\mathbf D_r^{-1}\mathbf v_r
=(\mathbf U^{\dagger}\mathbf v)^{\dagger}
(\mathbf U^{\dagger}\mathbf D\mathbf U)^{-1}
(\mathbf U^{\dagger}\mathbf v).
\label{eq:supp_sigma_projected}
\end{equation}
The projected dispersion equation is
\begin{equation}
\sqrt{k_t^2-k_0^2}
-\Gamma_0
-\Sigma_{\mathrm{sb}}^{\mathrm{proj}}
=0.
\label{eq:supp_disp_projected}
\end{equation}

For the explicit basis in Eq.~\eqref{eq:supp_U_explicit},
\begin{equation}
\mathbf v_r
=\frac{1}{\sqrt2}
\begin{pmatrix}
v_1+v_3\\
v_2+v_5\\
v_4+v_6
\end{pmatrix},
\qquad
\mathbf D_r
=\frac12
\begin{pmatrix}
D_1+D_3&0&0\\
0&D_2+D_5&0\\
0&0&D_4+D_6
\end{pmatrix}.
\label{eq:supp_explicit_vr_Dr}
\end{equation}
Therefore Eq.~\eqref{eq:supp_sigma_projected} can also be written as
\begin{equation}
\Sigma_{\mathrm{sb}}^{\mathrm{proj}}
=
\frac{|v_1+v_3|^2}{D_1+D_3}
+
\frac{|v_2+v_5|^2}{D_2+D_5}
+
\frac{|v_4+v_6|^2}{D_4+D_6}.
\label{eq:supp_sigma_projected_pairs}
\end{equation}
For an exactly mirror-symmetric geometry, with the patch centred
at the symmetry origin, $D_1=D_3$, $D_2=D_5$, and $D_4=D_6$.
In the adopted TE convention, mirror symmetry also gives
$v_1=v_3$, $v_2=v_5$, and $v_4=v_6$.
For each mirror-related pair $(p,\pi(p))$, the coupling to the
normalised symmetric coefficient combination is therefore
$(v_p+v_{\pi(p)})/\sqrt{2}=\sqrt{2}\,v_p$, whereas the coupling
to the complementary antisymmetric combination vanishes:
$(v_p-v_{\pi(p)})/\sqrt{2}=0$.
Equation~\eqref{eq:supp_sigma_projected_pairs} thus retains
the full self-energy contribution $2|v_p|^2/D_p$ of each
original pair, while the antisymmetric combination is
decoupled from the fundamental mode.

Whenever $\mathbf D_r$ is nonsingular, the ordinary
inverse in Eq.~\eqref{eq:supp_sigma_projected} applies.
To examine numerical rank, let
\begin{equation}
\mathbf D_r
=\mathbf Q\operatorname{diag}(d_1,d_2,d_3)\mathbf Q^{\dagger}.
\label{eq:supp_Dr_eigendecomposition}
\end{equation}
A Moore--Penrose inverse with a fixed machine-level cutoff can be defined by
\begin{equation}
\mathbf D_r^{+}
=\mathbf Q\operatorname{diag}(d_1^{+},d_2^{+},d_3^{+})
\mathbf Q^{\dagger},
\qquad
d_j^{+}=
\begin{cases}
d_j^{-1},&|d_j|>\tau_D D_{\mathrm{ref}},\\
0,&|d_j|\leq\tau_D D_{\mathrm{ref}},
\end{cases}
\label{eq:supp_pseudoinverse}
\end{equation}
where $D_{\mathrm{ref}}=\max_j|d_j|$ and $\tau_D$ is
fixed independently of geometry.
The cutoff provides a diagnostic of numerical rank loss.
At a true singularity of $\mathbf D_r$, the inverse-based
Schur expression is not applicable; the dispersion must
be assessed from the coupled system in
Eq.~\eqref{eq:supp_reduced_system}.
A singular channel must not be discarded through the
pseudoinverse and interpreted as an equivalent
dispersion condition.

Combining the preceding results, the complete calculation is
\begin{equation}
\begin{aligned}
\left\{
\overline{\overline C}_{\mathrm{eqv}},\partial\Sigma
\right\}
&\longrightarrow
\left\{
\Gamma_0,\Gamma_p,\eta_{0p},
T_p^{\mathrm{raw}},T_p,
\widetilde m_p(\mathbf k_t)
\right\}
\\
&\longrightarrow
\Sigma_{\mathrm{sb}}
\longrightarrow
R_0-\Sigma_{\mathrm{sb}}=0.
\end{aligned}
\label{eq:supp_complete_chain}
\end{equation}
The geometric factors and polarization projections
are evaluated at the same Bloch wavevector, with the
frequency dependence of the decay quantities retained.
The unprojected expression in
Eq.~\eqref{eq:supp_sigma_ind_expanded} is used when
the propagation-axis reflection is not a symmetry of
the complete patch-plus-lattice geometry.
When that reflection belongs to the little group of
the complete geometry, the projected self-energy in
Eq.~\eqref{eq:supp_sigma_projected} is used for
$\Sigma_{\mathrm{sb}}$ in the dispersion residual.
The projected expression retains the contributions of
all six original orders through three symmetry-adapted
channels.
The symmetry reduction introduces no additional
multiplicity factor or fitted coupling parameter.

\subsection{Long-wavelength consistency}
\label{subsec:supp_long_wavelength_consistency}

For fixed geometry with finite, nondispersive loading,
the projections $A_0$ and $A_p$ remain finite and
positive.
The expansion in Eq.~\eqref{eq:supp_Gamma_small} gives
\begin{equation}
\Gamma_0\simeq\frac{A_0k_0^2}{2},
\qquad
\Gamma_p\simeq\frac{A_pk_0^2}{2},
\qquad k_0A_0,\,k_0A_p\ll1.
\label{eq:supp_loaded_decay_low_frequency}
\end{equation}
Thus both sheet-loaded decays vanish quadratically
with frequency.
As $k_0,\mathbf k_t\to0$ along the lowest bound branch,
each retained nonzero reciprocal order satisfies
\begin{equation}
D_p=\gamma_{h_p}-\Gamma_p
\longrightarrow |\mathbf G_p|=G_0>0.
\label{eq:supp_Dp_low_frequency}
\end{equation}
The normalized overlaps obey
$|\eta_{0p}|\leq1$ by Eq.~\eqref{eq:supp_eta_bound},
and Eq.~\eqref{eq:supp_m_shell_mean} keeps the six
turning multipliers bounded.
The unprojected self-energy therefore satisfies
\begin{equation}
\frac{\Sigma_{\mathrm{sb}}}{\Gamma_0}
=
\sum_{p=1}^{6}
\frac{
\widetilde m_p^{\,2}(\mathbf k_t)\,
\Gamma_p\,|\eta_{0p}(\mathbf k_t)|^2
}{D_p}
\longrightarrow0.
\label{eq:supp_relative_self_energy_limit}
\end{equation}
After mirror projection,
$\mathbf D_r\to G_0\mathbf I_3$, so the same
relative suppression holds for
$\Sigma_{\mathrm{sb}}^{\mathrm{proj}}$.
The sideband correction consequently becomes negligible
relative to the local loading, recovering the leading
local long-wavelength dispersion.
The individual turning multipliers need not approach
unity or vanish at $\Gamma$.

\bibliographystyle{apsrev4-2}
\bibliography{PRX_cleaned_references}

\end{document}